\documentclass[fleqn,usenatbib]{mnras}

\usepackage{newtxtext,newtxmath}

\usepackage[T1]{fontenc}

\DeclareRobustCommand{\VAN}[3]{#2}
\let\VANthebibliography\thebibliography
\def\thebibliography{\DeclareRobustCommand{\VAN}[3]{##3}\VANthebibliography}

\usepackage{graphicx}	
\usepackage{amsmath}	
\usepackage{xcolor}
\usepackage{float}
\usepackage{upgreek}
\usepackage{subfig}

\usepackage[normalem]{ulem}
\usepackage{cancel}

\title[A search for planetary companions around pulsars]{A search for planetary companions around 800 pulsars from the Jodrell Bank pulsar timing programme}

\author[I. C. Ni\c{t}u et al.]{
Iuliana C. Ni\c{t}u,$^{1}$\thanks{E-mail: iuliana-camelia.nitu@manchester.ac.uk}
Michael J. Keith,$^{1}$
Ben W. Stappers,$^{1}$
Andrew G. Lyne,$^{1}$
Mitchell B. Mickaliger$^{1}$
\\
$^{1}$Jodrell Bank Centre for Astrophysics, Department of Physics and Astronomy, The University of Manchester, Manchester M13 9PL, UK\\
}

\date{Accepted XXX. Received YYY; in original form ZZZ}

\pubyear{2022}

\begin{document}
\label{firstpage}
\pagerange{\pageref{firstpage}--\pageref{lastpage}}
\maketitle

\begin{abstract}
We have searched for planetary companions around 800 pulsars monitored at the Jodrell Bank Observatory, with both circular and eccentric orbits of periods between $20$\,days and $17$\,years and inclination-dependent planetary masses from $10^{-4}$ to $100\,\mathrm{M}_{\oplus}$.
Using a Bayesian framework, we simultaneously model pulsar timing parameters and a stationary noise process with a power-law power spectral density.
We put limits on the projected masses of any planetary companions, which reach as low as 1/100th of the mass of the Moon ($\sim 10^{-4}\,\mathrm{M}_{\oplus}$). We find that two-thirds of our pulsars are highly unlikely to host any companions above $2-8\,\mathrm{M}_{\oplus}$. Our results imply that fewer than $0.5\%$ of pulsars could host terrestrial planets as large as those known to orbit PSR B1257$+$12 ($\sim4\,\mathrm{M}_{\oplus}$); however, the smaller planet in this system ($\sim0.02\,\mathrm{M}_{\oplus}$) would be undetectable in $95\%$ of our sample, hidden by both instrumental and intrinsic noise processes, although it is not clear if such tiny planets could exist in isolation.
We detect significant periodicities in 15 pulsars, however we find that intrinsic quasi-periodic magnetospheric effects can mimic the influence of a planet, and for the majority of these cases we believe this to be the origin of the detected periodicity.
Notably, we find that the highly periodic oscillations in PSR B0144$+$59 are correlated with changes in the pulse profile and therefore can be attributed to magnetospheric effects.
We believe the most plausible candidate for planetary companions in our sample is PSR J2007$+$3120.
\end{abstract}

\begin{keywords}
pulsars: general; pulsars: individual: PSR B0144$+$59, PSR J1947$+$1957, PSR J2007$+$3120; planets and satellites: detection; methods: data analysis
\end{keywords}







\section{Introduction}
The rapid and regular pulsations observed from pulsars allows us to track their rotation to a small fraction of the spin period over long timescales.
This technique of pulsar timing allows us to use them as tools for a wide range of physics and astrophysics applications, and makes our observations very sensitive to perturbations due to orbital motion, even from Earth or Moon mass planets \citep{Thorsett1992,Cordes1993}. 

Notably, the first extrasolar planets were discovered around the pulsar PSR B1257$+$12 \citep{Wolszczan1992,Wolszczan1994}, and are still among the lowest mass extrasolar planets known with masses of $0.020(2)\,\mathrm{M}_{\oplus}$, $4.3(2)\,\mathrm{M}_{\oplus}$ and $3.9(2)\,\mathrm{M}_{\oplus}$ \citep{Konacki2003}. Since then, almost $5000$ exoplanets have been discovered by various methods (see e.g. The Extrasolar Planets Encyclopaedia\footnote{Found at \url{exoplanet.eu}}). However, in contrast with this considerable increase in the number of exoplanets discovered around other types of stars, only five other pulsars have been confirmed to have planetary-mass companions. Of the pulsars with planetary-mass companions, four are ultra-low mass carbon white dwarfs, or ``diamond planets'', around pulsars J1719$-$1438 \citep{Bailes2011}, J0636$+$5128 \citep{Stovall2014}, J1311$-$3430 \citep{Romani2012, Pletsch2012} and J2322$-$2650 \citep{Spiewak2018}, respectively. Lastly, pulsar B1620$-$26 is part of a triple system, including a super-Jupiter mass planet; however, the location of this system in the globular cluster M4 suggests the planet was captured, rather than formed around the pulsar \citep{Sigurdsson2003}, a process unlikely to happen for pulsars outside of the high density of such a cluster.
PSR B1257$+$12 therefore remains as a unique example of a system of approximately Earth-mass planets orbiting a neutron star.

There have been several previous systematic searches for low-mass planetary companions around pulsars, none of which have reported any detections. \citet[][K15 hereafter]{Kerr2015_1} searched for planets around 151 (young) pulsars, a sample two orders of magnitude larger than previous efforts by~\citet{Thorsett1992}. K15 concluded that no planets more massive than $0.4\,\mathrm{M}_{\oplus}$ and with periods of less than one year are found around their sampled pulsars. 
Most recently, \citet{Behrens2020} used Pulsar Timing Array quality data to set mass constraints as low as a Moon mass ($\sim 0.01\,\mathrm{M}_{\oplus}$) for possible planetary masses around 45 millisecond pulsars (MSPs), for periods between 7 and 2000 days, also reporting no detectable planets.

The apparent rarity of systems like that of PSR B1257$+$12 may well be a consequence of the extreme conditions in which pulsars form. There have been several mechanisms proposed for the formation of planets around pulsars \citep[see e.g.][]{Podsiadlowski1993, Phillips1994}. Of these, three different types of scenarios can be distinguished. 

In some of the proposed scenarios, the planets had formed around a normal star and were then captured by a neutron star through the collision of the two stars \citep{Podsiadlowski1991} or survived the subsequent evolution of the initial system towards a neutron star system \citep{Bailes1991}. The resulting system would consist of a normal pulsar with planetary companions in eccentric orbits. These types of events would be very rare as they require fine tuning of environment conditions for the planets to survive or for the collision to take place outside of a globular cluster. 

In the majority of the proposed scenarios, planets form around pulsars from a disk of material around the neutron star. In the simplest case, some of the mass ejected from the supernova in which the pulsar was formed does not fully escape the gravitational pull of the star and falls back into a disk \citep{Bailes1991, Lin1991}; subsequently, a normal pulsar, surrounded by relatively small mass planets in circular orbits, is expected. In other models, a more massive circumstellar disk is a product of the neutron star distrupting its binary companion. If this companion was a white dwarf \citep{Podsiadlowski1991, vandenHeuvel1992}, we might expect to form a millisecond pulsar with planets in circular orbits.

A third type of mechanisms proposes that the planets are the final stage of evolution of binary millisecond pulsars. Depending on the model, the initial pulsar companion is either partially evaporated \citep{Bailes1991, Krolik1991, Rasio1992} or fully disrupted \citep{Stevens1992} by the influence of the neutron star, creating either a singular planet from the remains of the original companion (such as a ``diamond planet''), or a debris disk around the neutron star, which would then collapse to form planets in circular orbits.

Overall, there are a large number of proposed formation paths of planets around pulsars, and therefore large scale searches of planetary-mass companions and their orbital parameters are crucial to constraining and determining the feasibility of various models.
However, although the time of arrival (ToA) measurements of individual observations can be very precise, the detectability of planets around pulsars is also limited by the presence of so-called ``timing noise'' that manifests as a long-term red noise process in the rotation of the pulsar. Models of timing noise typically assume a stochastic stationary red noise process (e.g. \citealp{Shannon2010}),  but there is also strong evidence for quasi-periodic switching between spin-down states, which has been observed to be correlated to pulse shape changes in some pulsars \citep{Hobbs2010,Lyne2010}. This presents a further challenge in searching for planetary companions, as it can not only mask binary signatures, but also mimic them. A careful investigation is therefore necessary, as there have been previous claims of planet detections around pulsars \citep[e.g. PSR B0329$+$54;][]{Shabanova1995}, later proved to be caused by intrinsic pulsar timing noise \citep{Konacki1999}.

Likewise, previous attempts have been made to explain two periodicities seen in PSR B1828$-$11 using a planetary model, but \citet{Bailes1993} found that modelling this behaviour using only two simple sinusoids was not enough to fully account for the observed variation in the ToAs. The quasi-periodic behaviour in B1828$-$11 is now known to be of intrinsic magnetospheric origin \citep{Lyne2010, Stairs2019}.
To better understand the population of planetary systems in pulsars, it is therefore important to both look at a large sample of pulsars that have been observed for long time-spans, and also to carefully model the pulsar timing noise to identify any purely periodic signals arising from planetary companions.
The pulsar timing archive at the Jodrell Bank Observatory (JBO) is an ideal dataset for this project, consisting of regular observations of almost 800 pulsars with the 76-m Lovell Telescope. This represents nearly a third of all known pulsars, with observations spanning up to 50 years for each pulsar.

In this work we apply Bayesian techniques for modelling timing noise to the available JBO data, and conduct the largest systematic search for pulsar planetary companions to date, as well as set limits to any existing population of planetary-mass bodies around pulsars. 
This paper is structured as follows. In Section~\ref{sec:observations}, the observational properties are summarised. Section~\ref{sec:binary} introduces the parameters of a binary (planetary) system. The method and relevant pulsar parameters used in this analysis are further described in Section~\ref{sec:analysis}. The main results are presented and discussed in Section~\ref{sec:results}.

\section{Observations} \label{sec:observations}
The dataset used in this work is composed of observations of approximately 800 pulsars from the Jodrell Bank pulsar timing database. All pulsars have observations from the 76-m Lovell Telescope. Observations since April 2009 are recorded using a digital filterbank (a clone of the DFB described in \citealp{Manchester2013}), and processed using \textsc{psrchive} to form time and frequency averaged pulse profiles with a 200--400 MHz bandwidth centred at 1520 MHz (depending on RFI excision). RFI is excised by manual removal of small outlier segments of the band (``channels'') and portions of the observations (``sub-integrations''). ToAs are generated using \textsc{psrchive} with an analytic standard template derived from the sum of all DFB observations. Prior to 2009 the data used a range of receivers and were recorded with several generations of analogue systems, processed using legacy software to directly generate ToAs.
For 50 pulsars, daily observations with the 25-m Mark II Telescope at Jodrell Bank were used to fill intervals when the Lovell Telescope was undergoing upgrades and repairs. The recording and processing setup is identical to that used for the Lovell observations. For more details on the previous receivers, see \citet{Hobbs2004}.

For 16 pulsars these data are supplemented with observations made using the 13-m ``42-ft'' telescope at the Jodrell Bank Observatory. These consist of daily observations since 1990 (depending on the pulsar) with a 5~MHz bandwidth centred at 610 MHz. Since 2011 these observations used the ``cobra2'' coherent dedispersion backend. Additionally the earliest ToAs for 29 pulsars come from observations made by the NASA Deep Space Network (\citealp{Downs1983}; \citealp{Downs1986}).

In total the dataset consists of 800 pulsars, of which 104 are millisecond pulsars. In total there are $\sim 30\,000$ Lovell ToAs (with $1/3$ from the DFB backend) covering $\sim 17\,000$ years of pulsar rotational history, implying an average cadence of 20 days. However, it should be noted that the cadence varies significantly from pulsar to pulsar depending on the science case for any individual source.

Observations do not record absolute phase, so an initial timing solution is used to predict/estimate absolute phase.  With sufficient observational cadence, this can be done unambiguously and a pulse number assigned to each ToA.  Phase ambiguities arise at large glitches (instantaneous changes in spin-down), and these are left as free parameters as part of the pulsar timing model. Further details of the dataset and the study of the pulsar timing noise and astrometric parameters will be included in future publications.

\section{Modelling the influence of planets} \label{sec:binary}

When a pulsar is part of a binary system (either with a star or a planet), it revolves around the centre of mass of the system, moving with respect to the observer on Earth. This influences the arrival time of the signal to the observer, and is known as R\o{}mer delay. In the timing model, the R\o{}mer delay is expressed as \citep{Blandford1976}
\begin{equation}
    \Delta_\mathrm{RB} (t)= \frac{a_\mathrm{PSR}\,\sin{i}}{c} \, [\,(\cos{E(t)}-e)\,\sin{\omega} + \sin{E(t)}\,\sqrt{1-e^2}\,\cos{\omega}\,],
    \label{eq:roemer}
\end{equation}
where $t$ is the ToA; $i$ is the inclination angle of the plane of the orbit in the sky; $e$ is the eccentricity; $a_\mathrm{PSR}$ is the semi-major axis of the pulsar around the centre of mass, given by
\begin{equation}
    a_\mathrm{PSR} = \left[\frac{G \, m^3 \, P_{\mathrm{b}}^2}{4\pi^2 \,(m+M_\mathrm{PSR})^2}\right]^{1/3},
    \label{eq:asini}
\end{equation}
where $G$ is the gravitational constant, $M_\mathrm{PSR}$ is the mass of the pulsar, $m$ is the mass of the companion, and $P_\mathrm{b}$ is the orbital period; $\omega$ is the argument of the periapsis, i.e. the angle of the periapsis with respect to the plane of the sky; $E(t)$ is the eccentric anomaly at time $t$ and is related to the true anomaly, $A_\mathrm{T}(t)$, i.e. the angle of an object on the elliptical orbit with respect to the periapsis by \citep[see e.g.][]{FundAstro}
\begin{equation}
    \cos{E(t)} = \frac{e+\cos{A_\mathrm{T}(t)}}{1+e\cos{A_\mathrm{T}(t)}}.
    \label{eq:cosE}
\end{equation}

The mean anomaly ($M(t)$), which is the fraction of the orbit that has been completed at time $t$ since the orbiting body passed periapsis at time $t_0$, expressed as an angle between $0$ and $2\pi$, can also be expressed in terms of the eccentric anomaly by
\begin{equation}
    M(t) \equiv \frac{2\pi}{P_{\mathrm{b}}} \, (t-t_0) = E(t) - e\,\sin{E(t)}.
    \label{eq:meanan}
\end{equation}
If $t_0$ is known, the two expressions of $M(t)$ in Eq.~\ref{eq:meanan} can be used to find $E(t)$, which can then be used in Eq.~\ref{eq:roemer} to calculate the R\o{}mer delay at each ToA. However, if $P_{\mathrm{b}}$ is shorter than the time span of the data, there are multiple (degenerate) solutions for $t_0$, as there are multiple periapsis crossings in the observed time. To avoid this degeneracy, we considered a phase parameter ($\phi$), taking values between $0$ and $1$, such that
\begin{equation}
    \phi = \frac{A_\mathrm{T}(t_\mathrm{REF})}{2\pi},
    \label{eq:phi}
\end{equation}
where we chose a reference time $t_\mathrm{REF} = 55000\,\mathrm{MJD}$. The value of $\phi$ can be used to find a single $t_0$ solution (which is the periapsis crossing closest to $t_\mathrm{REF}$), as follows. Eq.~\ref{eq:phi} gives $A_\mathrm{T}(t_\mathrm{REF})$ straightforwardly from $\phi$. Then, we compute $\cos{E(t_\mathrm{REF})}$ using Eq.~\ref{eq:cosE} at $t \equiv t_\mathrm{REF}$ and the eccentricity value $e$. Lastly, equating the two expressions for $M(t_\mathrm{REF})$ as in Eq.~\ref{eq:meanan}, and given $E(t_\mathrm{REF})$ from the previous step, $e$ and $P_{\mathrm{b}}$, we can determine $t_0$. 

\section{The analysis} \label{sec:analysis}

\subsection{Method overview}

We implemented a Bayesian method to analyse pulsar timing data, by adapting the existing pulsar timing software \textsc{enterprise} \citep{Ellis2019}. This method consists of marginalising over parameters describing the deterministic timing model, and simultaneously fitting for white noise, red noise and a planet orbit. This approach is well-suited for a systematic search of planets around pulsars, for placing limits on the mass of any orbiting celestial bodies, and therefore for inferring statistically significant properties of the population of these planets. 

In practice, a function containing the Keplerian orbit parameters was added to the \textsc{enterprise} likelihood function, allowing for simultaneous fitting with the existing functionality; this is referred in this work as the ``extended \textsc{enterprise}'' and integrated into the \textsc{run\_enterprise} pulsar analysis toolset \citep{run_enterprise}.
The method used consists, broadly, of the following steps applied to the timing data of one pulsar:
\begin{enumerate}
    \item The orbital period prior is divided into equally-spaced bins in log-period, and the extended \textsc{enterprise} is run for each of these bins, fitting for white noise, red noise, and planet orbital parameters (see Section~\ref{sec:planetfit}). 
    \item For each period bin, 3-$\sigma$ ``detections'' in the mass posteriors (samples) are flagged for further investigation. To make our search for periodicities more thorough, and to counter the smearing effects of large period ranges, the mass posteriors in each of these period bins are divided into five equally-spaced period slices (``sub-bins''), and the 3-$\sigma$ check performed again for each of these slices. Note that, while these sub-bins are not guaranteed to each have enough mass samples to give reliable mass limits, we manually verify that any detections seen are genuine, and not due to a small number of mass samples. We therefore only use the mass posteriors in these sub-bins to check for detections, not to estimate mass limits.
    \item To further study the population of planets around pulsars, we also estimate limits for the mass of planets detectable with our method and data, as the 95-percentile in the mass posteriors in each period bin.

\end{enumerate}

\subsection{Timing analysis}
\subsubsection{Pulsar parameters}

There are a number of deterministic parameters that are known to affect the timing of most pulsars. For each pulsar in our analysis we fit for the spin frequency ($\nu$) and two frequency derivatives ($\dot{\nu}$ and $\ddot{\nu}$); astrometric parameters (i.e. position, proper motion); orbital parameters of known binary companions; and the clock offsets (``jumps'') between different telescope backend systems, if relevant. We then analytically marginalise over these parameters, using a design matrix computed by \textsc{tempo2} \citep{tempo2006}. Glitch parameters as given in \citet{Basu2021} were also used.

\subsubsection{Timing noise}
Pulsar timing is also affected by noise, which we model with a white-noise component, with parameters EFAC \citep{Hobbs2006} and EQUAD \citep{Liu2012}, and a power-law red-noise component. We characterise the red noise by the index $\gamma$ and the log-amplitude $\log_{10}(A_\mathrm{red})$, as used in the typical power-law Fourier-domain Gaussian-process model of timing noise \citep[see][]{Lentati2014,vanHaasteren2011},
\begin{equation}
P(f) = \frac{A_\mathrm{red}^2}{12\pi^2} \, \left(\frac{f}{1 \mathrm{yr}^{-1}}\right)^{-\gamma} \mathrm{yr}^3,
\label{eq:redpower}
\end{equation}
where $f$ is the frequency in the Fourier domain, and $P(f)$ is the one-sided power spectral density.

\subsection{Planet fitting} \label{sec:planetfit}

The set of orbital parameters we use to characterise a planet influence is: the pulsar mass~($M_\mathrm{PSR}$), the ``projected mass'' of the planet~($m\sin{i}$), the orbital period~($P_{\mathrm{b}}$), the eccentricity~($e$), the argument of periapsis~($\omega$) and the planet phase on its orbit at time $55000\,\mathrm{MJD}$~($\phi$).

We fix $M_\mathrm{PSR}$ at the representative value of $1.4$\,M$_{\odot}$ \citep[e.g. ][]{Lattimer2012}, and sample over the other orbital and noise parameters using the \textsc{emcee} ensemble MCMC sampler \citep{Foreman-Mackey2013,Foreman-Mackey2019} within \textsc{enterprise}. 
Uniform priors are used for the orbital parameters $\omega$, $e$, $\phi$,  allowing $\omega$ between $0$ and $2\pi$, $e$ between $0$ and $0.9$, $\phi$ between $0$ and $1$, and for each of the nine $P_{\mathrm{b}}$ bins. For $m\sin{i}$, a log-prior is used as we want to explore several orders of magnitude between $10^{-4}$ and $100\,\mathrm{M}_{\oplus}$. 
Note that $\omega$ and $\phi$ can be highly correlated, as they both depend strongly on the position of the periapsis, which is poorly defined for low-eccentricity orbits.

For each pulsar, the analysis is split into eight period ranges, uniform in log-space, as well as a narrower range around $P_{\mathrm{b}} = 1\,\mathrm{year}$, since fitting for the pulsar position in the timing model can absorb power at the orbital period of the Earth. Table~\ref{tab:pranges} shows the period ranges ($\min P_{\mathrm{b}} - \max P_{\mathrm{b}}$), as well as the corresponding initial number of samples per Markov chain used by the ensemble sampler ($N$). The number of samples for each period range was chosen  as a compromise between the running time of the sampler and having a sufficient number of samples such that injected planet signals were consistently found. 
For each run of $N$ samples, $64$ chains are used, which after ``burnin'' and ``thinning'' results in $4.8 N$ independent samples of the posterior. 

To check for 3-$\sigma$ detections and compute the 95-percentile mass limit, the mass posterior is re-weighted to reflect a uniform linear prior and detections are flagged if the mean is more than 3-$\sigma$ away from zero.

An exact treatment would require solving Eq.~\ref{eq:meanan} for the value of $E(t)$, at each ToA and each $e$. Due to the large amount of data to process, we instead interpolate over pre-computed values of $E(t)$ over a fixed grid of 1,000 mean anomaly values and 90,000 eccentricity values, chosen such that the maximum fractional error is less than $10^{-8}$. This approximation reduces the run-time by a factor of $\sim 6$.

\begin{table}
	\centering
	\caption{Table showing the period ranges and corresponding number of samples.}
	\label{tab:pranges}
	\begin{tabular}{ cccc} 
		\hline
		$\min P_{\mathrm{b}}$ [d] &$-$& $\max P_{\mathrm{b}}$ [d] & $N$\\
		\hline
		21.3 &$-$& 42.5 & $2\times 10^4$\\
		42.5 &$-$& 85 & $1\times 10^4$\\
		85 &$-$& 170 &  $8\times 10^3$\\
		170 &$-$& 340 & $6\times 10^3$\\
		340 &$-$& 390 & $6\times 10^3$\\
		390 &$-$& 780 & $4\times 10^3$\\
		780 &$-$& 1560 & $4\times 10^3$\\
		1560 &$-$& 3120 & $4\times 10^3$\\
		3120 &$-$& 6240 & $4\times 10^3$\\
		\hline
	\end{tabular}
\end{table}

\section{Results} \label{sec:results}

\subsection{Relevant examples}
 
\begin{figure}
    \centering
	\includegraphics[width=\columnwidth]{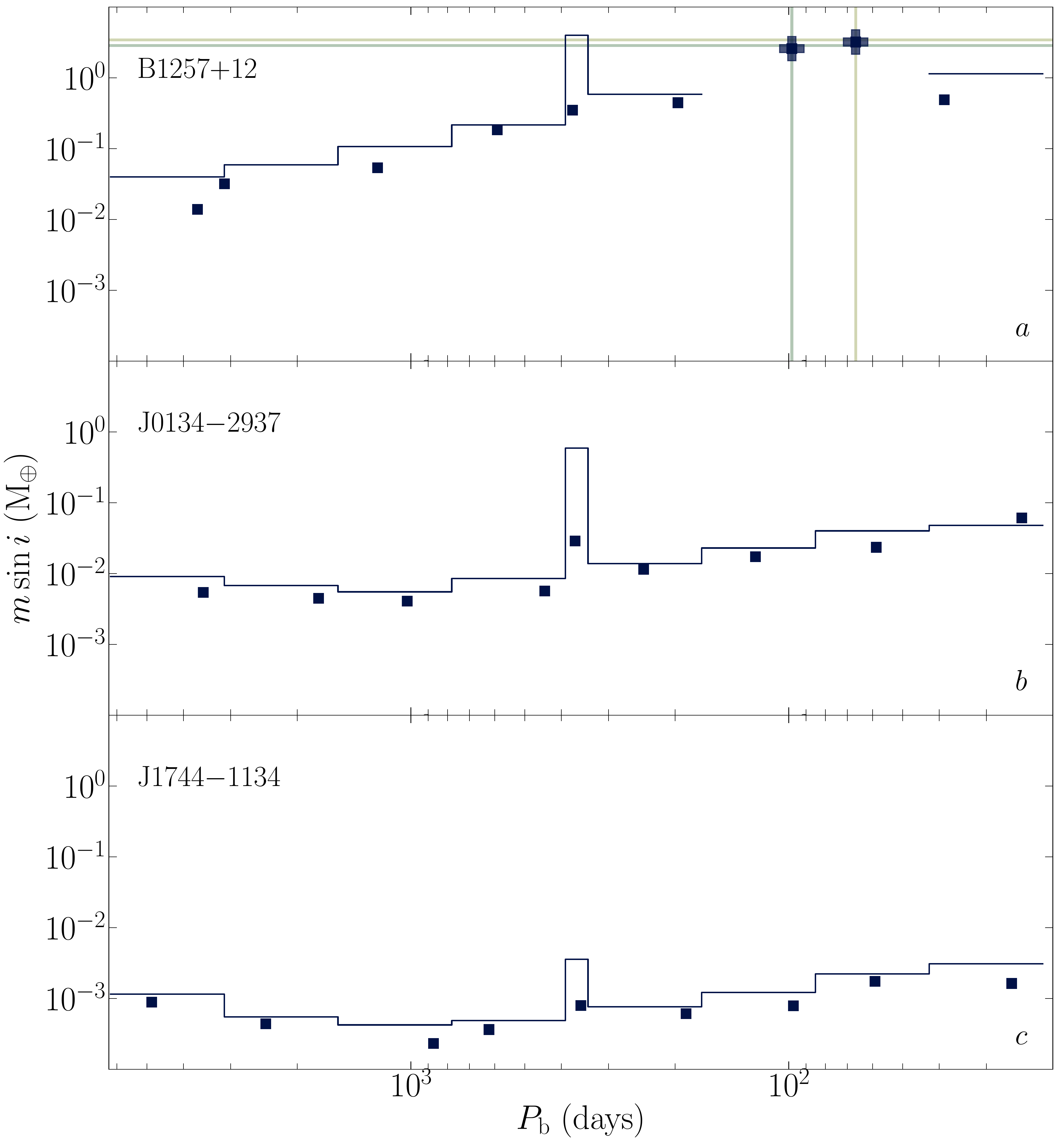}
    \caption{Example plots of outputs for three different pulsars after our analysis. Panel \textit{a} shows the results for B1257$+$12 without including its two known higher-mass planets of orbital periods $66.5$ and $98.2$\,days in the timing model; panel \textit{b} shows the results for J0134$-$2937, which gave the lowest mass limits for a normal pulsar; panel \textit{c} shows the results for J1744$-$1134, which gave the lowest mass limits for a millisecond pulsar. In each plot, the step line shows the 95\% mass limits, such that we consider masses above this line to be highly unlikely; the squares show the maximum-likelihood values of period and mass for each period bin. Furthermore, in panel \textit{a}, the horizontal and vertical dotted lines show the known values of the corresponding orbital parameters for the two known planets, while the ``plus'' markers show 3-$\sigma$ detections as a result of our analysis. Note that the x-axis increases to the left.}
    \label{fig:combined_masslims}
\end{figure}

As the only pulsar known to host Earth-mass planets, PSR B1257$+$12 provides a useful example for our analysis.
Two of these planets are in the orbital period range of our search - one at $P_{\mathrm{b}} = 66.5\,\mathrm{days}$, and the other at $P_{\mathrm{b}} = 98.2\,\mathrm{days}$ \citep{Konacki2003}. Panel~\textit{a} of Fig.~\ref{fig:combined_masslims} shows the results of our pipeline for B1257$+$12 when no planetary companions were included in the initial pulsar timing parameters. The two planets are easily detected at the expected orbital period and projected mass. Note that we have not included the upper limits in the period bins in which a detection is made to avoid confusion.

The constraints on the maximum mass of an undetected planet vary for each pulsar, depending on the length and cadence of the observations available, the presence of glitches or known binary companions, as well as on the amplitudes of any white and red noise present. The mass limits are generally better constrained for millisecond pulsars than for normal pulsars. This is illustrated in Fig.~\ref{fig:combined_masslims}, which shows the lowest mass limits for a normal pulsar (PSR J0134$-$2937; panel~\textit{b}), and a millisecond pulsar (PSR J1744$-$1134; panel~\textit{c}), respectively. At the most extreme, we can put mass limits as low as 1/100th of the mass of the Moon (or $\sim 10^{-4} \,\mathrm{M}_{\oplus}$).

\subsection{3-$\sigma$ detections} \label{sec:3sigma}
\begin{figure}
    \centering
	\includegraphics[width=\columnwidth]{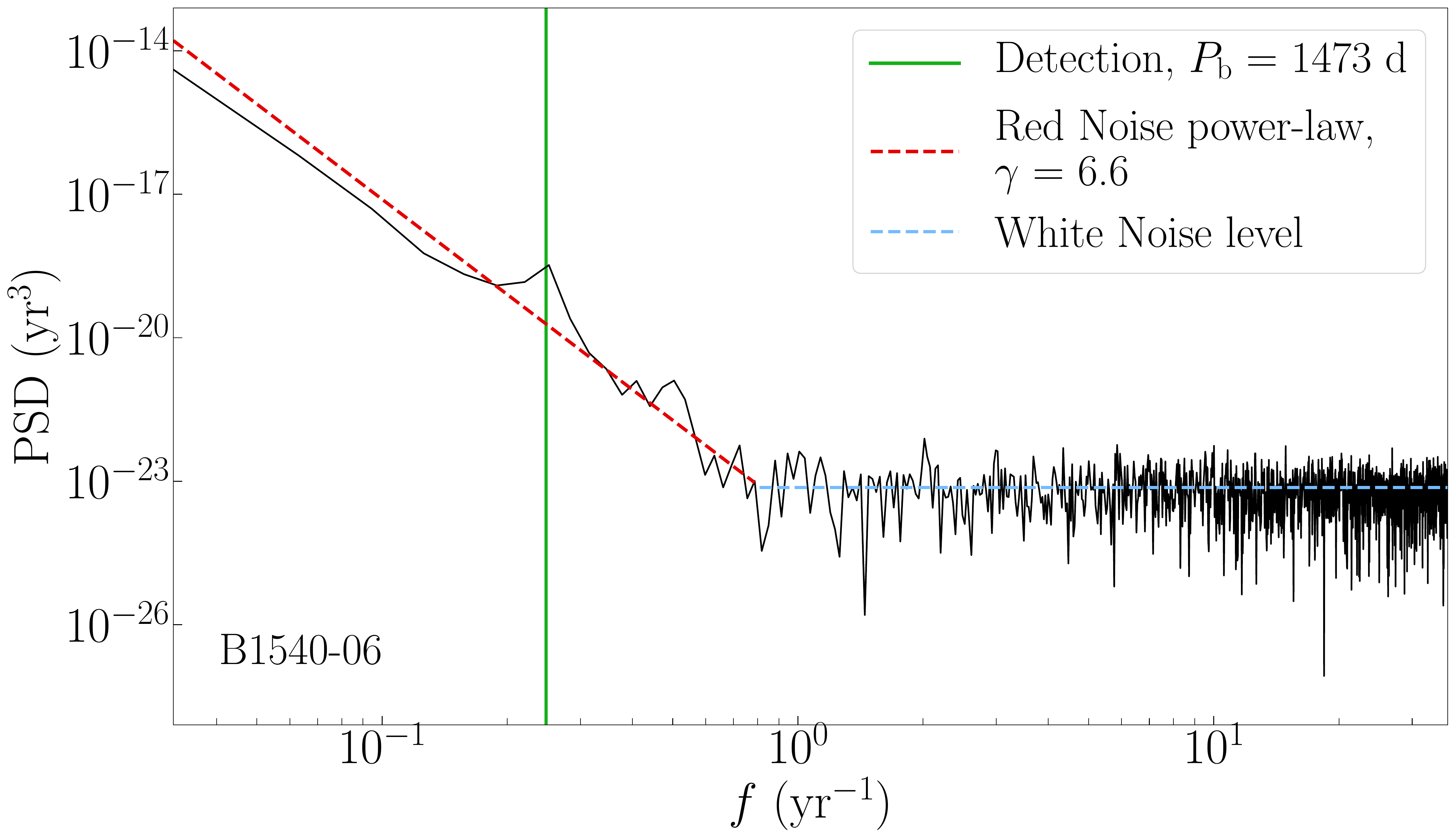}
    \caption{The power spectral density (PSD) against frequency for PSR B1540$-$06 is shown by the black continuous line. The fitted power-law, without including a planet influence, and as given by Eq.~\ref{eq:redpower}, is shown by the dashed red line, with the value of the index $\gamma$ given in the legend. The white noise level is shown by the horizontal blue dashed line. The detected periodicity, as given in Table~\ref{tab:pmresults}, is shown by the vertical green line, and a peak in the PSD can be seen at its approximate position.}
    \label{fig:PSD_example}
\end{figure}

Excluding B1257$+$12, we found 15 pulsars in our sample for which the analysis indicates a greater than 3-$\sigma$ detection in at least one of the searched period bins. None of these are millisecond pulsars. An algorithmic detection on its own, however, is not necessarily evidence for an orbiting planet, and further inspection is warranted. 
The methodology described in \citet{Coles2011} was used to estimate the power spectra from our data for each of the pulsars showing detections. These were used to visually inspect the agreement between the noisy power spectra and the power-law model, as well as the behaviour at the detected period. An example is given in Fig.~\ref{fig:PSD_example}, which shows the estimated power spectral density (PSD) as a function of frequency, as well as the fitted power-law model without including a planet influence. The equivalent plots for the other 14 pulsars can be seen in Appendix~\ref{app:PSD}.
As many of these pulsars are dominated by red timing noise, we found it useful to additionally visualise residuals after subtracting the lowest frequency components of the timing noise model. We limit this to frequencies below $1/2P_{\mathrm{b}}$ so that we avoid falsely emphasising the detected periodicity. Additionally, in order to better understand the significance of the planetary signal, for any bin in which such a detection was made, we repeated the extended \textsc{enterprise} analysis using a nested sampler (\textsc{dynesty}; \citealp{Speagle2020}). This nested sampler was used to estimate the evidence and therefore the Bayes factor between models with and without the planetary companion included.
We consider a log-Bayes factor larger than $\sim 3$ (i.e. odds ratio larger than $\sim 20$) to be ``strong'' evidence \citep{logBayes}.
We categorise and discuss these detections in the sections below, as well as summarise the main parameters of the fitted planets found in our analysis in Table~\ref{tab:pmresults}.

\begin{table}
	\centering
	\caption{Table showing the orbital period, mass and eccentricity values of the fitted planets found in our search. The values in brackets correspond to one standard deviation. The values of eccentricity given as an upper limit (e.g. < 0.14) represent $95\%$ upper limits of values consistent with 0. The last column shows which section in this paper discusses the respective pulsar.}
	\label{tab:pmresults}
	\begin{tabular}{llllc} 
		\hline
		PSR name & $P_{\mathrm{b}} [\mathrm{d}]$ & $m\sin{i} [\mathrm{M}_{\oplus}]$ & $e$ & Sect.\\
		\hline
		B1540$-$06 & 1473(14) & 1.1(2) & 0.12(4) & \ref{sec:B1540}\\
		B1714$-$34 & 1417(7) & 6.3(6) & 0.14(3) & \ref{sec:B1540}\\
		B1826$-$17 & 1102(5) & 2.6(4) & 0.35(6) & \ref{sec:B1540}\\
		B1828$-$11 (a) & 231(6) & 1.3(2) & < 0.14 & \ref{sec:B1540}\\ 
		B1828$-$11 (b) & 498(2) & 6(1) & 0.23(5) & \ref{sec:B1540}\\
		B0144$+$59 & 319(1) & 0.060(8) & < 0.45 & \ref{sec:B0144}\\ 
		B1727$-$33 & 350(1) & 3.5(5) & < 0.26 &\ref{sec:moreQP}\\ 
		B2053$+$36 & 1013(9) & 0.09(2) & < 0.40 & \ref{sec:moreQP}\\ 
		J1758$-$1931 & 719(8) & 6.1(9) & < 0.43 & \ref{sec:moreQP}\\ 
		J1843$-$0744 & 650(7) & 1.0(2) & 0.4(1) & \ref{sec:moreQP}\\
		J1904$+$0800 & 946(15) & 1.0(2) & < 0.18 &\ref{sec:moreQP}\\ 
		J2216$+$5759 & 117(9) & 3.5(6) & < 0.41 & \ref{sec:moreQP}\\ 
		J2007$+$3120 & 723(8) & 2.3(3) & < 0.38 & \ref{sec:J2007}\\ 
		J1947$+$1957 & 1070(9) & 3.7(3) & 0.56(6) & \ref{sec:J1947}\\
		B1931$+$24 & 5180(160) & 56(8) & 0.25(7) & \ref{sec:B1931}\\
		B0823$+$26 & 28.0(1) & 0.08(2) & 0.37(2) & \ref{sec:B0823}\\
		\hline
	\end{tabular}
\end{table}

\subsubsection{Pulsars with known quasi-periodic spin-down}\label{sec:B1540}

\begin{figure*}
	\centering
	\includegraphics[width=.49\linewidth]{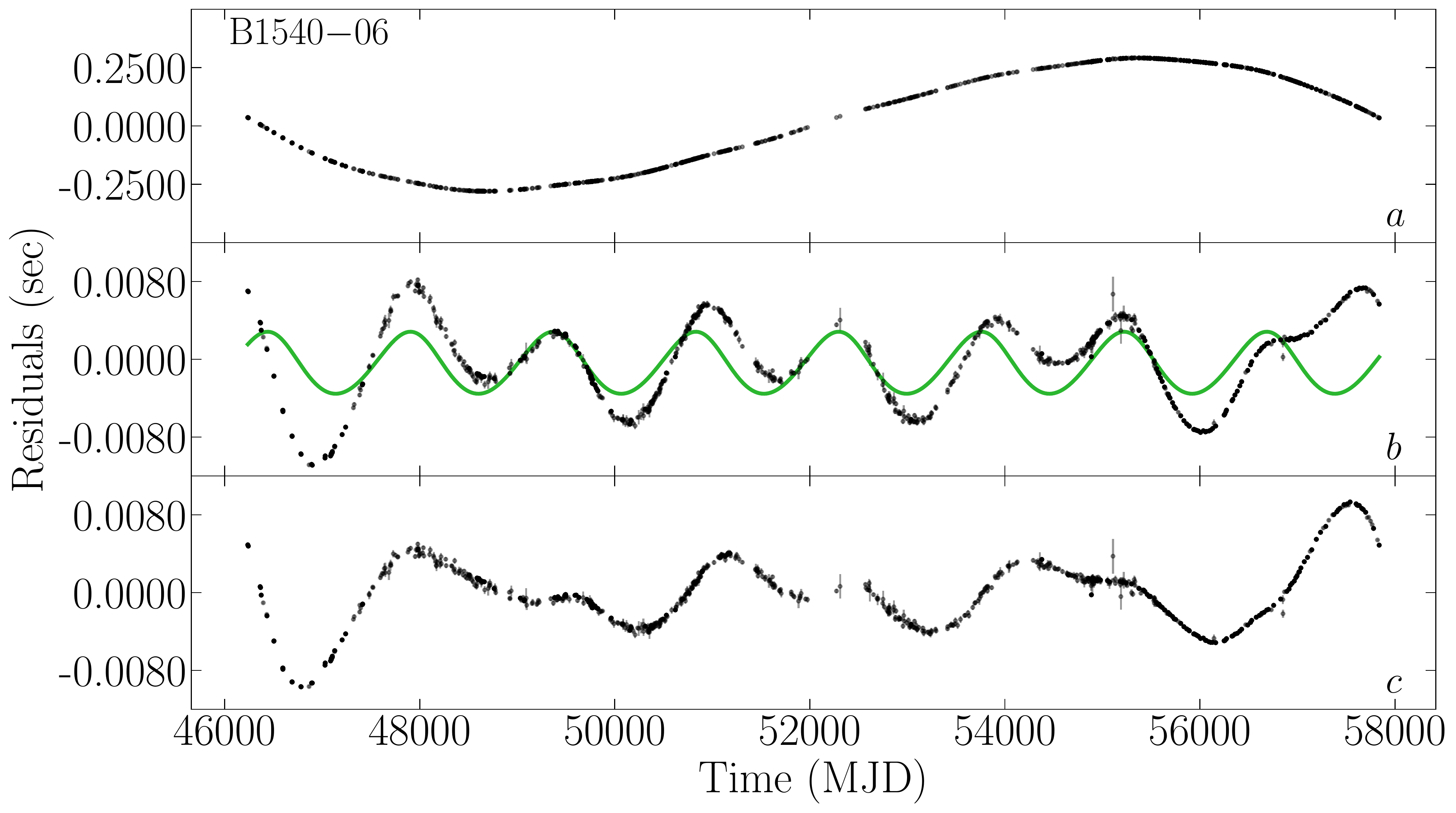}
	\includegraphics[width=.49\linewidth]{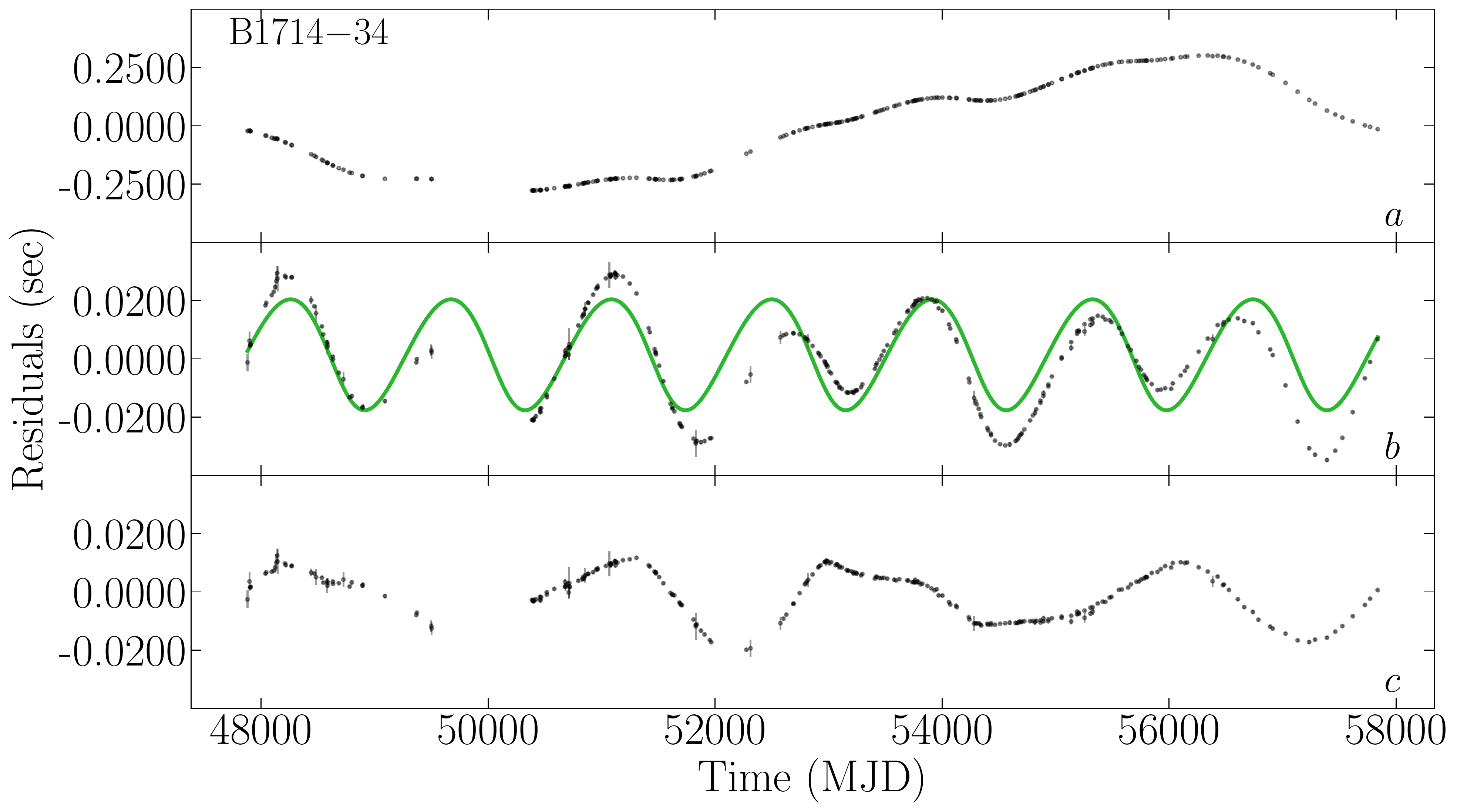} \\
	\includegraphics[width=.49\linewidth]{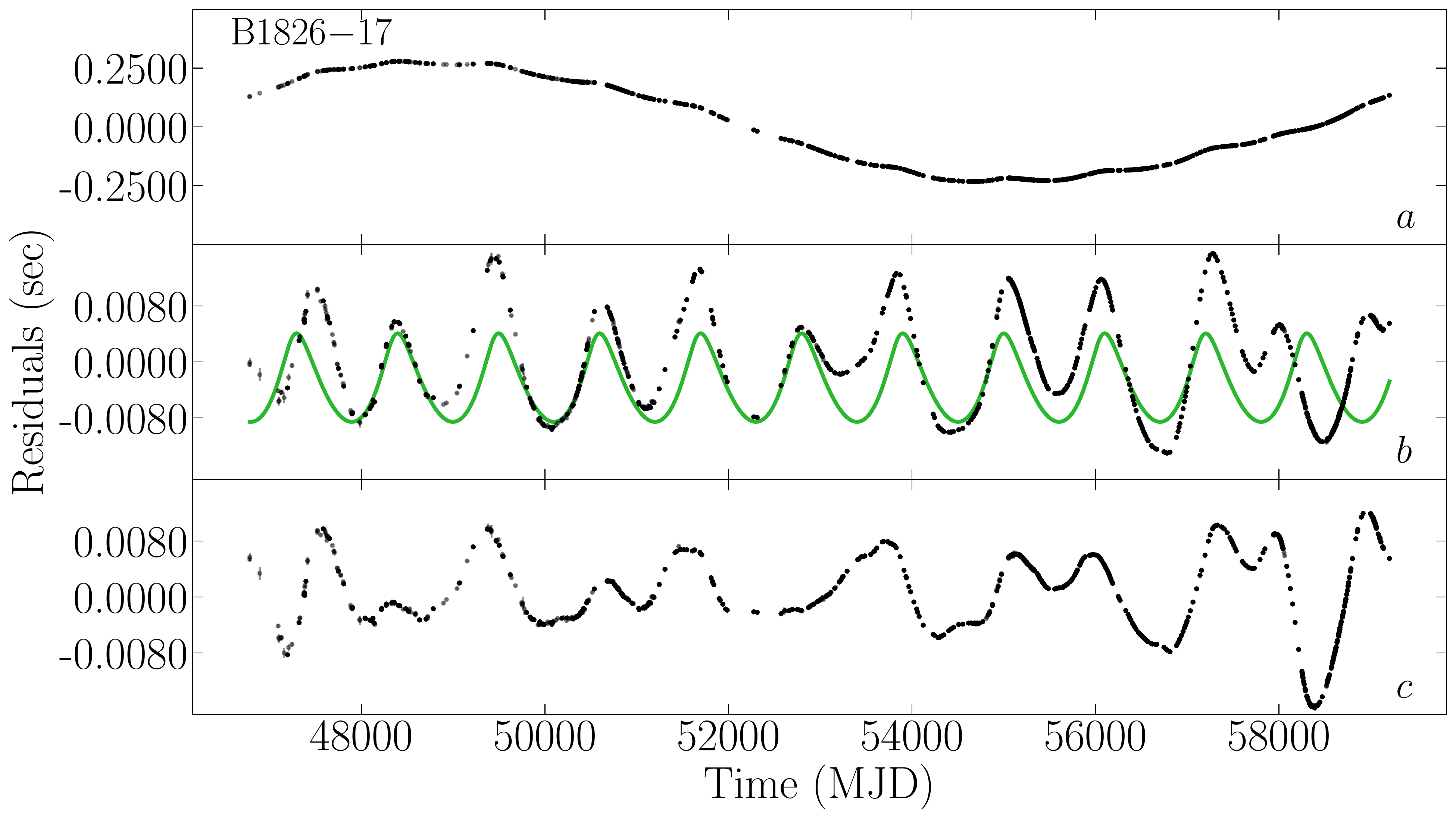}
	\includegraphics[width=.49\linewidth]{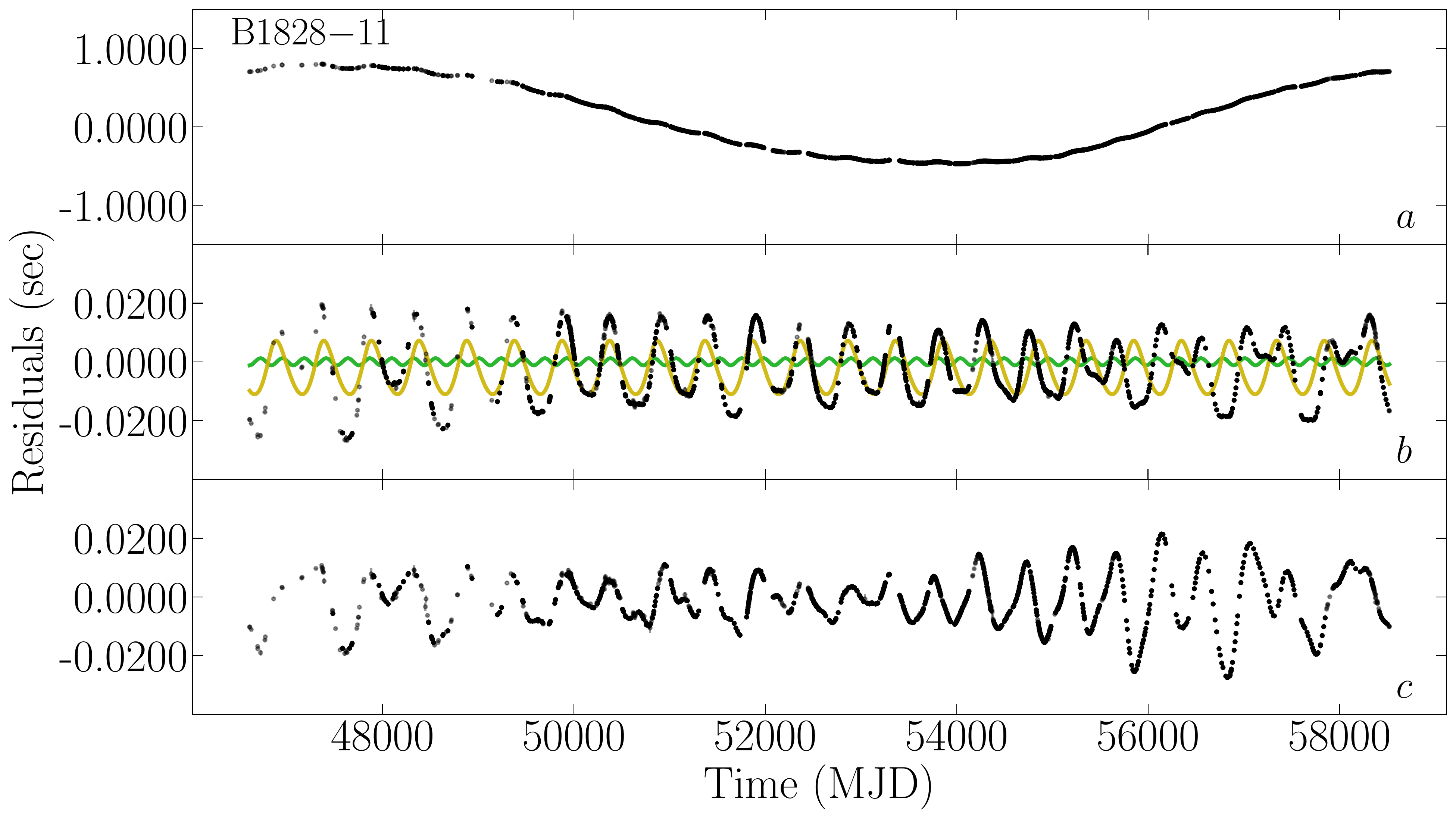}
	
	\caption{The residuals of PSR B1540$-$06, B1714$-$34, B1826$-$17 and B1828$-$11 are shown. The top panel (\textit{a}) shows the residuals after only removing the effect of deterministic timing parameters, as they are usually seen when using e.g. \textsc{tempo2}. The middle panel (\textit{b}) shows the result of also removing the low frequency red-noise components, up to half the frequency of the fitted planet; the green solid line in the same panel is the R\o{}mer delay that the fitted planet would cause. Finally, the bottom panel (\textit{c}) is the result of removing the planet effect from the residuals in the middle panel. Note that the timing and noise parameters used here were those found by our analysis, which included a planetary companion.}
	\label{fig:knownQP}
\end{figure*}

Quasi-periodic (QP) timing irregularities have been observed in pulsar data as variations in their frequency ($\nu$) or frequency derivative ($\dot{\nu}$). \citet{Lyne2010} characterised several pulsars showing QP behaviours, and concluded that these were caused by changes in the pulsars' magnetosphere, as they were correlated with pulse profile changes. PSRs B1540$-$06, B1714$-$34 and B1826$-$17 were among those found by \citet{Lyne2010} to show QP timing noise, and the frequencies of their periodicities were estimated as $0.24(2)\,\mathrm{yr}^{-1}$, $0.26(4)\,\mathrm{yr}^{-1}$ and $0.33(2)\,\mathrm{yr}^{-1}$, respectively, where the numbers in brackets represent the 1-$\sigma$ error in the last digit.

Similarly, PSR B1828$-$11 is known to show strong QP oscillations in $\dot{\nu}$, at a frequency of $0.73(2)\,\mathrm{yr}^{-1}$. The second harmonic of this is also clearly detectable \citep{Lyne2010, Stairs2019}. 
In our analysis, we detected an oscillation consistent with a planet for each of PSRs B1540$-$06, B1714$-$34 and B1826$-$17, with orbital periods which correspond to frequencies indistinguishable from those measured by \citet{Lyne2010}. Note that the orbits found for these planets are eccentric, such that the oscillations in the residuals are asymmetrical. For B1828$-$11, we find that, within our model, the residuals are best described by the influence of two planets, at the expected orbital periods corresponding to the first and second harmonics as estimated by \citet{Lyne2010}.

Fig.~\ref{fig:knownQP} shows the residuals for PSRs B1540$-$06, B1714$-$34, B1826$-$17 and B1828$-$11, respectively. Panel \textit{b} in each pulsar figure, showing the residuals
without the low-frequency red noise components, as well as the corresponding R\o{}mer delay of the fitted planet(s), illustrates that the behaviour is not perfectly periodic, and the purely periodic orbital solution only removes a fraction of the power on these time scales. While a model fitting for one (two) periodic term(s) better describes our data than the model without it (as also supported by the log-Bayes factors, which are larger than $4$ in all cases), a quasi-periodic term, caused by intrinsic properties of the pulsars, would be more appropriate, and would rule out the planetary-like detections for these pulsars.

\subsubsection{B0144$+$59} \label{sec:B0144}
\begin{figure}
    \centering
	\includegraphics[width=\columnwidth]{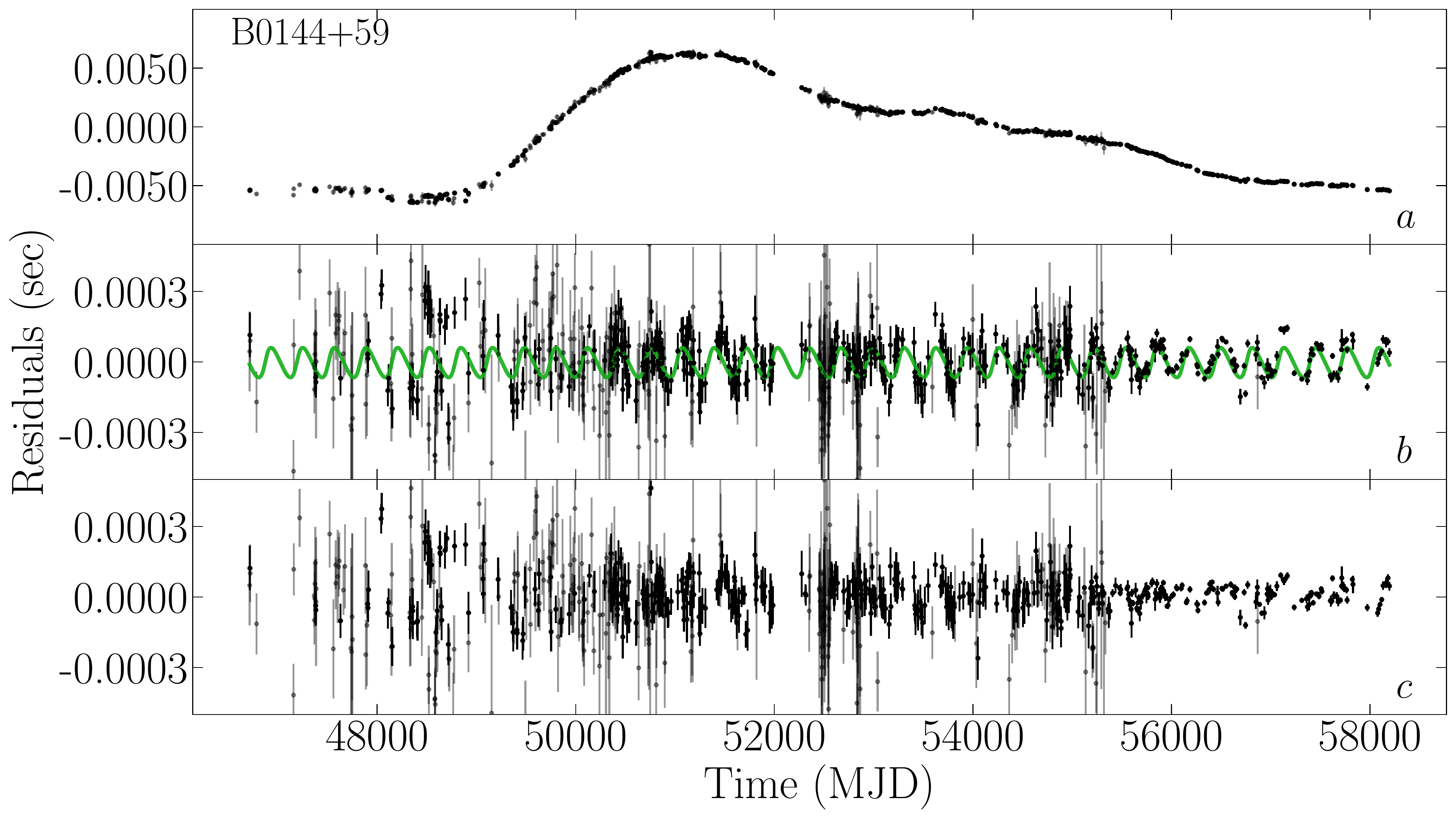}
    \caption{The residuals of PSR B0144$+$59 are shown. For a detailed description see the caption of Fig.~\ref{fig:knownQP}.}
    \label{fig:B0144_res}
\end{figure}

\begin{figure}
    \centering
    \includegraphics[width=\columnwidth]{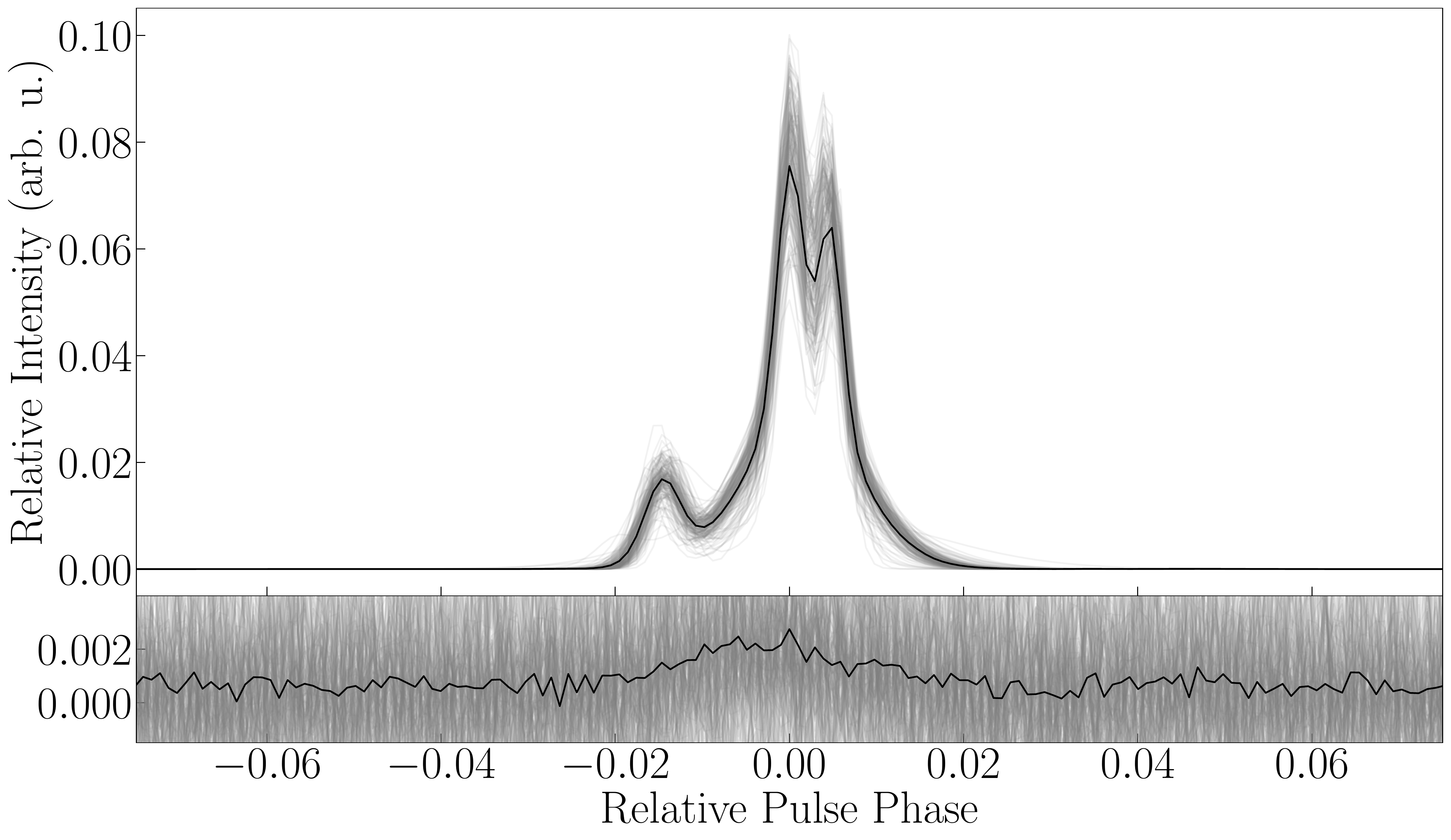}
    \caption{The different main pulse profile shapes for B0144$+$59 are shown in the top panel, including an average profile in black. The average shape of the interpulse is also shown in the bottom panel, while there are large variations between different interpulse shapes due to the low signal-to-noise.}
    \label{fig:B0144_pprofile}
\end{figure}

\begin{figure}
    \centering
    \includegraphics[width=\columnwidth]{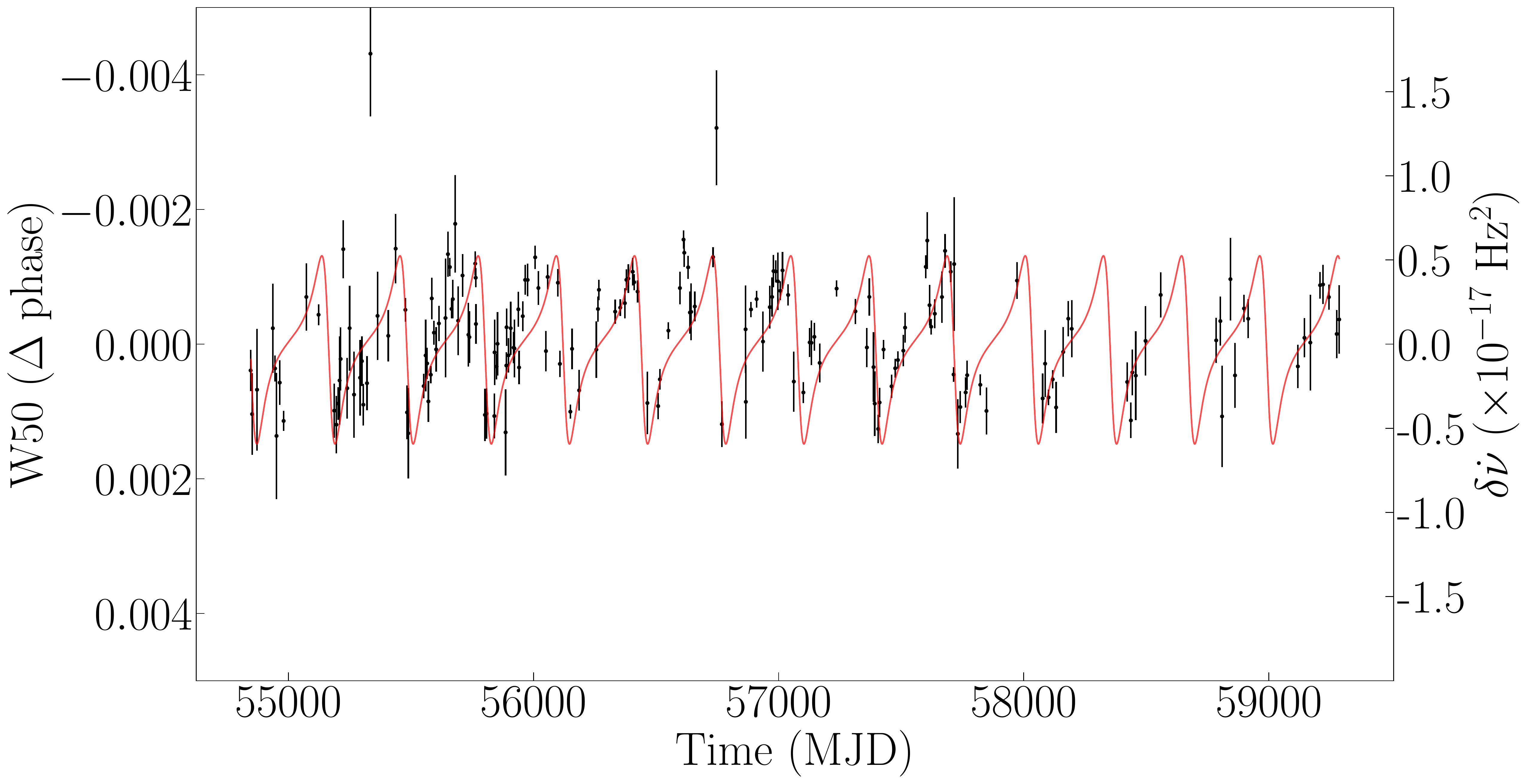}
    \caption{The estimated full-width-half-maximum (W50) values characterising the pulse profile shape variation of PSR B0144$+$59 are shown by the black data points. The red line shows the $\dot{\nu}$ estimated using our fitted planet model.}
    \label{fig:B0144_pvar}
\end{figure}

PSR B0144$+$59 is dominated by red timing noise, but we detect a significant 319(1) day periodicity in our analysis.
Removing the lowest frequency red-noise components reveals a highly periodic signal in the residuals, especially clear in the more recent DFB data where the precision of the observations is higher, as can be seen in Fig.~\ref{fig:B0144_res}. The existence of this periodic signal is also supported by the Bayesian evidence, as calculated using the software \textsc{dynesty}: our model has a log-Bayes factor of ${9.2\pm0.5}$ compared to the standard \textsc{enterprise} model, which does not include a planet.

Given the detections in pulsars with known correlations between profile shape and spin-down presented in Section~\ref{sec:B1540}, we looked for pulse shape changes in PSR B0144$+$59. As shown in Figure~\ref{fig:B0144_pprofile}, PSR B0144$+$59 exhibits an interpulse, but only the main pulse is strong enough to detect observation-to-observation variations. 
We used \textsc{psrsalsa}~\citep{Weltevrede2016} to fit the components of the pulse profiles using von Mises functions and to estimate the full-width half-maximum (FWHM) of the main pulse at each observation. Fig.~\ref{fig:B0144_pvar} shows the measured width overlaid with the $\dot{\nu}$ derived from our best-fit orbital model. The strong correlation, with correlation coefficient of $\sim 0.7$, is strong evidence that the observed periodic variations in PSR B0144$+$59 are magnetospheric in origin, similar to that of the pulsars in \citet{Lyne2010}.
Therefore, we conclude that the highly-periodic behaviour found in the residuals of PSR B0144$+$59 is likely not due to an external orbiting body, but to an intrinsic magnetospheric effect of the pulsar.

PSR B0144$+$59 exhibits very periodic spin-down oscillations, however this is not uncommon amongst the pulsars that exhibit QP behaviours (see e.g. the pulsars in Section~\ref{sec:B1540}). Our analysis searches for purely periodic variability and therefore selects for those with the most stable periodicity. Nevertheless, this does emphasise the somewhat surprising result that pulsars can exhibit extremely stable variability with periods of tens or hundreds of days.

\subsubsection{Additional quasi-periodic pulsars}  \label{sec:moreQP}

\begin{figure*}
	\centering
	\includegraphics[width=.49\linewidth]{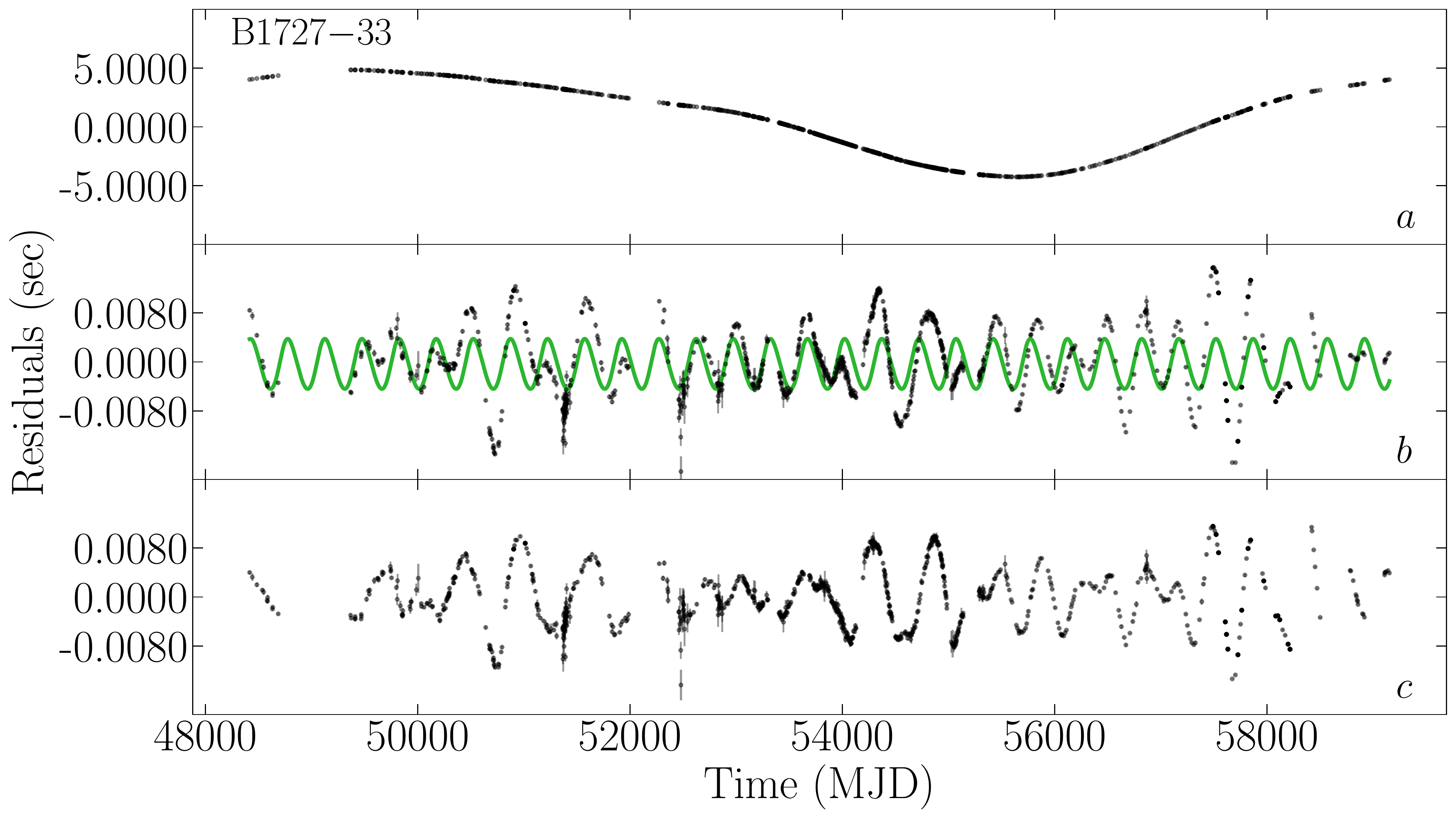}
	\includegraphics[width=.49\linewidth]{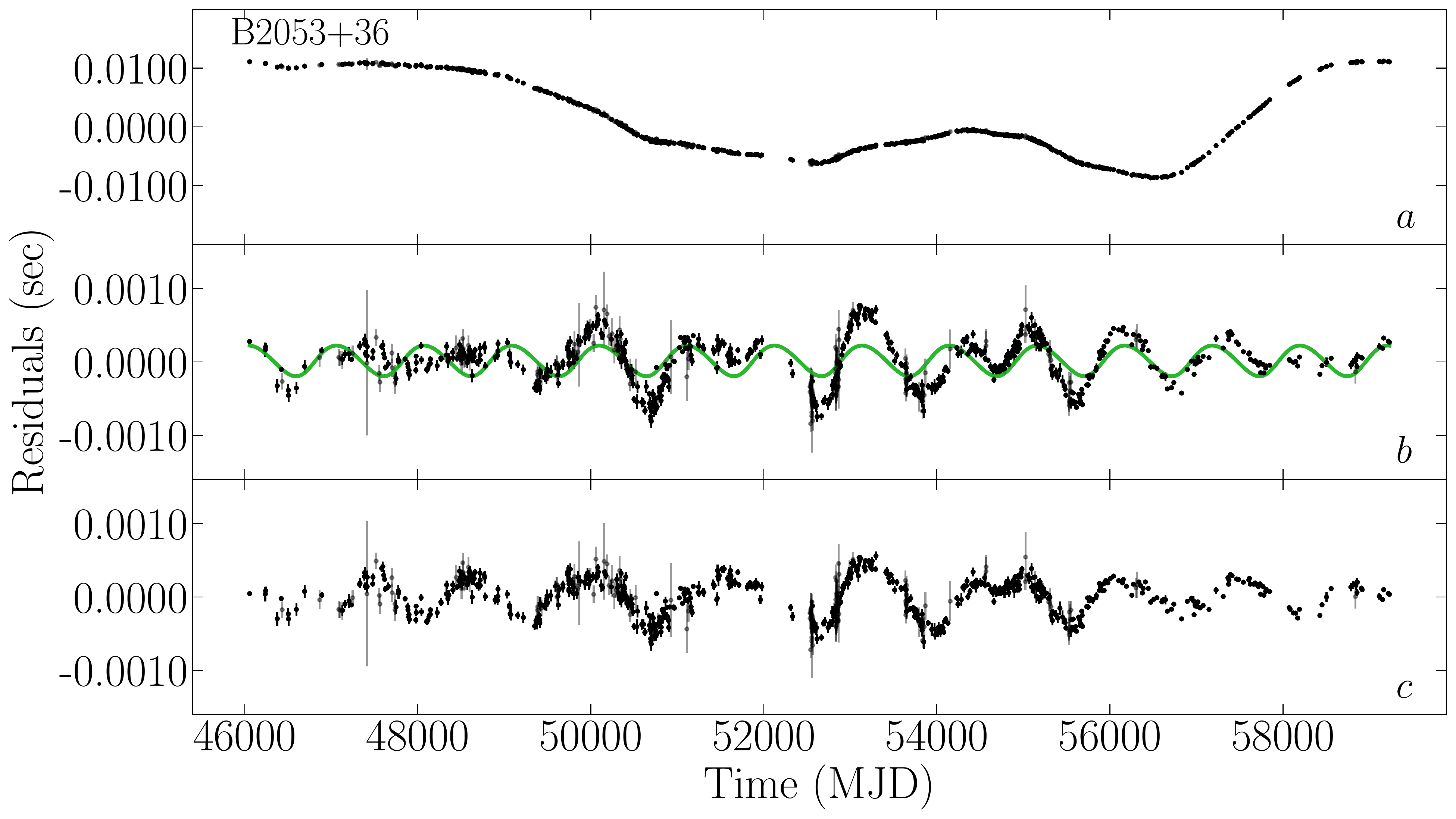} \\
	\includegraphics[width=.49\linewidth]{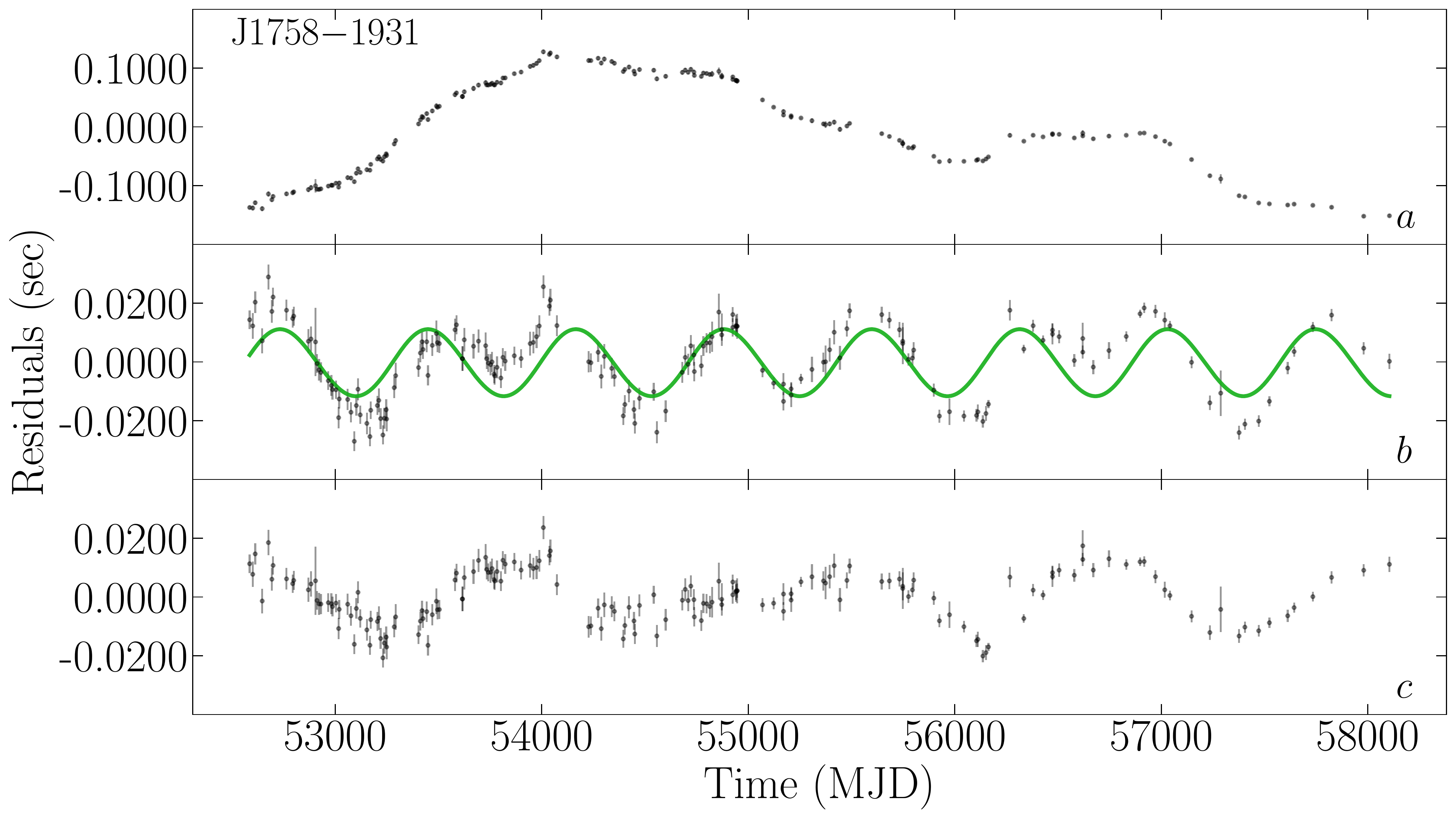}
	\includegraphics[width=.49\linewidth]{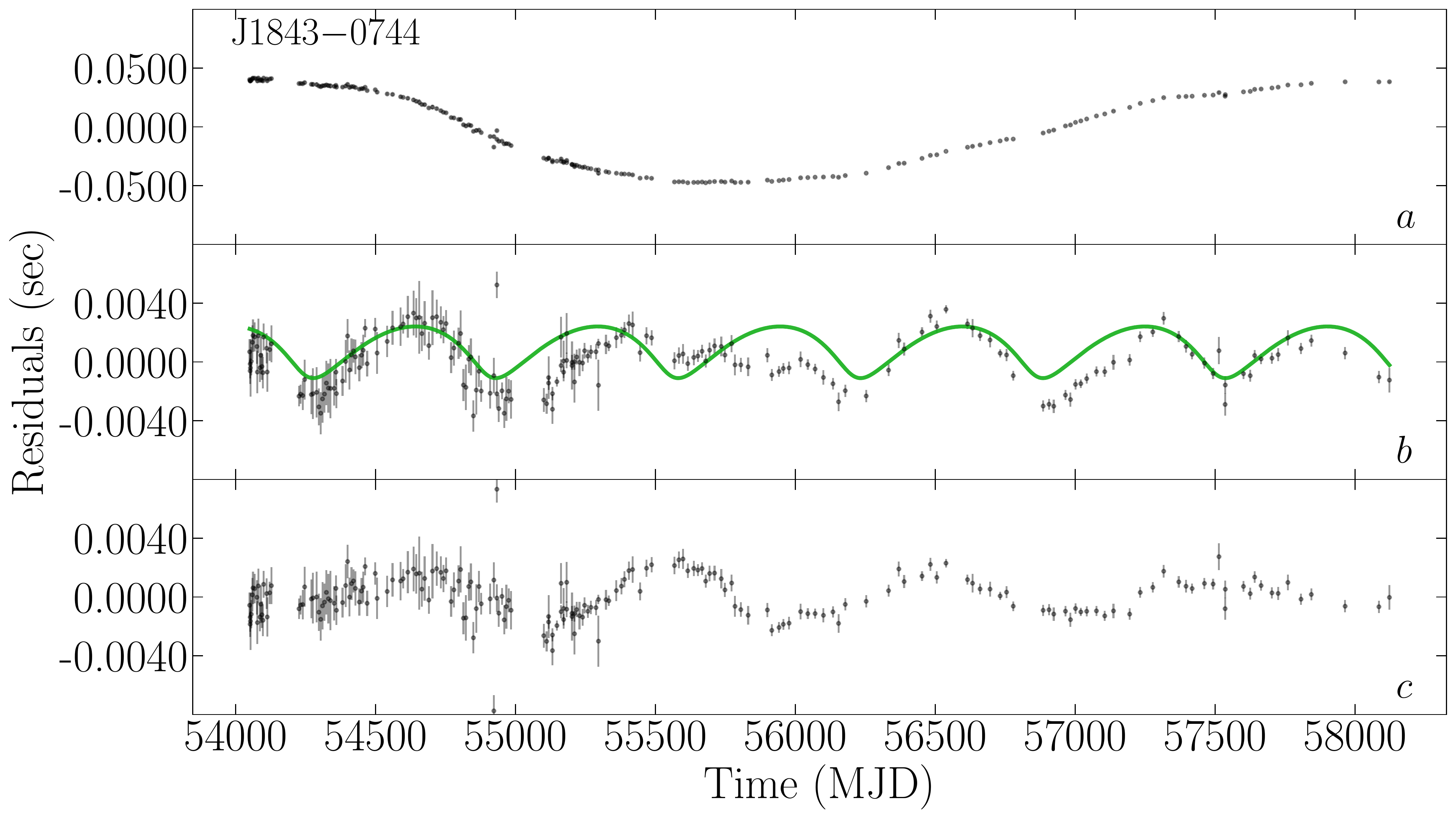} \\
	\includegraphics[width=.49\linewidth]{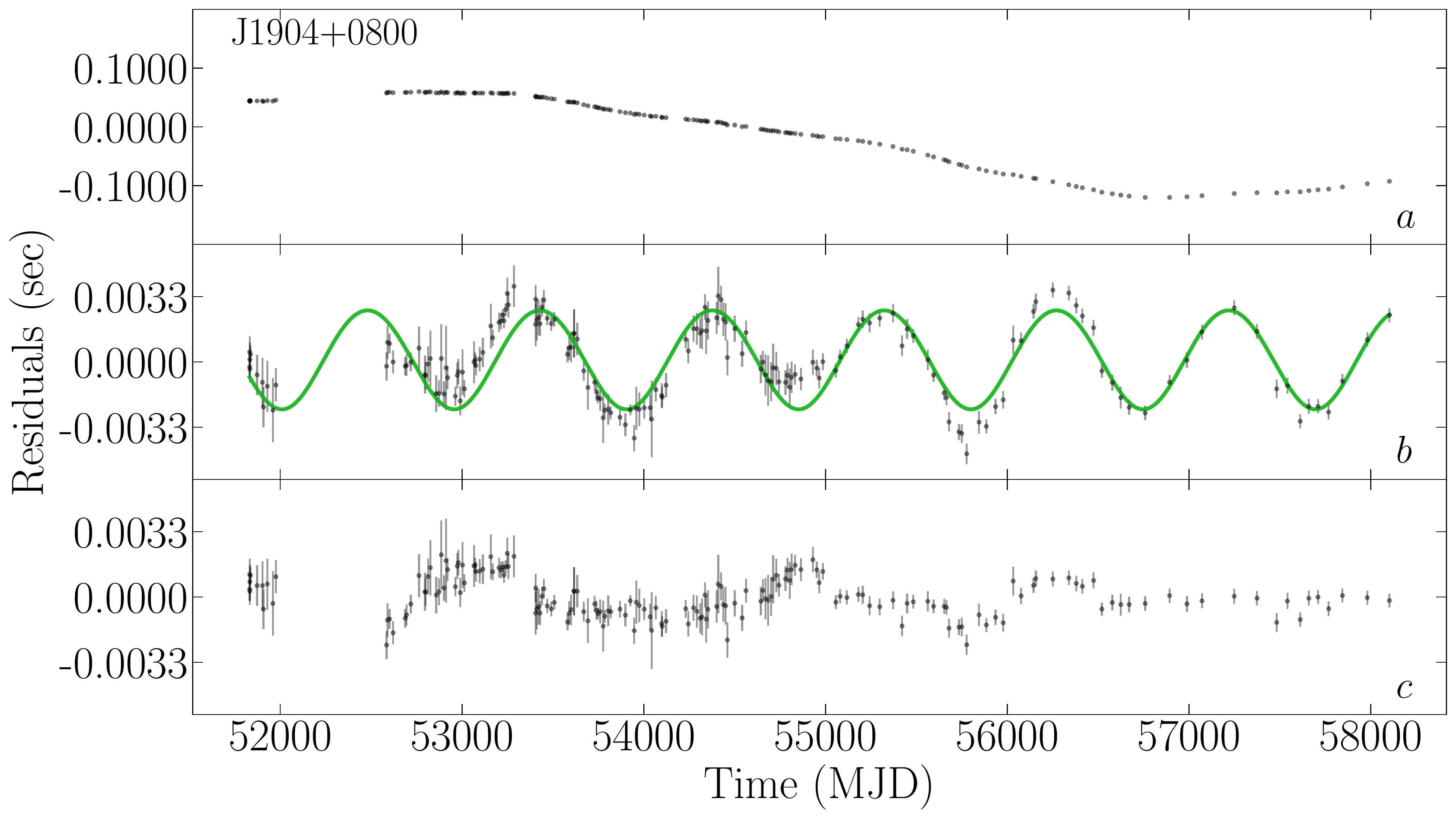}
	\includegraphics[width=.49\linewidth]{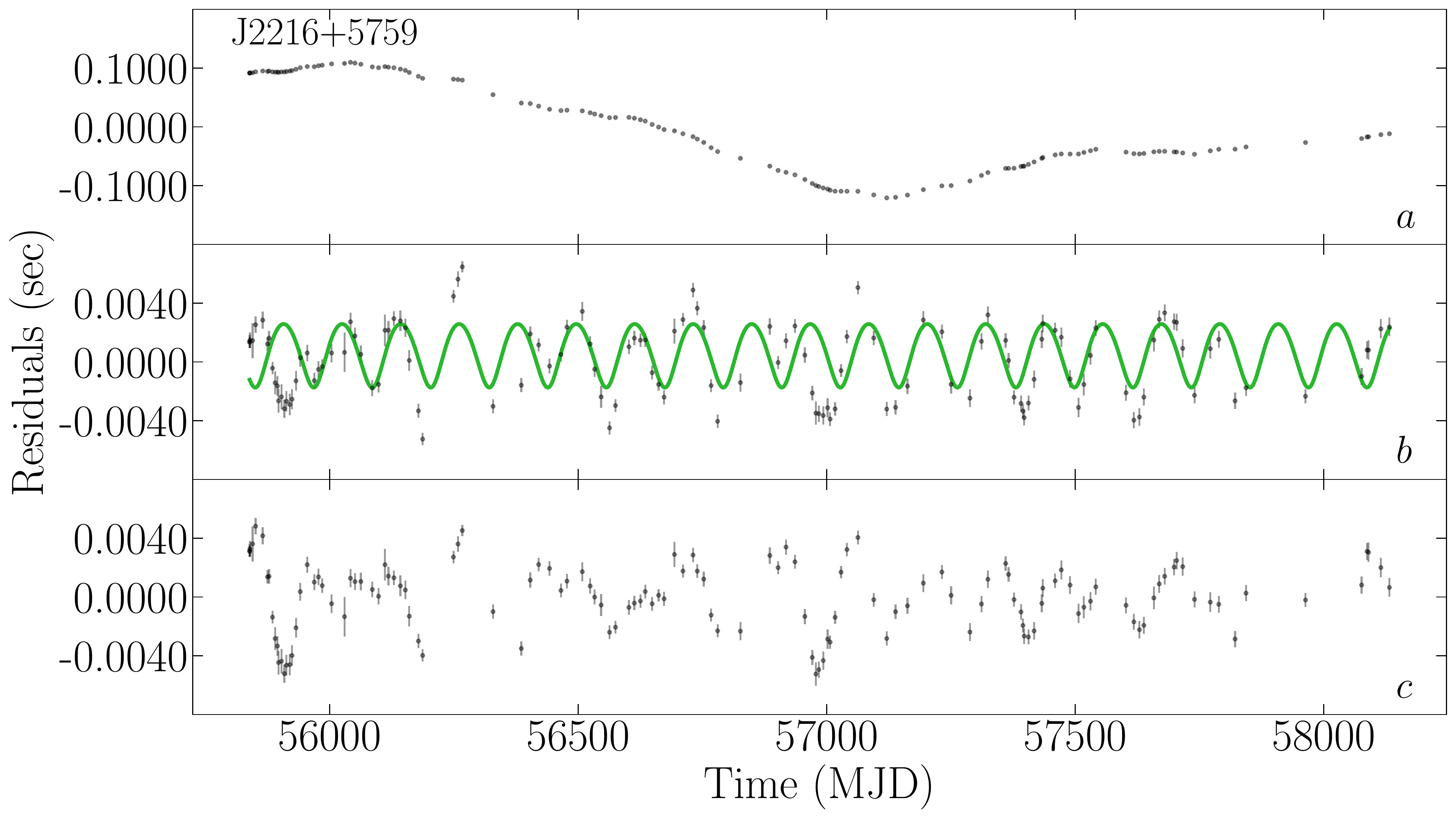}

	\caption{The residuals of PSR B1727$-$33, B2053$+$36, J1758$-$1931, J1843$-$0744, J1904$+$0800 and J2216$+$5759 are shown. For a detailed description see the caption of Fig.~\ref{fig:knownQP}.}
	\label{fig:posQP}
\end{figure*}

In this section, we discuss six of the pulsars for which we detected a planet-like periodicity, as they show similar behaviours, namely PSRs B1727$-$33, B2053$+$36, J1758$-$1931, J1843$-$0744, J1904$+$0800 and J2216$+$5759. 

Fig.~\ref{fig:posQP} shows the residuals of these pulsars. The systematic behaviours of the residuals after subtracting the planet-like influence (shown in panels~\textit{c}) suggest that these R\o{}mer delays are not enough to account for the shape of the oscillating residuals of these pulsars. Furthermore, we do not expect these residuals to be caused by extra companions in the system, since their shape does not indicate a different periodic oscillation. Unlike PSR B0144$+$59, the signal-to-noise of the observations of these pulsars is insufficient for us to reliably search for pulse profile variations.

PSR B1727$-$33 has two known glitches, at observation times 52\,107\,MJD and 55\,930\,MJD, respectively \citep[Jodrell Bank Glitch Catalogue\footnote{\href{http://www.jb.man.ac.uk/pulsar/glitches.html}{www.jb.man.ac.uk/pulsar/glitches}}; ][]{Espinoza2011}. We include the standard glitch fitting parameters, as well as the \textsc{tempo2} glitch recovery parameters (GLF0D and GLTD), in the fit for our analysis. Note that, although the period of the fitted planet is close to one year ($P_\mathrm{b} = 350(1)\,\mathrm{d}$), we can rule out the effect of the rotation of the Earth as a source of this oscillation, as our analysis includes a fit for the position and proper motion of the pulsar simultaneously with all other parameters.

We therefore find that the pulsars in this section seem more likely characterised by similar QP behaviours to the pulsars we discussed in Section~\ref{sec:B1540} than orbital motions, but without further data we cannot be certain of the origins of the detections in PSRs B1727$-$33, B2053$+$36, J1758$-$1931, J1843$-$0744, J1904$+$0800 and J2216$+$5759.

\subsubsection{J2007$+$3120} \label{sec:J2007}
\begin{figure}
    \centering
	\includegraphics[width=\columnwidth]{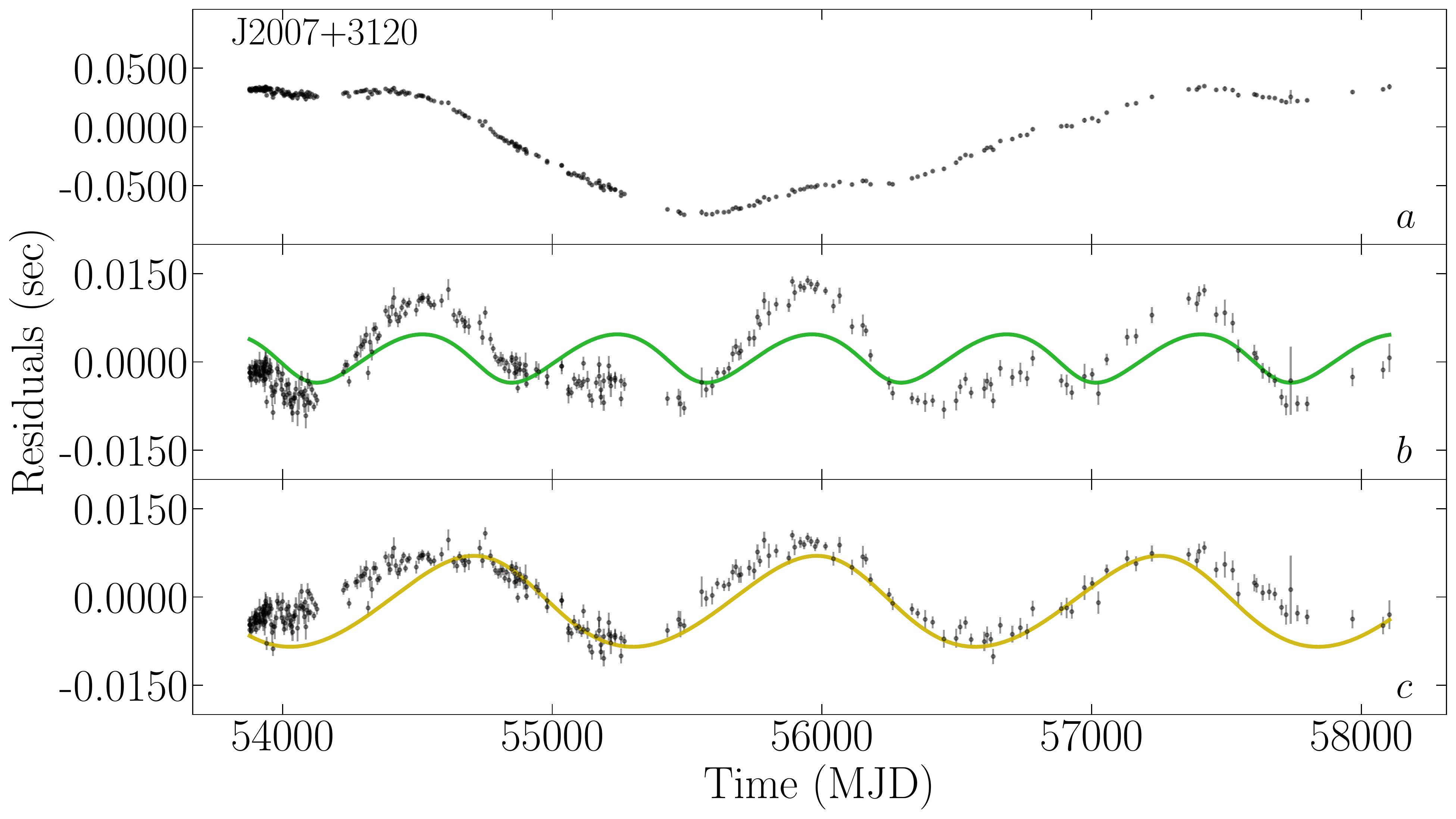}
    \caption{The residuals of PSR J2007$+$3120 are shown. For a detailed description see the caption of Fig.~\ref{fig:knownQP}. A second planet-like periodicity fit is shown by the solid line in panel \textit{c}.}
    \label{fig:J2007_res}
\end{figure} 

Our initial analysis of PSR J2007$+$3120 revealed an oscillation consistent with a planetary companion of orbital period $723(8)$\,days. The corresponding R\o{}mer delay is shown in green in panel \textit{b} of Fig.~\ref{fig:J2007_res}, on top of the residuals of this pulsar. After removing the influence of this planet, the residuals (shown in panel \textit{c} of Fig.~\ref{fig:J2007_res}) further show a periodic behaviour. When repeating our analysis to fit for two planets accordingly, the parameters of the second potential companion were less constrained, giving an orbital period of $1297(76)$\,days.

The log-Bayes factors between models with zero, one and two planets suggest that, while the existence of the first planet (of shorter period) is strongly preferred (by ${5.8\pm0.5}$), there is not a strong preference between a red noise component and the second planet, since the two models have indistinguishable evidences.    

While the residuals of J2007$+$3120 appear highly periodic, the time span of our current data only allows for a few oscillations. It follows that this is not sufficient to attest whether (any of) these two planet-like influences are indeed due to planetary companions to the pulsar, or are due to other effects, such as intrinsic QP noise. The signal-to-noise in the observations of J2007$+$3120 is very low, which prevents us from reliably searching for pulse profile variations.

The two periodicities found in our analysis are consistent, within two standard deviations, to a 2:1 ratio. On one hand, this ratio is characteristic of a harmonic relationship shown by QP noise, as is found for PSR B1828$-$11. On the other hand, an orbital resonance with a 2:1 period ratio has been found to be fairly common in extrasolar planetary systems \citep{Lissauer2011}. If the periodic behaviour is indeed due to planetary companions, detecting the effect of a (near) resonance in the form of a predictable variation of the orbital parameters would firmly prove the existence of these companions around J2007$+$3120. This is the same technique used to confirm the first two planetary companions of B1257$+$12, which are also in resonance, with a period ratio of 3:2 \citep{Rasio1992,Wolszczan1994}. Following the three-body analysis of \citet{Malhotra1993}, we estimate that for the parameters of our two found planets around J2007$+$3120, the corresponding resonant gravitational interactions would introduce deviations in the residuals of the pulsar as an oscillation of period of a few tens of years and amplitude of $\sim 1 \upmu\mathrm{s}$. This is undetectable with our current data, as the residuals are characterised by a white noise level of $\sim 1000 \upmu\mathrm{s}$. However, long term observations using a highly sensitive instrument such as FAST, CHIME or the Square Kilometre Array might be able to detect/rule out these oscillations, and therefore settle the cause of the periodicities seen in J2007$+$3120.

\subsubsection{J1947$+$1957} \label{sec:J1947}
\begin{figure}
    \centering
	\includegraphics[width=\columnwidth]{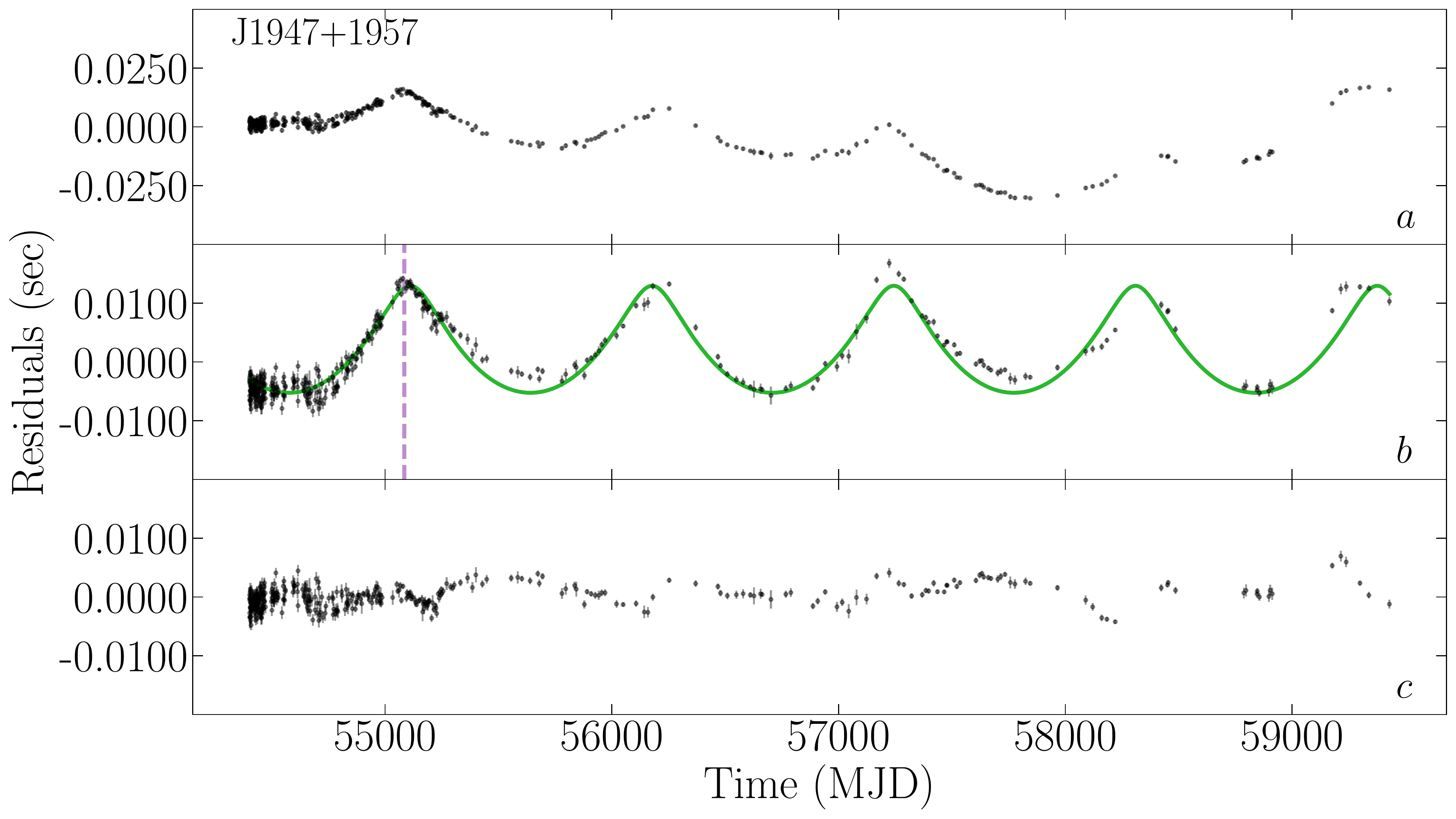}
    \caption{The residuals of PSR J1947$+$1957 are shown. The dotted vertical line corresponds to the glitch found by \citet{Nice2013}. For a detailed description see the caption of Fig.~\ref{fig:knownQP}.}
    \label{fig:J1947_res}
\end{figure}

PSR J1947$+$1957 was discovered by \citet{Nice2013}, who at the same time also claimed the detection of a glitch in the ToAs at MJD 55\,085. In our analysis of J1947$+$1957, we found a planet-like influence, characterised by an orbital period of $3$\,years, projected mass of $4\,\mathrm{M}_{\oplus}$ and a highly eccentric orbit ($e = 0.6$); this model was preferred to the standard \textsc{enterprise} fitting by a large log-Bayes factor, of ${7.9\pm0.5}$.

The residuals of PSR J1947$+$1957 (see Fig.~\ref{fig:J1947_res}) are dominated by five peaked oscillations. The purely periodic model is successful at removing a good deal of the structure in the timing residuals, though the remaining residuals in panel \textit{c} still have structures that appear correlated with each periastron passage. Although in theory the pulsar wind or tidal effects could perturb the orbit at periastron, it is worth noting that the closest approach of $\sim 1\,\mathrm{AU}$ for the planetary companion is not especially small, and such effects could also be caused by variability in the periodicity of the oscillation.

Alternatively, the cusps observed in the timing residuals could be due to glitches, as originally proposed by \citet{Nice2013} for the event at 55\,085\,MJD. There is no indication in our data of the presence of a glitch recovery for this proposed glitch. The subsequent four features are much less well sampled, so it is hard to tell if they exhibit the sudden change of gradient expected for a pulsar glitch. Although not common, some frequently glitching pulsars have been observed to exhibit glitches at regular intervals \citep[e.g.][]{Yu2013, Fuentes2019, Basu2020}. However, note that the inferred glitch size for J1947$+$1957 would be much smaller than those of other observed quasi-periodically glitching pulsars. Another potential origin of the periodic behaviour in J1947$+$1957 could be an intrinsic QP effect in the pulsar. The very low signal-to-noise of our observations of this pulsar makes any study of the pulse profile variability extremely difficult.

It is notable that the argument of periapsis found for the planetary-like signal in pulsar data was $\omega = (80\pm20)\deg$, while the high-eccentricity, yet symmetrical shape of the cusps (as seen in Fig.~\ref{fig:J1947_res}) also suggests that the planetary orbit would have to have a special, symmetrical orientation in the sky, such that the observer, the periastron and the apastron would be aligned, i.e. $\omega\sim 90\deg$. However, this symmetry is typical in QP magnetospheric pulsars. 

\subsubsection{B1931$+$24} \label{sec:B1931}
\begin{figure}
    \centering
	\includegraphics[width=\columnwidth]{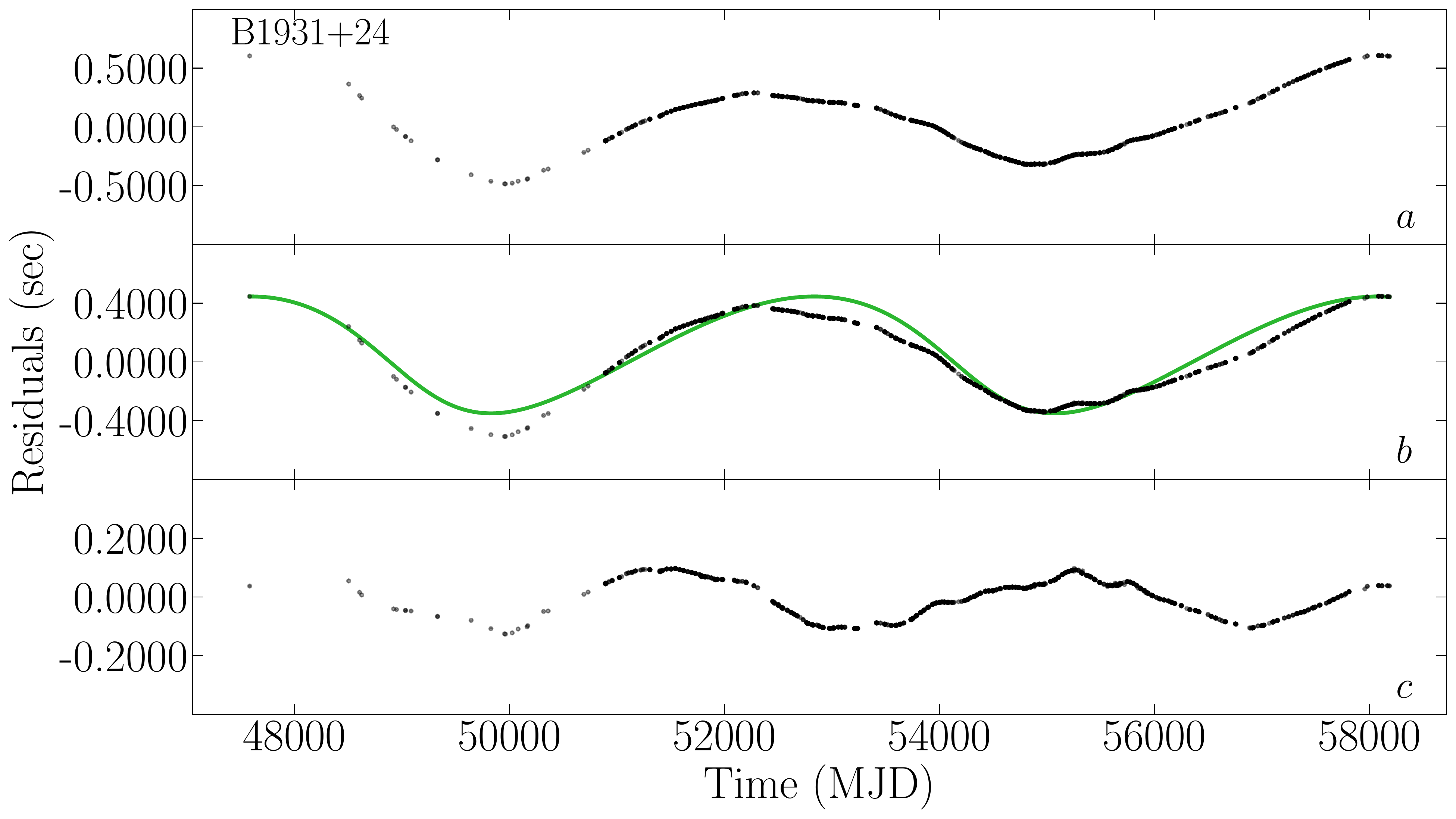}
    \caption{The residuals of PSR B1931$+$24 are shown. For a detailed description see the caption of Fig.~\ref{fig:knownQP}.}
    \label{fig:B1931_res}
\end{figure}

PSR B1931$+$24 was the first intermittent pulsar discovered; it is known to show normal pulsar behaviour for $5-10$\,days, and then undergo extreme nulling events for $25-35$\,days, making it undetectable \citep{Kramer2006}. The possibility that this observed QP behaviour is due to a binary companion of B1931$+$24 has been studied by \citet{Rea2008} and \citet{Mottez2013}; both concluded that the influence of a companion would not be enough to account for all the properties of the intermittent signal.

In our analysis of this pulsar, we do not find that a single planet-like periodicity in the range $21.3$ - $42.5$ days is a good model to describe the residuals. This is because, as described by \citet{Young2013}, the nulling events do not have an exact periodicity, but multiple, narrowly-spaced periodicities in the range discussed. 

However, we find a different planet-like influence, of period $\sim 14$\,years. Fig.~\ref{fig:B1931_res} shows the only two oscillations that correspond to our current data span. The shape of the residuals in panel \textit{b} as compared to the R\o{}mer delay and the clearly systematic behaviour of the residuals in panel \textit{c} of Fig.~\ref{fig:B1931_res} suggest that this model does not fully describe the behaviour seen, and therefore this is likely not actually caused by a binary companion. Nonetheless, the log-Bayes factor between accounting for a planet-like periodicity, and not, is ${7.9 \pm 0.5}$. This relatively large value suggests that there should be a better model to describe the current residuals than the purely power-law description of the red noise used in this analysis. However, with only two oscillations seen in these residuals, we cannot determine if this pulsar will continue showing the same periodic behaviour, and therefore what the cause of this is.

\subsubsection{B0823$+$26} \label{sec:B0823}
PSR B0823$+$26 is one of the brightest known radio pulsars in the northern sky, and has been observed for nearly $50$\,years. This pulsar is known to exhibit emission phenomena such as subpulse drifting and mode-changing \citep{Sobey2015,Basu2019}, as well as nulling over a range of timescales from minutes to hours, often in clustered groups and with a fraction of $\sim 7\%$ \citep{Herfindal2009,Redman2009,Young2012}. 

Our initial analysis flagged a planet-like signature in the residuals of B0823$+$26, with a period of $28(2)$ days. The residuals of this pulsar show that the influence of the fitted planet would be very small compared to the general white noise in the pulsar timing. On further inspection, the power spectrum (shown in Fig.~\ref{fig:app_PSD}) of the timing data for this pulsar appears more like a broken power-law with a steep red component transitioning to a flatter red component at a frequency close to that found by our analysis; this is likely to be a consequence of the complicated nature of its emission. The power spectrum of the noise from B0823$+$26 was therefore fitted poorly by a single power-law model, resulting in a false planet detection.

\subsection{Discussion on the results for the full data set}
\begin{figure}
    \centering
	\includegraphics[width=\columnwidth]{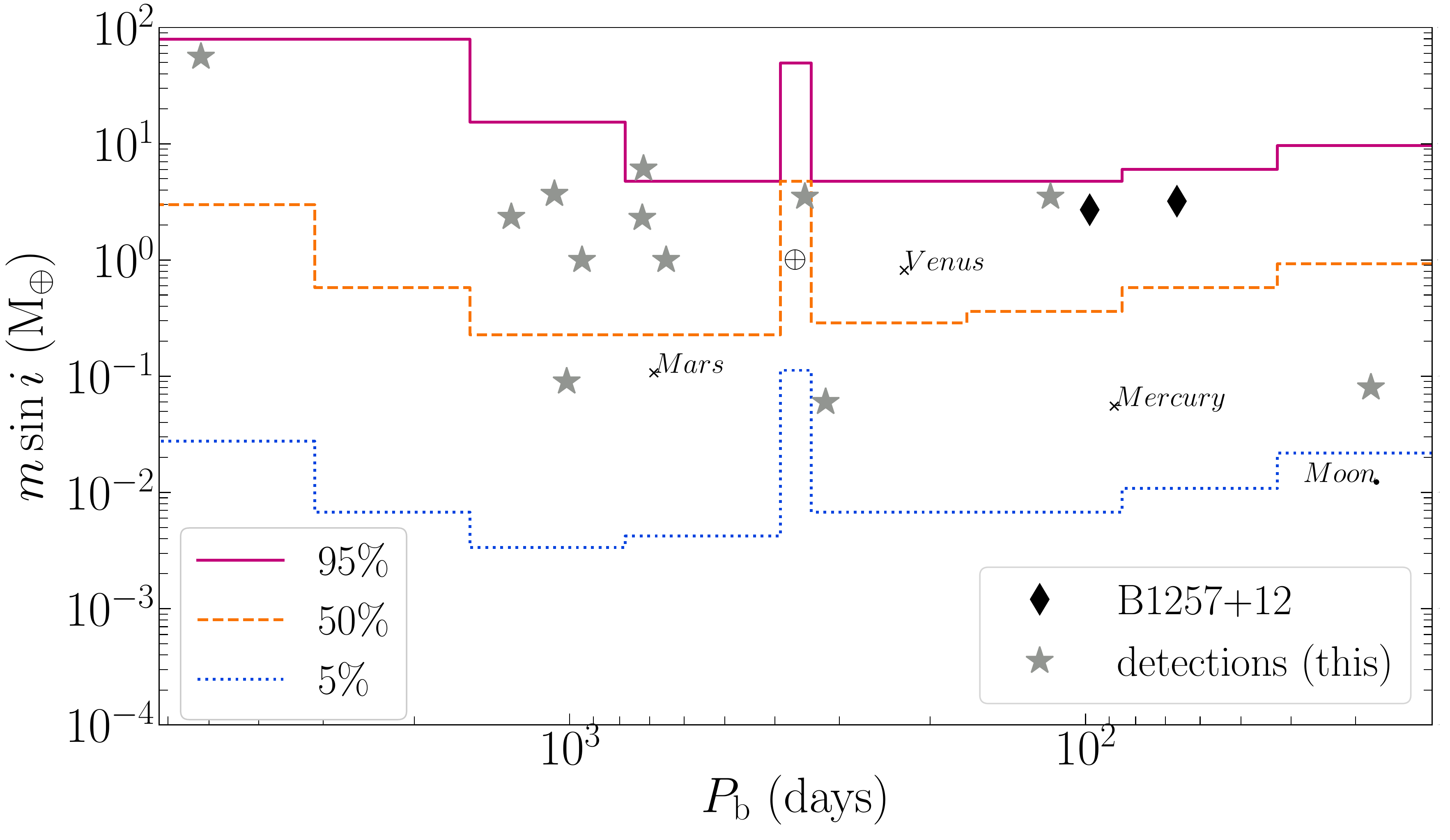}
    \caption{Summary of 95-percentile mass limits for all pulsars in the analysis. Note that the x-axis increases to the left. The colours used represent the fraction of pulsar mass limits lower than the respective mass, at each corresponding period bin. The diamonds represent the two known planets of B1257$+$12 whose orbits are within our parameter space, for reference. The stars show the detections flagged by this analysis, excluding the known QP pulsars presented in \citet{Lyne2010}.}
    \label{fig:2D_masslim}
\end{figure}

We show a summary of the 95\% mass limits of all the pulsars in our sample in Fig.~\ref{fig:2D_masslim}. The different contours illustrate the fraction of the pulsars in our sample that allow a planet below the respective projected mass within the 95\% limit, for each period bin. Note that we can only estimate the projected mass on the sky with respect to the observer ($m\sin{i}$); therefore the actual physical mass limits are on average $\sim20\%$ larger, assuming a random orientation of binaries in the sky (since the mean of $\sin{i}$ is $0.785$). 

In general, the mass limits as presented in Fig.~\ref{fig:2D_masslim} depend on the properties of the pulsars, as well as on the properties of the observations. Mass limits close to a $1$-year orbital period are noticeably worse than all the rest due to the required fitting for the proper motion and distance to the pulsar, which vary periodically with the Earth's orbit and therefore reduce our sensitivity. Similarly, any known binary companions reduce the sensitivity of our search in the respective period bin. At the long-period end the sensitivity is mainly limited by the amount of red noise in the residuals. At periods larger than 5-10 years, the total timespan of the observations can also limit the sensitivity of our search, especially for more recently discovered pulsars. In the short-period limit the white noise is the dominant factor in limiting our sensitivity.

Fig.~\ref{fig:2D_masslim} also shows a summary of the planet detections found in our analysis, with the exception of the known QP pulsars included in \citet{Lyne2010}. As discussed in the preceding sections, we suspect that the majority of these are likely to be spurious detections caused by QP noise in the pulsar residuals. Interestingly, these show up in a pure periodicity search, emphasising that pulsars can exhibit highly stable periodic variability on timescales of hundreds of days. A detailed re-analysis of the sources in \citet{Lyne2010} is ongoing (Shaw et al., in prep).

It is notable that in all cases where the eccentricity is well constrained the eccentricities are between 0.1 and 0.6, and are certainly not the circular orbits that might be expected from formation in a fall-back debris disk. This further supports the hypothesis that these detections are more likely due to intrinsic spin variation rather than orbital motion.

In the period range of $0.2 - 4.3$\,years, with the exception of the aforementioned small interval around $1$\,year, about 70\% of the pulsars in our sample have mass limits lower than $1\,\mathrm{M}_{\oplus}$. This implies that the majority of our pulsars are generally unlikely to host any planets larger than this. Furthermore, 5-10\% of all the pulsars in our data set have mass limits lower than one Moon mass (or $\sim 0.01\,\mathrm{M}_{\oplus}$ ) in the same period range, ruling out any substantial planetary companions in our search range. Note that these mass limits are similar to the total mass of the asteroid belt model around PSR B1937$+$21 as estimated by \citet{Shannon2013}. The largest known mass of an asteroid is, however, two orders of magnitude smaller \citep[$\sim10^{-4}\,\mathrm{M}_{\oplus}$; ][]{Baer2008}.

\begin{figure}
    \centering
	\includegraphics[width=\columnwidth]{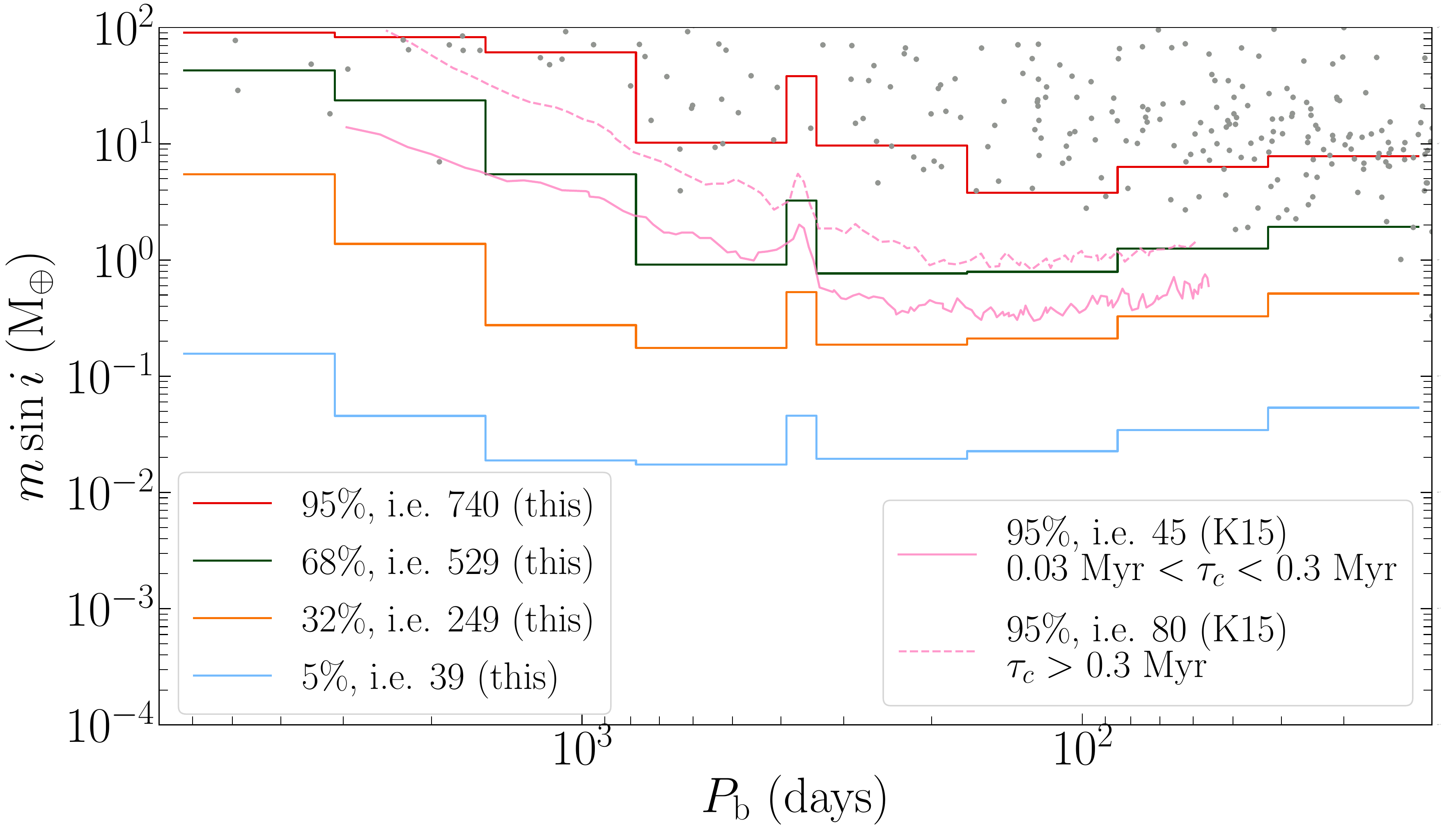}
    \caption{Joint period-mass posterior for all pulsars. The red, green, orange and blue lines show the $95\%$, $68\%$, $32\%$ and $5\%$ mass limits, respectively, of the mass posterior formed by adding the mass posteriors of all the pulsars, except those that showed a detection. This is done separately for each period bin, as can be seen. The number shown in the caption (e.g. ``i.e. 740'') represents the number of pulsars corresponding to the respective percentage (e.g. $95\%$). Two curves from K15 are also included for comparison, showing the $95\%$ mass limits for the pulsars in their sample of characteristic ages above and below $0.3$ Myr. The grey dots represent known exoplanets, from the Exoplanet Database \citep{Wright2011}. Note that the x-axis increases to the left.}
    \label{fig:all_post}
\end{figure}

Fig.~\ref{fig:all_post} shows another way of illustrating the mass limits for our entire pulsar data set. This is an approach similar to that of Fig.~4 of K15, and we include some of their results for comparison. To obtain the curves shown in Fig.~\ref{fig:all_post}, the mass posteriors of all the pulsars in our data set, except those that showed detections, were added for each period bin to create a mass posterior of the data set. The $95\%$, $68\%$, $32\%$ and $5\%$ limits were then calculated and are shown by the red, green, orange and blue lines in Fig.~\ref{fig:all_post}. Note that our mass sampling is re-weighted into a uniform prior between $10^{-4}$ and $100\,\mathrm{M}_{\oplus}$.

Comparing our results with the limits from K15, we note that our data set includes more than 5 times the number of pulsars in the K15 sample (consisting of 151 pulsars). Therefore, although for most period bins the $95\%$ mass limits of K15 give better constrains than our $95\%$ mass limits, they represent a much smaller number of pulsars, i.e. 50-100 for K15 as opposed to 740 for this analysis. 
We also note that K15 specifically used pulsars of high spin-down luminosity and with regular cadence observations, whereas our sample consists of essentially all pulsars observed by the Jodrell Bank timing programme, and includes a number of pulsars which have been infrequently and irregularly observed.

The mass limits shown can be interpreted as the fraction of pulsars in the data set that do not show a detection below the respective projected mass. For example, we can conclude that roughly two-thirds (or $68\%$) of the pulsars in our data set (i.e. $\sim 530$ pulsars) do not show any planets above $2\,\mathrm{M}_{\oplus}$ and with orbital period smaller than $1$\,year, as well as any planets above $8\,\mathrm{M}_{\oplus}$ and with periods larger than $1$\,year. Likewise, for about $40$ of our pulsars, we can rule out planetary companions of projected masses above $0.2\,\mathrm{M}_{\oplus}$.
Despite the large increase in sampled pulsars, our mass limits are better than those of K15 for large orbital periods. This is a consequence of our longer-span data sets, which improve our potential for detecting longer-period orbits, and often allow us to better distinguish between red noise processes and truly periodic behaviours.

This large sample study shows that although planets around pulsars are rare, there is still a large population of pulsars for which we cannot rule out planets of order $\sim1\,\mathrm{M}_{\oplus}$. The prevalence of QP timing noise in pulsars exacerbates this problem as it can lead to spurious signals that greatly increase the detection threshold. 
Nevertheless, we can certainly rule out a population of planets greater than $\sim10 \,\mathrm{M}_{\oplus}$ from our pulsars, even those with extensive timing noise. This also rules out a population of Jupiter-mass companions (including so-called ``diamond planets'') with orbital periods in our search range.

As seen in Fig.~\ref{fig:all_post}, most of the known exoplanets could be ruled out around our sample of pulsars. Recent results suggest that the distribution of exoplanet-to-host-star mass ratio ($q$) follows a broken power-law, with maxima at $q\sim3\times10^{-5}$ for systems with host stars being ``G dwarfs'' \citep{Pascucci2018}, and at $q\sim2\times10^{-4}$ for ``microlensing'' systems \citep{Suzuki2016, Udalski2018}, respectively. 
One of the pulsar planet formation models requires the planet(s) to have initially formed around a massive star and survive the subsequent supernova state \citep[see e.g.][]{Bailes1991, Podsiadlowski1993}. If we were to assume the shape of the microlensing $q$-distribution holds when extrapolated to host stars massive enough to be pulsar progenitors, i.e. more than $\sim8\,\mathrm{M}_{\oplus}$ \citep[see e.g.][]{Woosley1986}, the most likely planets would have masses much above $100\,\mathrm{M}_{\oplus}$, and anything below $10\,\mathrm{M}_{\oplus}$ would be extremely unlikely.  Therefore, as also argued by \citet{Martin2016}, our analysis would support the idea that fossil (or remnant) planets are, at most, rare.
If, on the other hand, we were to make the even more speculative assumption that planets forming around pulsars would follow similar $q$-distributions as more typical star systems, we would also expect that these exoplanets were $10-100\,\mathrm{M}_{\oplus}$, which are absent from our search.

Alternatively we might consider our results in the context of PSR B1257$+$12, which is the only pulsar known to host terrestrial (i.e. $\sim$ Earth-mass) planets.
The joint posterior shown in Fig.~\ref{fig:all_post} implies that the probability that a system with a planet of mass similar to that of the larger planets of B1257$+$12 ($\sim 4\,\mathrm{M}_{\oplus}$) would have gone undetected in our sample of pulsars is less than 0.3.
This implies a reasonable upper bound of 3 undetected B1257$+$12-like planetary companions amongst our 800 pulsars, with a probability of less than 0.03.
Given our broad sample, this implies that no more than 0.5\% of pulsars host terrestrial planets as large as those in PSR B1257$+$12.
We therefore confirm the hypothesis that the formation of planets around pulsars is rare, and PSR B1257$+$12 is a special case.
It is worth noting that the smaller planet in PSR B1257$+$12 ($0.02\,\mathrm{M}_{\oplus}$) would be undetectable in 95\% of our sample, and we cannot rule out a large population of tiny planets or asteroids.
However, given the unique case of PSR B1257$+$12, it is unclear if such tiny planets are likely to form without the presence of larger planets.
Further, such planets may be part of larger debris disks or asteroid belts and exhibit noise-like timing behaviour, rather than the purely periodic signal considered in this work \citep{Jennings2020}.
Further study into the existence of a population of these systems with only small planets would help determine how common a system such as that of B1257$+$12 truly is.

\section{Conclusions}

In this work, we have presented the largest search for planets orbiting pulsars to date, analysing JBO observations of 800 pulsars. We confirm that PSR B1257$+$12 must have an unusual formation mechanism, placing an upper bound of $0.5\%$ of pulsars exhibiting similar planets. We rule out a population of undetected planetary companions greater than $\sim10\,\mathrm{M}_{\oplus}$.
The timing noise present in most pulsars means that we cannot rule out a substantial population of tiny ($<0.1\,\mathrm{M}_{\oplus}$) planets, though it is not clear if such planets would exist in isolation.

For most pulsars in our data set, no planet-like periodicity was detected. Of the few that did show these detections, several are known to exhibit QP spin-noise which is correlated with variations in their pulse profile. We have identified that PSR B0144$+$59 also exhibits this behaviour, by looking at the variation of its pulse profile shape. We suspect that many of the detections made in our analysis are caused by similar effects, but the signal-to-noise is insufficient to see pulse profile changes. The prevalence of QP noise in pulsar residuals makes searches for low-mass planetary companions to pulsars very challenging and further work is needed for robust pulsar timing analysis in the presence of such noise.

\section*{Acknowledgements}
Pulsar research at Jodrell Bank is supported by a consolidated grant from the UK Science and Technology Facilities Council (STFC). ICN is also supported by the STFC doctoral training grant ST/T506291/1.

\section*{Data Availability}
The data underlying the work in this paper are available upon reasonable request.
Linearised mass posterior distributions and additional figures showing the mass limits for individual pulsars are available at \href{https://doi.org/10.5281/zenodo.5751995}{https://doi.org/10.5281/zenodo.5751995}.
 



\bibliographystyle{mnras}
\bibliography{bibliography} 

\begin{thebibliography}{}
\makeatletter
\relax
\def\mn@urlcharsother{\let\do\@makeother \do\$\do\&\do\#\do\^\do\_\do\%\do\~}
\def\mn@doi{\begingroup\mn@urlcharsother \@ifnextchar [ {\mn@doi@}
  {\mn@doi@[]}}
\def\mn@doi@[#1]#2{\def\@tempa{#1}\ifx\@tempa\@empty \href
  {http://dx.doi.org/#2} {doi:#2}\else \href {http://dx.doi.org/#2} {#1}\fi
  \endgroup}
\def\mn@eprint#1#2{\mn@eprint@#1:#2::\@nil}
\def\mn@eprint@arXiv#1{\href {http://arxiv.org/abs/#1} {{\tt arXiv:#1}}}
\def\mn@eprint@dblp#1{\href {http://dblp.uni-trier.de/rec/bibtex/#1.xml}
  {dblp:#1}}
\def\mn@eprint@#1:#2:#3:#4\@nil{\def\@tempa {#1}\def\@tempb {#2}\def\@tempc
  {#3}\ifx \@tempc \@empty \let \@tempc \@tempb \let \@tempb \@tempa \fi \ifx
  \@tempb \@empty \def\@tempb {arXiv}\fi \@ifundefined
  {mn@eprint@\@tempb}{\@tempb:\@tempc}{\expandafter \expandafter \csname
  mn@eprint@\@tempb\endcsname \expandafter{\@tempc}}}

\bibitem[\protect\citeauthoryear{{Baer} \& {Chesley}}{{Baer} \&
  {Chesley}}{2008}]{Baer2008}
{Baer} J.,  {Chesley} S.~R.,  2008, \mn@doi [Celestial Mechanics and Dynamical
  Astronomy] {10.1007/s10569-007-9103-8}, \href
  {https://ui.adsabs.harvard.edu/abs/2008CeMDA.100...27B} {100, 27}

\bibitem[\protect\citeauthoryear{{Bailes}, {Lyne}  \& {Shemar}}{{Bailes}
  et~al.}{1991}]{Bailes1991}
{Bailes} M.,  {Lyne} A.~G.,   {Shemar} S.~L.,  1991, \mn@doi [\nat]
  {10.1038/352311a0}, \href
  {https://ui.adsabs.harvard.edu/abs/1991Natur.352..311B} {352, 311}

\bibitem[\protect\citeauthoryear{{Bailes}, {Lyne}  \& {Shemar}}{{Bailes}
  et~al.}{1993}]{Bailes1993}
{Bailes} M.,  {Lyne} A.~G.,   {Shemar} S.~L.,  1993, in {Phillips} J.~A.,
  {Thorsett} S.~E.,   {Kulkarni} S.~R.,  eds,  Astronomical Society of the
  Pacific Conference Series Vol. 36, Planets Around Pulsars. pp 19--30

\bibitem[\protect\citeauthoryear{{Bailes} et~al.,}{{Bailes}
  et~al.}{2011}]{Bailes2011}
{Bailes} M.,  et~al., 2011, \mn@doi [Science] {10.1126/science.1208890}, \href
  {https://ui.adsabs.harvard.edu/abs/2011Sci...333.1717B} {333, 1717}

\bibitem[\protect\citeauthoryear{{Basu} \& {Mitra}}{{Basu} \&
  {Mitra}}{2019}]{Basu2019}
{Basu} R.,  {Mitra} D.,  2019, \mn@doi [\mnras] {10.1093/mnras/stz1590}, \href
  {https://ui.adsabs.harvard.edu/abs/2019MNRAS.487.4536B} {487, 4536}

\bibitem[\protect\citeauthoryear{{Basu}, {Joshi}, {Krishnakumar},
  {Bhattacharya}, {Nandi}, {Bandhopadhay}, {Char}  \& {Manoharan}}{{Basu}
  et~al.}{2020}]{Basu2020}
{Basu} A.,  {Joshi} B.~C.,  {Krishnakumar} M.~A.,  {Bhattacharya} D.,  {Nandi}
  R.,  {Bandhopadhay} D.,  {Char} P.,   {Manoharan} P.~K.,  2020, \mn@doi
  [\mnras] {10.1093/mnras/stz3230}, \href
  {https://ui.adsabs.harvard.edu/abs/2020MNRAS.491.3182B} {491, 3182}

\bibitem[\protect\citeauthoryear{{Basu} et~al.,}{{Basu}
  et~al.}{2021}]{Basu2021}
{Basu} A.,  et~al., 2021, \mn@doi [\mnras] {10.1093/mnras/stab3336}, \href
  {https://ui.adsabs.harvard.edu/abs/2021MNRAS.tmp.3056B} {}

\bibitem[\protect\citeauthoryear{{Behrens} et~al.,}{{Behrens}
  et~al.}{2020}]{Behrens2020}
{Behrens} E.~A.,  et~al., 2020, \mn@doi [\apjl] {10.3847/2041-8213/ab8121},
  \href {https://ui.adsabs.harvard.edu/abs/2020ApJ...893L...8B} {893, L8}

\bibitem[\protect\citeauthoryear{{Blandford} \& {Teukolsky}}{{Blandford} \&
  {Teukolsky}}{1976}]{Blandford1976}
{Blandford} R.,  {Teukolsky} S.~A.,  1976, \mn@doi [\apj] {10.1086/154315},
  \href {https://ui.adsabs.harvard.edu/abs/1976ApJ...205..580B} {205, 580}

\bibitem[\protect\citeauthoryear{{Coles}, {Hobbs}, {Champion}, {Manchester}  \&
  {Verbiest}}{{Coles} et~al.}{2011}]{Coles2011}
{Coles} W.,  {Hobbs} G.,  {Champion} D.~J.,  {Manchester} R.~N.,   {Verbiest}
  J.~P.~W.,  2011, \mn@doi [MNRAS] {10.1111/j.1365-2966.2011.19505.x}, \href
  {https://ui.adsabs.harvard.edu/abs/2011MNRAS.418..561C} {418, 561}

\bibitem[\protect\citeauthoryear{{Cordes}}{{Cordes}}{1993}]{Cordes1993}
{Cordes} J.~M.,  1993, in {Phillips} J.~A.,  {Thorsett} S.~E.,   {Kulkarni}
  S.~R.,  eds,  Astronomical Society of the Pacific Conference Series Vol. 36,
  Planets Around Pulsars. pp 43--60

\bibitem[\protect\citeauthoryear{{Downs} \& {Krause-Polstorff}}{{Downs} \&
  {Krause-Polstorff}}{1986}]{Downs1986}
{Downs} G.~S.,  {Krause-Polstorff} J.,  1986, \mn@doi [\apjs] {10.1086/191134},
  \href {https://ui.adsabs.harvard.edu/abs/1986ApJS...62...81D} {62, 81}

\bibitem[\protect\citeauthoryear{{Downs} \& {Reichley}}{{Downs} \&
  {Reichley}}{1983}]{Downs1983}
{Downs} G.~S.,  {Reichley} P.~E.,  1983, \mn@doi [\apjs] {10.1086/190890},
  \href {https://ui.adsabs.harvard.edu/abs/1983ApJS...53..169D} {53, 169}

\bibitem[\protect\citeauthoryear{{Edwards}, {Hobbs}  \& {Manchester}}{{Edwards}
  et~al.}{2006}]{tempo2006}
{Edwards} R.~T.,  {Hobbs} G.~B.,   {Manchester} R.~N.,  2006, \mn@doi [MNRAS]
  {10.1111/j.1365-2966.2006.10870.x}, \href
  {https://ui.adsabs.harvard.edu/abs/2006MNRAS.372.1549E} {372, 1549}

\bibitem[\protect\citeauthoryear{{Ellis}, {Vallisneri}, {Taylor}  \&
  {Baker}}{{Ellis} et~al.}{2019}]{Ellis2019}
{Ellis} J.~A.,  {Vallisneri} M.,  {Taylor} S.~R.,   {Baker} P.~T.,  2019,
  {ENTERPRISE: Enhanced Numerical Toolbox Enabling a Robust PulsaR Inference
  SuitE} (\mn@eprint {ascl} {1912.015})

\bibitem[\protect\citeauthoryear{{Espinoza}, {Lyne}, {Stappers}  \&
  {Kramer}}{{Espinoza} et~al.}{2011}]{Espinoza2011}
{Espinoza} C.~M.,  {Lyne} A.~G.,  {Stappers} B.~W.,   {Kramer} M.,  2011,
  \mn@doi [\mnras] {10.1111/j.1365-2966.2011.18503.x}, \href
  {https://ui.adsabs.harvard.edu/abs/2011MNRAS.414.1679E} {414, 1679}

\bibitem[\protect\citeauthoryear{{Foreman-Mackey}, {Hogg}, {Lang}  \&
  {Goodman}}{{Foreman-Mackey} et~al.}{2013}]{Foreman-Mackey2013}
{Foreman-Mackey} D.,  {Hogg} D.~W.,  {Lang} D.,   {Goodman} J.,  2013, \mn@doi
  [\pasp] {10.1086/670067}, \href
  {https://ui.adsabs.harvard.edu/abs/2013PASP..125..306F} {125, 306}

\bibitem[\protect\citeauthoryear{{Foreman-Mackey} et~al.,}{{Foreman-Mackey}
  et~al.}{2019}]{Foreman-Mackey2019}
{Foreman-Mackey} D.,  et~al., 2019, \mn@doi [The Journal of Open Source
  Software] {10.21105/joss.01864}, \href
  {https://ui.adsabs.harvard.edu/abs/2019JOSS....4.1864F} {4, 1864}

\bibitem[\protect\citeauthoryear{{Fuentes}, {Espinoza}  \&
  {Reisenegger}}{{Fuentes} et~al.}{2019}]{Fuentes2019}
{Fuentes} J.~R.,  {Espinoza} C.~M.,   {Reisenegger} A.,  2019, \mn@doi [\aap]
  {10.1051/0004-6361/201935939}, \href
  {https://ui.adsabs.harvard.edu/abs/2019A&A...630A.115F} {630, A115}

\bibitem[\protect\citeauthoryear{{Herfindal} \& {Rankin}}{{Herfindal} \&
  {Rankin}}{2009}]{Herfindal2009}
{Herfindal} J.~L.,  {Rankin} J.~M.,  2009, \mn@doi [\mnras]
  {10.1111/j.1365-2966.2008.14119.x}, \href
  {https://ui.adsabs.harvard.edu/abs/2009MNRAS.393.1391H} {393, 1391}

\bibitem[\protect\citeauthoryear{{Hobbs}, {Lyne}, {Kramer}, {Martin}  \&
  {Jordan}}{{Hobbs} et~al.}{2004}]{Hobbs2004}
{Hobbs} G.,  {Lyne} A.~G.,  {Kramer} M.,  {Martin} C.~E.,   {Jordan} C.,  2004,
  \mn@doi [\mnras] {10.1111/j.1365-2966.2004.08157.x}, \href
  {https://ui.adsabs.harvard.edu/abs/2004MNRAS.353.1311H} {353, 1311}

\bibitem[\protect\citeauthoryear{{Hobbs}, {Edwards}  \& {Manchester}}{{Hobbs}
  et~al.}{2006}]{Hobbs2006}
{Hobbs} G.~B.,  {Edwards} R.~T.,   {Manchester} R.~N.,  2006, \mn@doi [\mnras]
  {10.1111/j.1365-2966.2006.10302.x}, \href
  {https://ui.adsabs.harvard.edu/abs/2006MNRAS.369..655H} {369, 655}

\bibitem[\protect\citeauthoryear{{Hobbs}, {Lyne}  \& {Kramer}}{{Hobbs}
  et~al.}{2010}]{Hobbs2010}
{Hobbs} G.,  {Lyne} A.~G.,   {Kramer} M.,  2010, \mn@doi [MNRAS]
  {10.1111/j.1365-2966.2009.15938.x}, \href
  {https://ui.adsabs.harvard.edu/abs/2010MNRAS.402.1027H} {402, 1027}

\bibitem[\protect\citeauthoryear{{Jennings}, {Cordes}  \&
  {Chatterjee}}{{Jennings} et~al.}{2020}]{Jennings2020}
{Jennings} R.~J.,  {Cordes} J.~M.,   {Chatterjee} S.,  2020, \mn@doi [\apj]
  {10.3847/1538-4357/abc178}, \href
  {https://ui.adsabs.harvard.edu/abs/2020ApJ...904..191J} {904, 191}

\bibitem[\protect\citeauthoryear{{Karttunen}, {Kr\"{o}ger}, {Oja}, {Poutanen}
  \& {Donner}}{{Karttunen} et~al.}{2007}]{FundAstro}
{Karttunen} H.,  {Kr\"{o}ger} P.,  {Oja} H.,  {Poutanen} M.,   {Donner} K.~J.,
  2007, {Fundamental Astronomy}, fifth edn.
Springer

\bibitem[\protect\citeauthoryear{Kass \& Raftery}{Kass \&
  Raftery}{1995}]{logBayes}
Kass R.~E.,  Raftery A.~E.,  1995, Journal of the American Statistical
  Association, 90, 773

\bibitem[\protect\citeauthoryear{Keith, Niţu  \& Liu}{Keith
  et~al.}{2022}]{run_enterprise}
Keith M.~J.,  Niţu I.~C.,   Liu Y.,  2022, run\_enterprise,
  \mn@doi{10.5281/zenodo.6046212}

\bibitem[\protect\citeauthoryear{{Kerr}, {Johnston}, {Hobbs}  \&
  {Shannon}}{{Kerr} et~al.}{2015}]{Kerr2015_1}
{Kerr} M.,  {Johnston} S.,  {Hobbs} G.,   {Shannon} R.~M.,  2015, \mn@doi [The
  Astrophysical Journal Letters] {10.1088/2041-8205/809/1/L11}, \href
  {https://ui.adsabs.harvard.edu/abs/2015ApJ...809L..11K} {809, L11}

\bibitem[\protect\citeauthoryear{{Konacki} \& {Wolszczan}}{{Konacki} \&
  {Wolszczan}}{2003}]{Konacki2003}
{Konacki} M.,  {Wolszczan} A.,  2003, \mn@doi [\apjl] {10.1086/377093}, \href
  {https://ui.adsabs.harvard.edu/abs/2003ApJ...591L.147K} {591, L147}

\bibitem[\protect\citeauthoryear{{Konacki}, {Lewandowski}, {Wolszczan},
  {Doroshenko}  \& {Kramer}}{{Konacki} et~al.}{1999}]{Konacki1999}
{Konacki} M.,  {Lewandowski} W.,  {Wolszczan} A.,  {Doroshenko} O.,   {Kramer}
  M.,  1999, \mn@doi [\apjl] {10.1086/312089}, \href
  {https://ui.adsabs.harvard.edu/abs/1999ApJ...519L..81K} {519, L81}

\bibitem[\protect\citeauthoryear{{Kramer}, {Lyne}, {O'Brien}, {Jordan}  \&
  {Lorimer}}{{Kramer} et~al.}{2006}]{Kramer2006}
{Kramer} M.,  {Lyne} A.~G.,  {O'Brien} J.~T.,  {Jordan} C.~A.,   {Lorimer}
  D.~R.,  2006, \mn@doi [Science] {10.1126/science.1124060}, \href
  {https://ui.adsabs.harvard.edu/abs/2006Sci...312..549K} {312, 549}

\bibitem[\protect\citeauthoryear{{Krolik}}{{Krolik}}{1991}]{Krolik1991}
{Krolik} J.~H.,  1991, \mn@doi [\nat] {10.1038/353829a0}, \href
  {https://ui.adsabs.harvard.edu/abs/1991Natur.353..829K} {353, 829}

\bibitem[\protect\citeauthoryear{{Lattimer}}{{Lattimer}}{2012}]{Lattimer2012}
{Lattimer} J.~M.,  2012, \mn@doi [Annual Review of Nuclear and Particle
  Science] {10.1146/annurev-nucl-102711-095018}, \href
  {https://ui.adsabs.harvard.edu/abs/2012ARNPS..62..485L} {62, 485}

\bibitem[\protect\citeauthoryear{{Lentati}, {Alexander}, {Hobson}, {Feroz},
  {van Haasteren}, {Lee}  \& {Shannon}}{{Lentati} et~al.}{2014}]{Lentati2014}
{Lentati} L.,  {Alexander} P.,  {Hobson} M.~P.,  {Feroz} F.,  {van Haasteren}
  R.,  {Lee} K.~J.,   {Shannon} R.~M.,  2014, \mn@doi [MNRAS]
  {10.1093/mnras/stt2122}, \href
  {https://ui.adsabs.harvard.edu/abs/2014MNRAS.437.3004L} {437, 3004}

\bibitem[\protect\citeauthoryear{{Lin}, {Woosley}  \& {Bodenheimer}}{{Lin}
  et~al.}{1991}]{Lin1991}
{Lin} D.~N.~C.,  {Woosley} S.~E.,   {Bodenheimer} P.~H.,  1991, \mn@doi [\nat]
  {10.1038/353827a0}, \href
  {https://ui.adsabs.harvard.edu/abs/1991Natur.353..827L} {353, 827}

\bibitem[\protect\citeauthoryear{{Lissauer} et~al.,}{{Lissauer}
  et~al.}{2011}]{Lissauer2011}
{Lissauer} J.~J.,  et~al., 2011, \mn@doi [\apjs] {10.1088/0067-0049/197/1/8},
  \href {https://ui.adsabs.harvard.edu/abs/2011ApJS..197....8L} {197, 8}

\bibitem[\protect\citeauthoryear{{Liu}, {Keane}, {Lee}, {Kramer}, {Cordes}  \&
  {Purver}}{{Liu} et~al.}{2012}]{Liu2012}
{Liu} K.,  {Keane} E.~F.,  {Lee} K.~J.,  {Kramer} M.,  {Cordes} J.~M.,
  {Purver} M.~B.,  2012, \mn@doi [\mnras] {10.1111/j.1365-2966.2011.20041.x},
  \href {https://ui.adsabs.harvard.edu/abs/2012MNRAS.420..361L} {420, 361}

\bibitem[\protect\citeauthoryear{{Lyne}, {Hobbs}, {Kramer}, {Stairs}  \&
  {Stappers}}{{Lyne} et~al.}{2010}]{Lyne2010}
{Lyne} A.,  {Hobbs} G.,  {Kramer} M.,  {Stairs} I.,   {Stappers} B.,  2010,
  \mn@doi [Science] {10.1126/science.1186683}, \href
  {https://ui.adsabs.harvard.edu/abs/2010Sci...329..408L} {329, 408}

\bibitem[\protect\citeauthoryear{{Malhotra}}{{Malhotra}}{1993}]{Malhotra1993}
{Malhotra} R.,  1993, in {Phillips} J.~A.,  {Thorsett} S.~E.,   {Kulkarni}
  S.~R.,  eds,  Astronomical Society of the Pacific Conference Series Vol. 36,
  Planets Around Pulsars. pp 89--106

\bibitem[\protect\citeauthoryear{{Manchester} et~al.,}{{Manchester}
  et~al.}{2013}]{Manchester2013}
{Manchester} R.~N.,  et~al., 2013, \mn@doi [\pasa] {10.1017/pasa.2012.017},
  \href {https://ui.adsabs.harvard.edu/abs/2013PASA...30...17M} {30, e017}

\bibitem[\protect\citeauthoryear{{Martin}, {Livio}  \& {Palaniswamy}}{{Martin}
  et~al.}{2016}]{Martin2016}
{Martin} R.~G.,  {Livio} M.,   {Palaniswamy} D.,  2016, \mn@doi [\apj]
  {10.3847/0004-637X/832/2/122}, \href
  {https://ui.adsabs.harvard.edu/abs/2016ApJ...832..122M} {832, 122}

\bibitem[\protect\citeauthoryear{{Mottez}, {Bonazzola}  \&
  {Heyvaerts}}{{Mottez} et~al.}{2013}]{Mottez2013}
{Mottez} F.,  {Bonazzola} S.,   {Heyvaerts} J.,  2013, \mn@doi [\aap]
  {10.1051/0004-6361/201321182}, \href
  {https://ui.adsabs.harvard.edu/abs/2013A&A...555A.125M} {555, A125}

\bibitem[\protect\citeauthoryear{{Nice} et~al.,}{{Nice}
  et~al.}{2013}]{Nice2013}
{Nice} D.~J.,  et~al., 2013, \mn@doi [\apj] {10.1088/0004-637X/772/1/50}, \href
  {https://ui.adsabs.harvard.edu/abs/2013ApJ...772...50N} {772, 50}

\bibitem[\protect\citeauthoryear{{Pascucci}, {Mulders}, {Gould}  \&
  {Fernandes}}{{Pascucci} et~al.}{2018}]{Pascucci2018}
{Pascucci} I.,  {Mulders} G.~D.,  {Gould} A.,   {Fernandes} R.,  2018, \mn@doi
  [\apjl] {10.3847/2041-8213/aab6ac}, \href
  {https://ui.adsabs.harvard.edu/abs/2018ApJ...856L..28P} {856, L28}

\bibitem[\protect\citeauthoryear{{Phillips} \& {Thorsett}}{{Phillips} \&
  {Thorsett}}{1994}]{Phillips1994}
{Phillips} J.~A.,  {Thorsett} S.~E.,  1994, \mn@doi [\apss]
  {10.1007/BF00984513}, \href
  {https://ui.adsabs.harvard.edu/abs/1994Ap&SS.212...91P} {212, 91}

\bibitem[\protect\citeauthoryear{{Pletsch} et~al.,}{{Pletsch}
  et~al.}{2012}]{Pletsch2012}
{Pletsch} H.~J.,  et~al., 2012, \mn@doi [Science] {10.1126/science.1229054},
  \href {https://ui.adsabs.harvard.edu/abs/2012Sci...338.1314P} {338, 1314}

\bibitem[\protect\citeauthoryear{{Podsiadlowski}}{{Podsiadlowski}}{1993}]{Podsiadlowski1993}
{Podsiadlowski} P.,  1993, in {Phillips} J.~A.,  {Thorsett} S.~E.,   {Kulkarni}
  S.~R.,  eds,  Astronomical Society of the Pacific Conference Series Vol. 36,
  Planets Around Pulsars. pp 149--165

\bibitem[\protect\citeauthoryear{{Podsiadlowski}, {Pringle}  \&
  {Rees}}{{Podsiadlowski} et~al.}{1991}]{Podsiadlowski1991}
{Podsiadlowski} P.,  {Pringle} J.~E.,   {Rees} M.~J.,  1991, \mn@doi [\nat]
  {10.1038/352783a0}, \href
  {https://ui.adsabs.harvard.edu/abs/1991Natur.352..783P} {352, 783}

\bibitem[\protect\citeauthoryear{{Rasio}, {Nicholson}, {Shapiro}  \&
  {Teukolsky}}{{Rasio} et~al.}{1992}]{Rasio1992}
{Rasio} F.~A.,  {Nicholson} P.~D.,  {Shapiro} S.~L.,   {Teukolsky} S.~A.,
  1992, \mn@doi [\nat] {10.1038/355325a0}, \href
  {https://ui.adsabs.harvard.edu/abs/1992Natur.355..325R} {355, 325}

\bibitem[\protect\citeauthoryear{{Rea} et~al.,}{{Rea} et~al.}{2008}]{Rea2008}
{Rea} N.,  et~al., 2008, \mn@doi [\mnras] {10.1111/j.1365-2966.2008.13795.x},
  \href {https://ui.adsabs.harvard.edu/abs/2008MNRAS.391..663R} {391, 663}

\bibitem[\protect\citeauthoryear{{Redman} \& {Rankin}}{{Redman} \&
  {Rankin}}{2009}]{Redman2009}
{Redman} S.~L.,  {Rankin} J.~M.,  2009, \mn@doi [\mnras]
  {10.1111/j.1365-2966.2009.14632.x}, \href
  {https://ui.adsabs.harvard.edu/abs/2009MNRAS.395.1529R} {395, 1529}

\bibitem[\protect\citeauthoryear{{Romani}, {Filippenko}, {Silverman}, {Cenko},
  {Greiner}, {Rau}, {Elliott}  \& {Pletsch}}{{Romani}
  et~al.}{2012}]{Romani2012}
{Romani} R.~W.,  {Filippenko} A.~V.,  {Silverman} J.~M.,  {Cenko} S.~B.,
  {Greiner} J.,  {Rau} A.,  {Elliott} J.,   {Pletsch} H.~J.,  2012, \mn@doi
  [\apjl] {10.1088/2041-8205/760/2/L36}, \href
  {https://ui.adsabs.harvard.edu/abs/2012ApJ...760L..36R} {760, L36}

\bibitem[\protect\citeauthoryear{{Shabanova}}{{Shabanova}}{1995}]{Shabanova1995}
{Shabanova} T.~V.,  1995, \mn@doi [\apj] {10.1086/176440}, \href
  {https://ui.adsabs.harvard.edu/abs/1995ApJ...453..779S} {453, 779}

\bibitem[\protect\citeauthoryear{{Shannon} \& {Cordes}}{{Shannon} \&
  {Cordes}}{2010}]{Shannon2010}
{Shannon} R.~M.,  {Cordes} J.~M.,  2010, \mn@doi [\apj]
  {10.1088/0004-637X/725/2/1607}, \href
  {https://ui.adsabs.harvard.edu/abs/2010ApJ...725.1607S} {725, 1607}

\bibitem[\protect\citeauthoryear{{Shannon} et~al.,}{{Shannon}
  et~al.}{2013}]{Shannon2013}
{Shannon} R.~M.,  et~al., 2013, \mn@doi [\apj] {10.1088/0004-637X/766/1/5},
  \href {https://ui.adsabs.harvard.edu/abs/2013ApJ...766....5S} {766, 5}

\bibitem[\protect\citeauthoryear{{Sigurdsson}, {Richer}, {Hansen}, {Stairs}  \&
  {Thorsett}}{{Sigurdsson} et~al.}{2003}]{Sigurdsson2003}
{Sigurdsson} S.,  {Richer} H.~B.,  {Hansen} B.~M.,  {Stairs} I.~H.,
  {Thorsett} S.~E.,  2003, \mn@doi [Science] {10.1126/science.1086326}, \href
  {https://ui.adsabs.harvard.edu/abs/2003Sci...301..193S} {301, 193}

\bibitem[\protect\citeauthoryear{{Sobey} et~al.,}{{Sobey}
  et~al.}{2015}]{Sobey2015}
{Sobey} C.,  et~al., 2015, \mn@doi [\mnras] {10.1093/mnras/stv1066}, \href
  {https://ui.adsabs.harvard.edu/abs/2015MNRAS.451.2493S} {451, 2493}

\bibitem[\protect\citeauthoryear{{Speagle}}{{Speagle}}{2020}]{Speagle2020}
{Speagle} J.~S.,  2020, \mn@doi [\mnras] {10.1093/mnras/staa278}, \href
  {https://ui.adsabs.harvard.edu/abs/2020MNRAS.493.3132S} {493, 3132}

\bibitem[\protect\citeauthoryear{{Spiewak} et~al.,}{{Spiewak}
  et~al.}{2018}]{Spiewak2018}
{Spiewak} R.,  et~al., 2018, \mn@doi [\mnras] {10.1093/mnras/stx3157}, \href
  {https://ui.adsabs.harvard.edu/abs/2018MNRAS.475..469S} {475, 469}

\bibitem[\protect\citeauthoryear{{Stairs} et~al.,}{{Stairs}
  et~al.}{2019}]{Stairs2019}
{Stairs} I.~H.,  et~al., 2019, \mn@doi [\mnras] {10.1093/mnras/stz647}, \href
  {https://ui.adsabs.harvard.edu/abs/2019MNRAS.485.3230S} {485, 3230}

\bibitem[\protect\citeauthoryear{{Stevens}, {Rees}  \&
  {Podsiadlowski}}{{Stevens} et~al.}{1992}]{Stevens1992}
{Stevens} I.~R.,  {Rees} M.~J.,   {Podsiadlowski} P.,  1992, \mn@doi [\mnras]
  {10.1093/mnras/254.1.19P}, \href
  {https://ui.adsabs.harvard.edu/abs/1992MNRAS.254P..19S} {254, 19P}

\bibitem[\protect\citeauthoryear{{Stovall} et~al.,}{{Stovall}
  et~al.}{2014}]{Stovall2014}
{Stovall} K.,  et~al., 2014, \mn@doi [\apj] {10.1088/0004-637X/791/1/67}, \href
  {https://ui.adsabs.harvard.edu/abs/2014ApJ...791...67S} {791, 67}

\bibitem[\protect\citeauthoryear{{Suzuki} et~al.,}{{Suzuki}
  et~al.}{2016}]{Suzuki2016}
{Suzuki} D.,  et~al., 2016, \mn@doi [\apj] {10.3847/1538-4357/833/2/145}, \href
  {https://ui.adsabs.harvard.edu/abs/2016ApJ...833..145S} {833, 145}

\bibitem[\protect\citeauthoryear{{Thorsett} \& {Phillips}}{{Thorsett} \&
  {Phillips}}{1992}]{Thorsett1992}
{Thorsett} S.~E.,  {Phillips} J.~A.,  1992, \mn@doi [The Astrophysical Journal
  Letters] {10.1086/186307}, \href
  {https://ui.adsabs.harvard.edu/abs/1992ApJ...387L..69T} {387, L69}

\bibitem[\protect\citeauthoryear{{Udalski} et~al.,}{{Udalski}
  et~al.}{2018}]{Udalski2018}
{Udalski} A.,  et~al., 2018, \mn@doi [\actaa] {10.32023/0001-5237/68.1.1},
  \href {https://ui.adsabs.harvard.edu/abs/2018AcA....68....1U} {68, 1}

\bibitem[\protect\citeauthoryear{{Van Haasteren} et~al.,}{{Van Haasteren}
  et~al.}{2011}]{vanHaasteren2011}
{Van Haasteren} R.,  et~al., 2011, \mn@doi [\mnras]
  {10.1111/j.1365-2966.2011.18613.x}, \href
  {https://ui.adsabs.harvard.edu/abs/2011MNRAS.414.3117V} {414, 3117}

\bibitem[\protect\citeauthoryear{{Van den Heuvel}}{{Van den
  Heuvel}}{1992}]{vandenHeuvel1992}
{Van den Heuvel} E.~P.~J.,  1992, \mn@doi [\nat] {10.1038/356668b0}, \href
  {https://ui.adsabs.harvard.edu/abs/1992Natur.356..668V} {356, 668}

\bibitem[\protect\citeauthoryear{{Weltevrede}}{{Weltevrede}}{2016}]{Weltevrede2016}
{Weltevrede} P.,  2016, \mn@doi [\aap] {10.1051/0004-6361/201527950}, \href
  {https://ui.adsabs.harvard.edu/abs/2016A&A...590A.109W} {590, A109}

\bibitem[\protect\citeauthoryear{{Wolszczan}}{{Wolszczan}}{1994}]{Wolszczan1994}
{Wolszczan} A.,  1994, \mn@doi [Science] {10.1126/science.264.5158.538}, \href
  {https://ui.adsabs.harvard.edu/abs/1994Sci...264..538W} {264, 538}

\bibitem[\protect\citeauthoryear{{Wolszczan} \& {Frail}}{{Wolszczan} \&
  {Frail}}{1992}]{Wolszczan1992}
{Wolszczan} A.,  {Frail} D.~A.,  1992, \mn@doi [Nature] {10.1038/355145a0},
  \href {https://ui.adsabs.harvard.edu/abs/1992Natur.355..145W} {355, 145}

\bibitem[\protect\citeauthoryear{{Woosley} \& {Weaver}}{{Woosley} \&
  {Weaver}}{1986}]{Woosley1986}
{Woosley} S.~E.,  {Weaver} T.~A.,  1986, \mn@doi [\araa]
  {10.1146/annurev.aa.24.090186.001225}, \href
  {https://ui.adsabs.harvard.edu/abs/1986ARA&A..24..205W} {24, 205}

\bibitem[\protect\citeauthoryear{{Wright} et~al.,}{{Wright}
  et~al.}{2011}]{Wright2011}
{Wright} J.~T.,  et~al., 2011, \mn@doi [\pasp] {10.1086/659427}, \href
  {https://ui.adsabs.harvard.edu/abs/2011PASP..123..412W} {123, 412}

\bibitem[\protect\citeauthoryear{{Young}, {Stappers}, {Weltevrede}, {Lyne}  \&
  {Kramer}}{{Young} et~al.}{2012}]{Young2012}
{Young} N.~J.,  {Stappers} B.~W.,  {Weltevrede} P.,  {Lyne} A.~G.,   {Kramer}
  M.,  2012, \mn@doi [\mnras] {10.1111/j.1365-2966.2012.21934.x}, \href
  {https://ui.adsabs.harvard.edu/abs/2012MNRAS.427..114Y} {427, 114}

\bibitem[\protect\citeauthoryear{{Young}, {Stappers}, {Lyne}, {Weltevrede},
  {Kramer}  \& {Cognard}}{{Young} et~al.}{2013}]{Young2013}
{Young} N.~J.,  {Stappers} B.~W.,  {Lyne} A.~G.,  {Weltevrede} P.,  {Kramer}
  M.,   {Cognard} I.,  2013, \mn@doi [\mnras] {10.1093/mnras/sts532}, \href
  {https://ui.adsabs.harvard.edu/abs/2013MNRAS.429.2569Y} {429, 2569}

\bibitem[\protect\citeauthoryear{{Yu} et~al.,}{{Yu} et~al.}{2013}]{Yu2013}
{Yu} M.,  et~al., 2013, \mn@doi [\mnras] {10.1093/mnras/sts366}, \href
  {https://ui.adsabs.harvard.edu/abs/2013MNRAS.429..688Y} {429, 688}

\makeatother
\end{thebibliography}




\appendix

\section{Power spectral density plots} \label{app:PSD}

The plots of power spectral density against frequency for the pulsars showing detections are shown in Fig.~\ref{fig:app_PSD} (except PSR B1540$-$06, see Fig.~\ref{fig:PSD_example}). The frequency range shown is limited at the low end by the timespan of our observations, and at the high end by our chosen value of $0.1\,\mathrm{d}^{-1}$. Notably, the intermittent PSRs B1931$+$24 and B0823$+$26 show that the model described by a single red noise power-law and a flat white noise component does not describe the data well. Specifically in the case of B0823$+$26, the extra, unmodelled power at high frequencies leads to what we consider a false periodicity detection at $28$ days.

The remaining plots show that, in general, our periodicity detections correspond to peaks in the power spectra, unmodelled by the red noise power-law, as expected. However, note that this power-law model is not a perfect description of the variations in the power spectral density, and therefore a more involved analysis is necessary, as described in the main body of this work.

\begin{figure*}
	\centering
	\includegraphics[width=.49\linewidth]{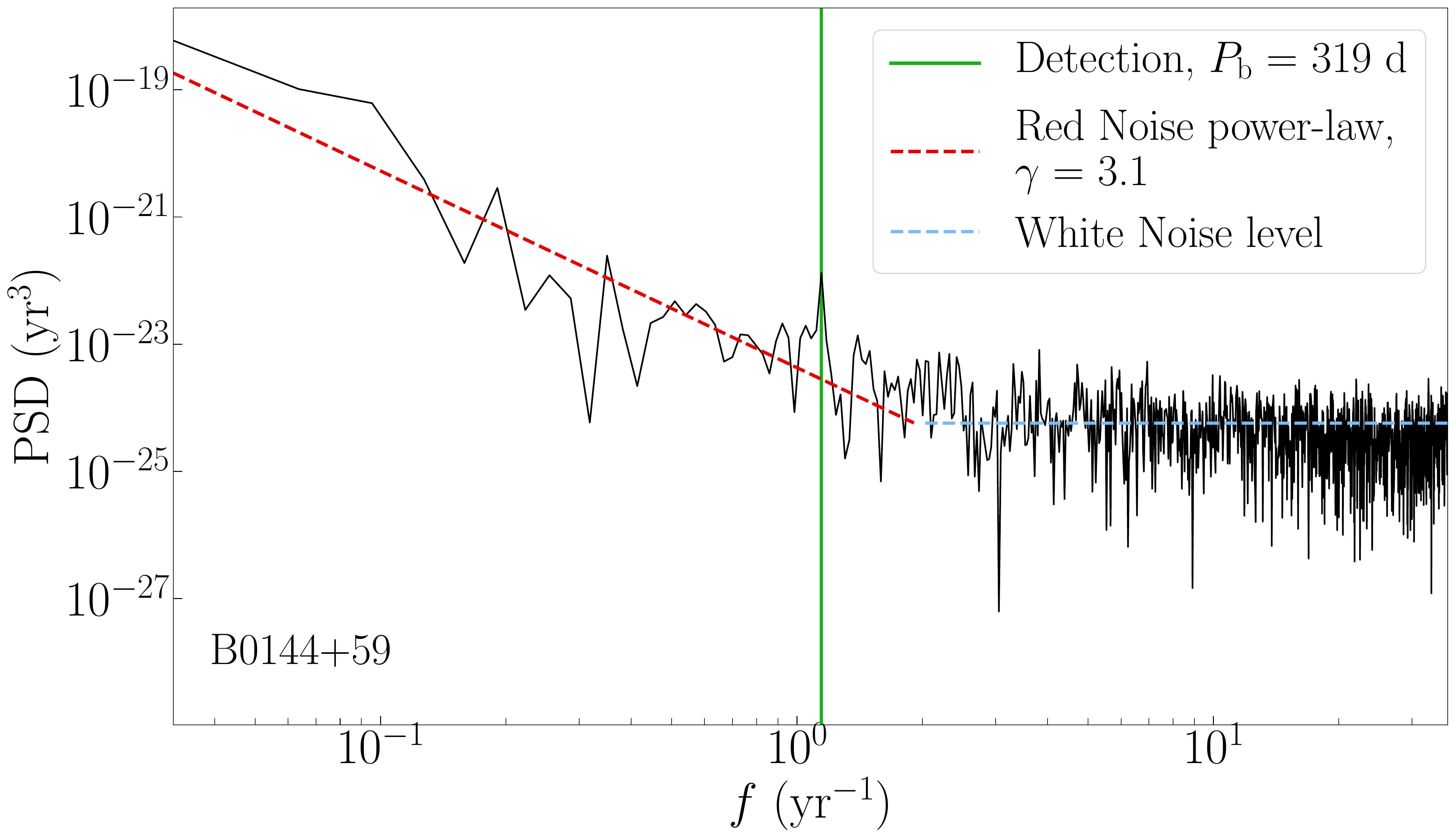}
	\includegraphics[width=.49\linewidth]{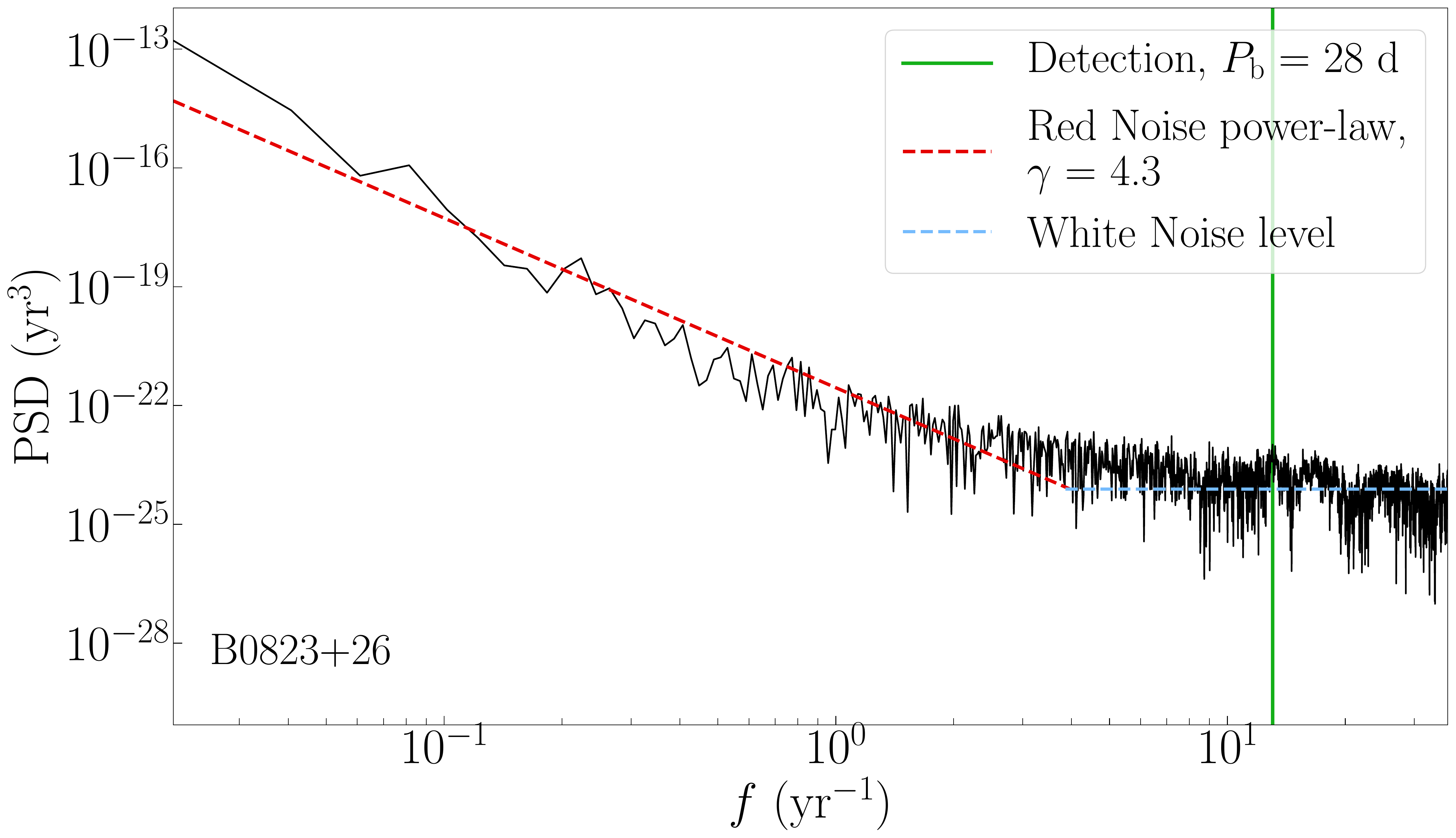}
    \\
	\includegraphics[width=.49\linewidth]{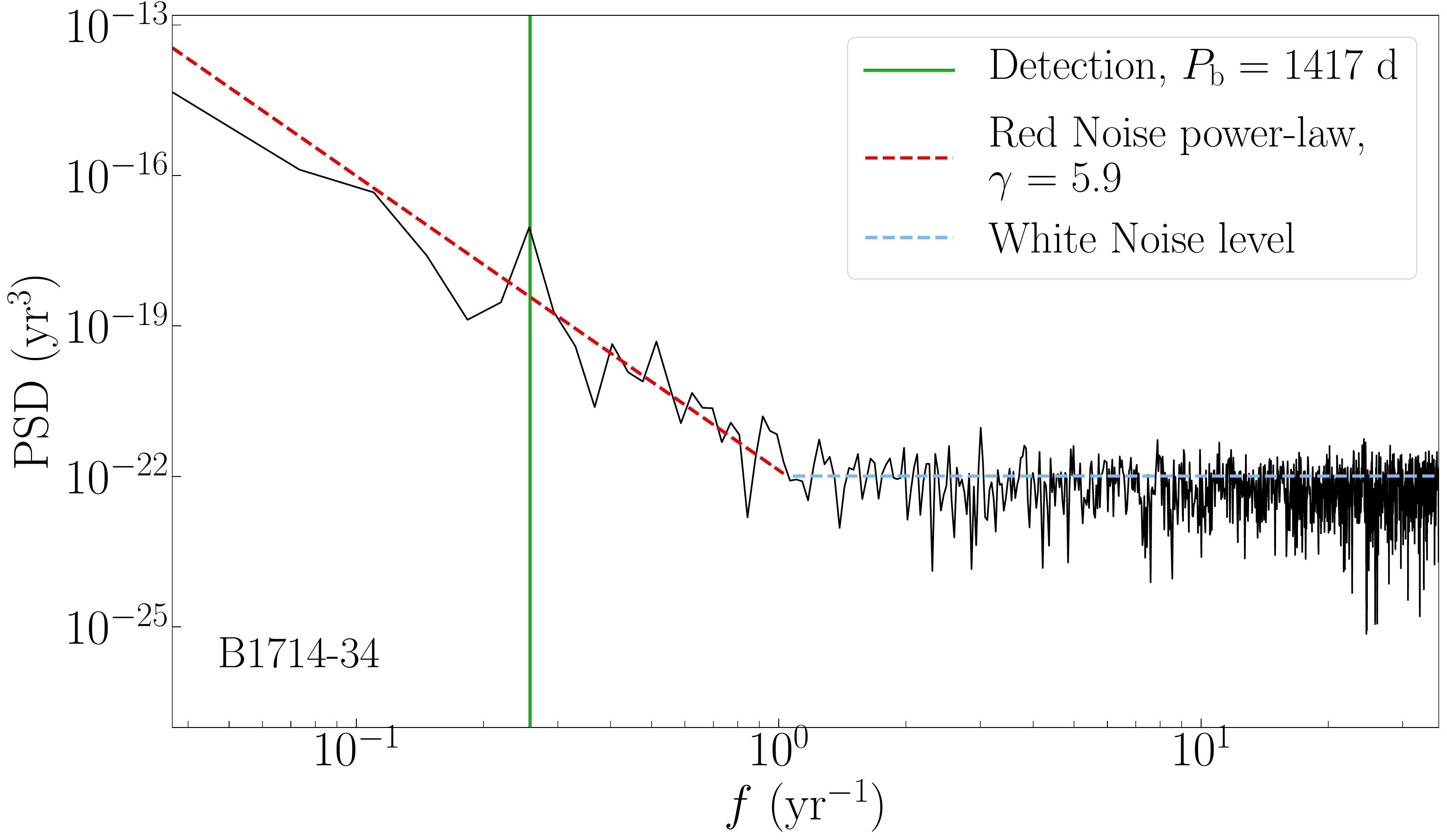}
    \includegraphics[width=.49\linewidth]{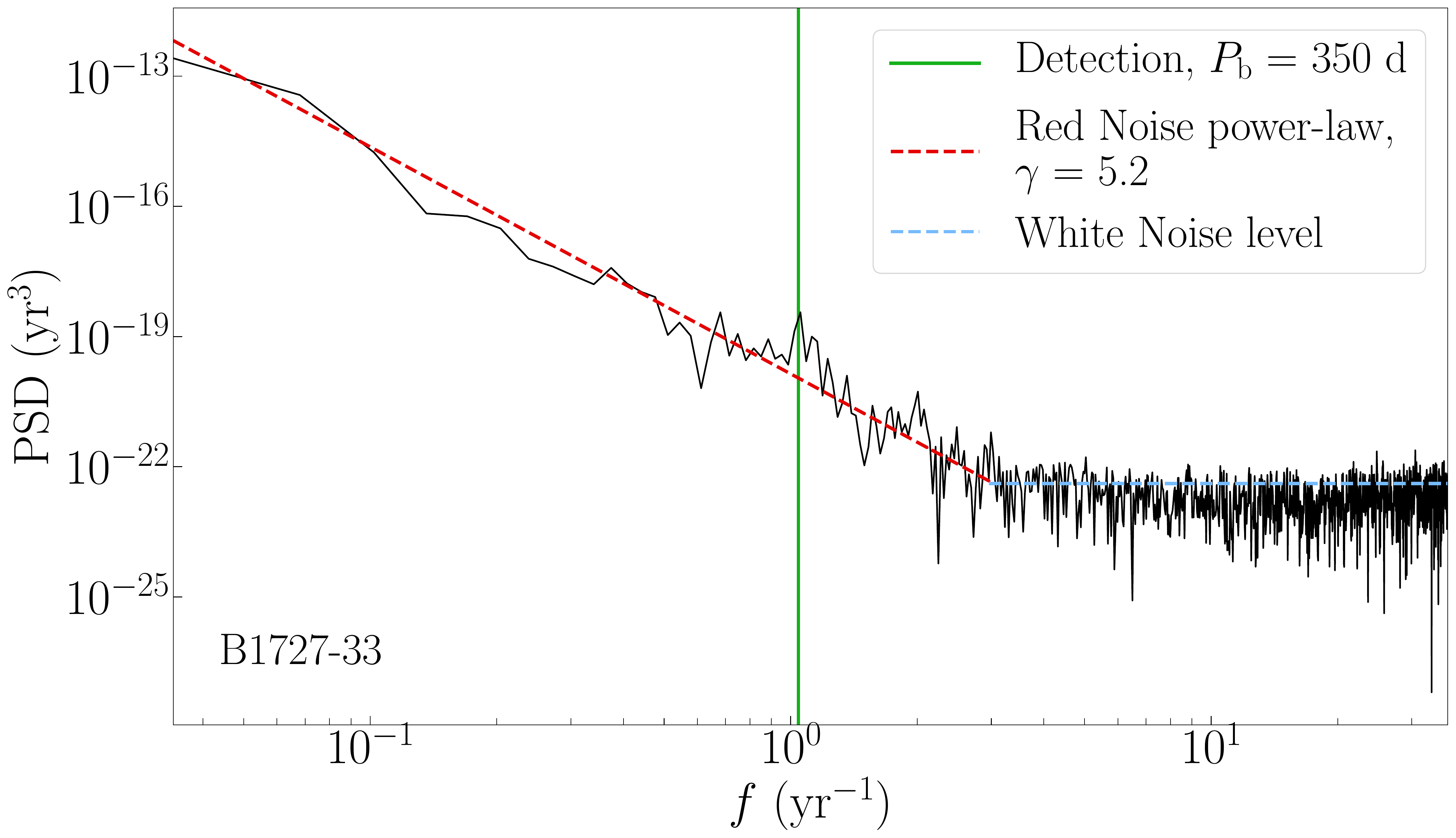}
    \\
    \includegraphics[width=.49\linewidth]{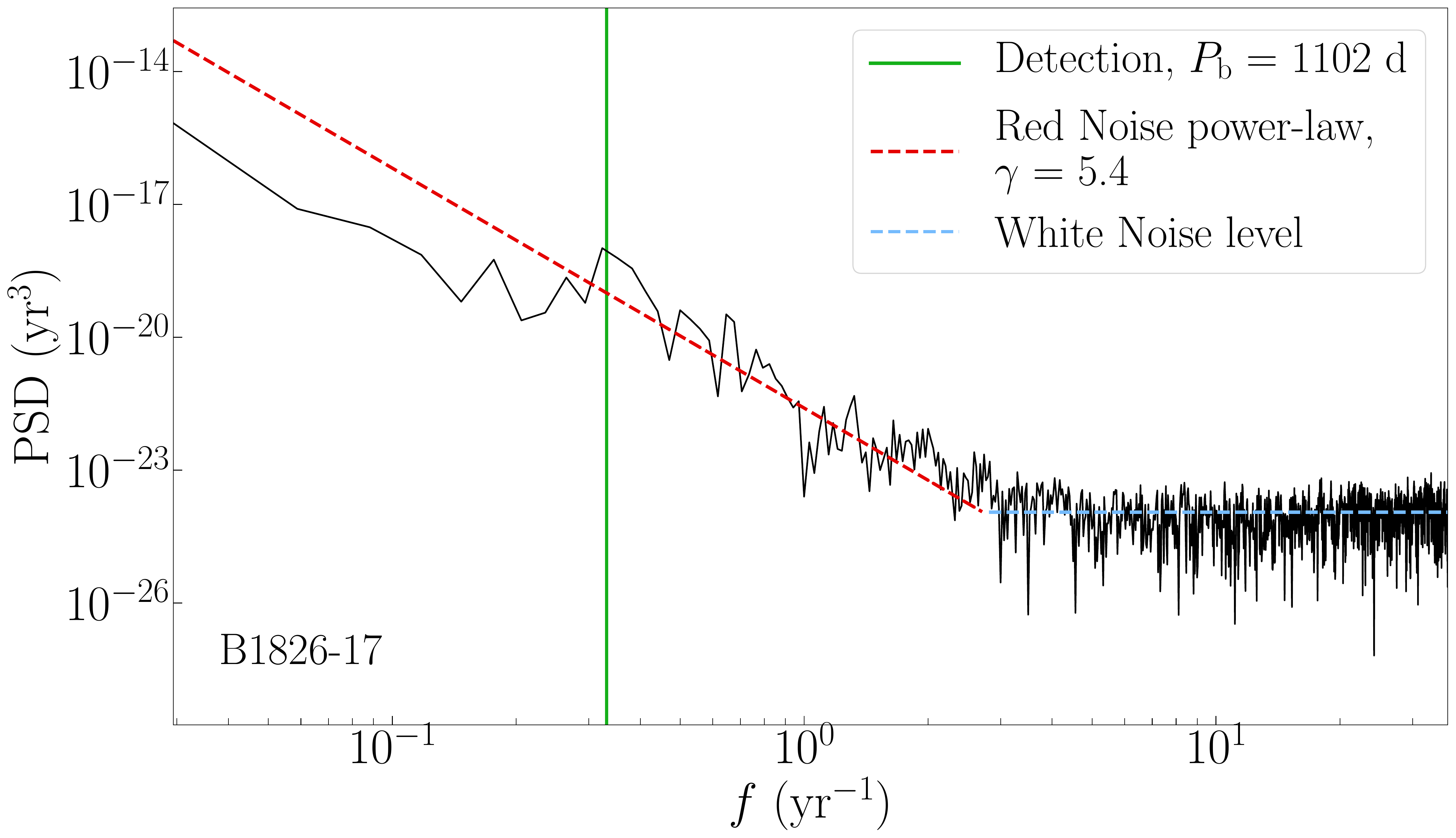} \includegraphics[width=.49\linewidth]{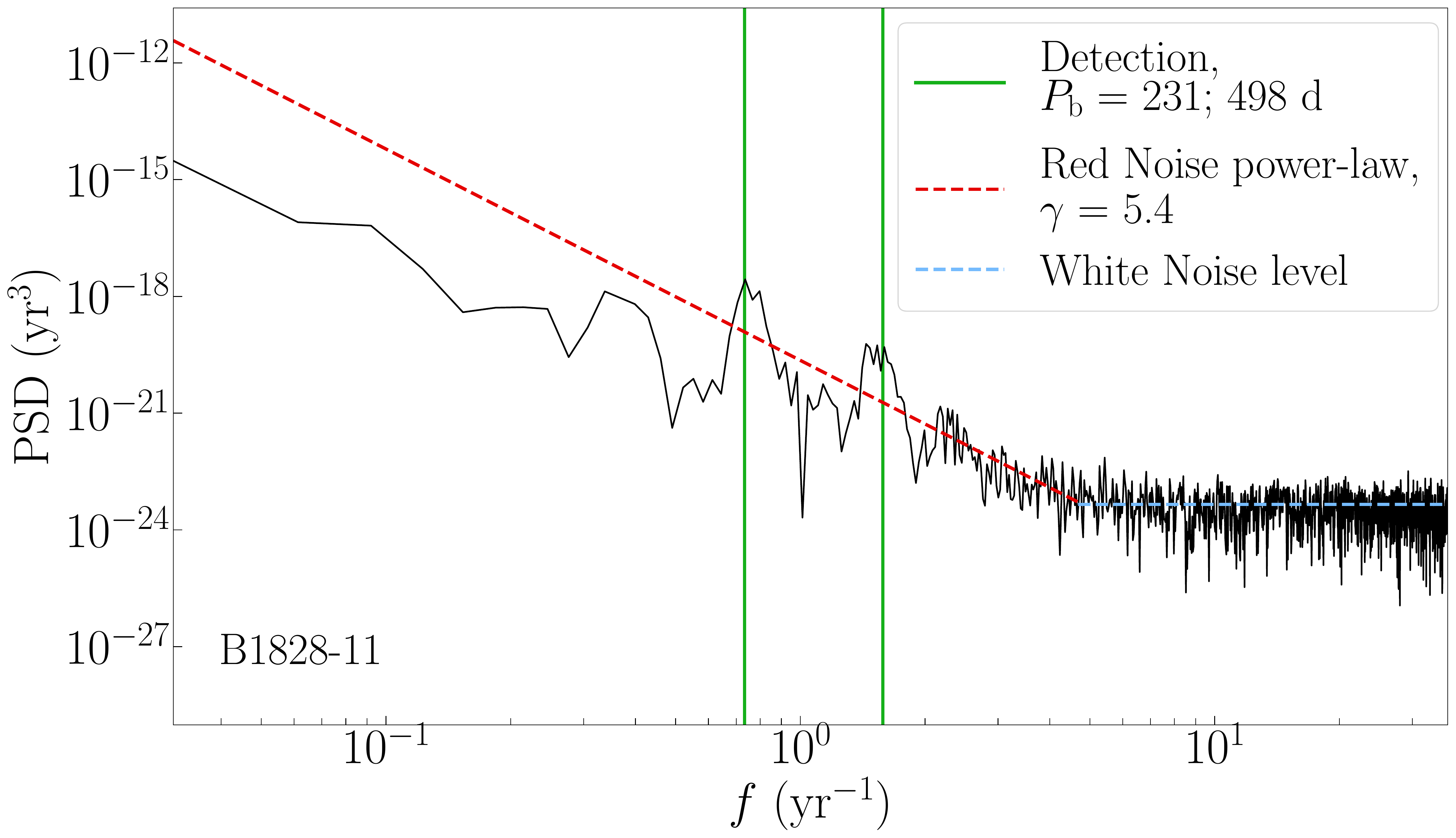}
	\\
	\includegraphics[width=.49\linewidth]{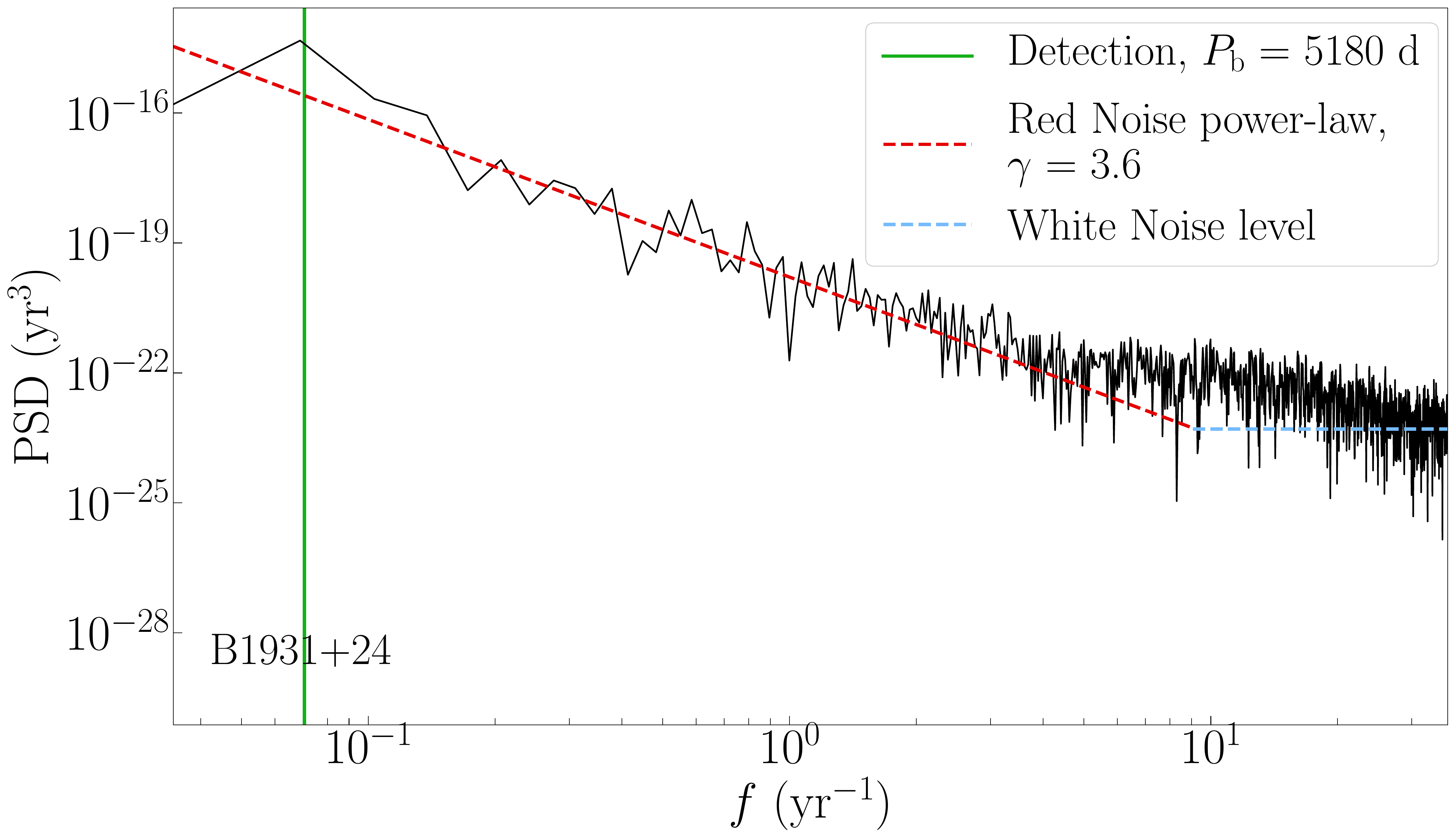}
	\includegraphics[width=.49\linewidth]{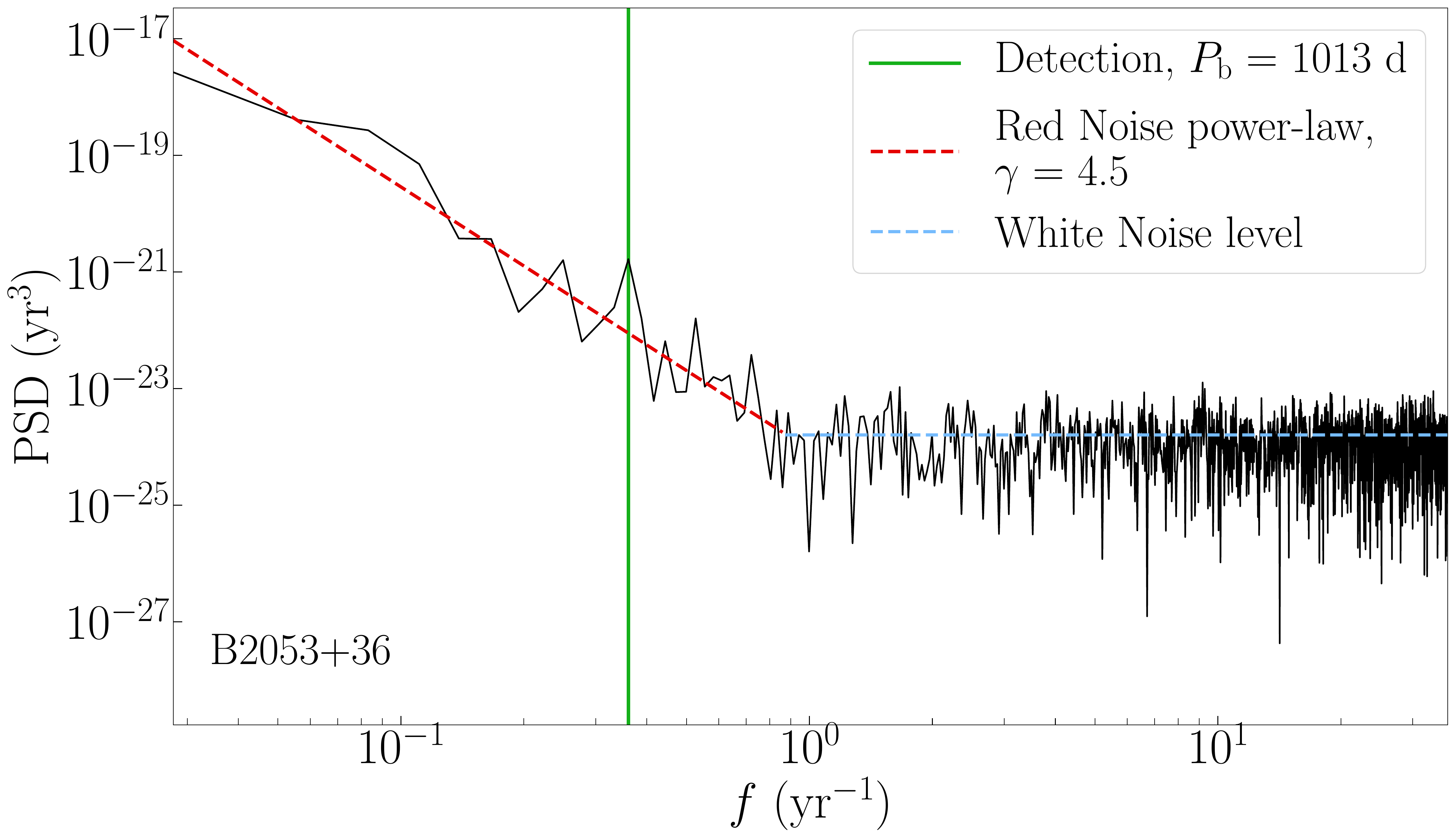}
	
	\caption{The power spectral density against frequency for the pulsars that showed detections, as summarised in Table~\ref{tab:pmresults}. For more details see the caption of Fig.~\ref{fig:PSD_example}}.
	\label{fig:app_PSD}
\end{figure*}

\begin{figure*}\ContinuedFloat
	\centering
	\includegraphics[width=.49\linewidth]{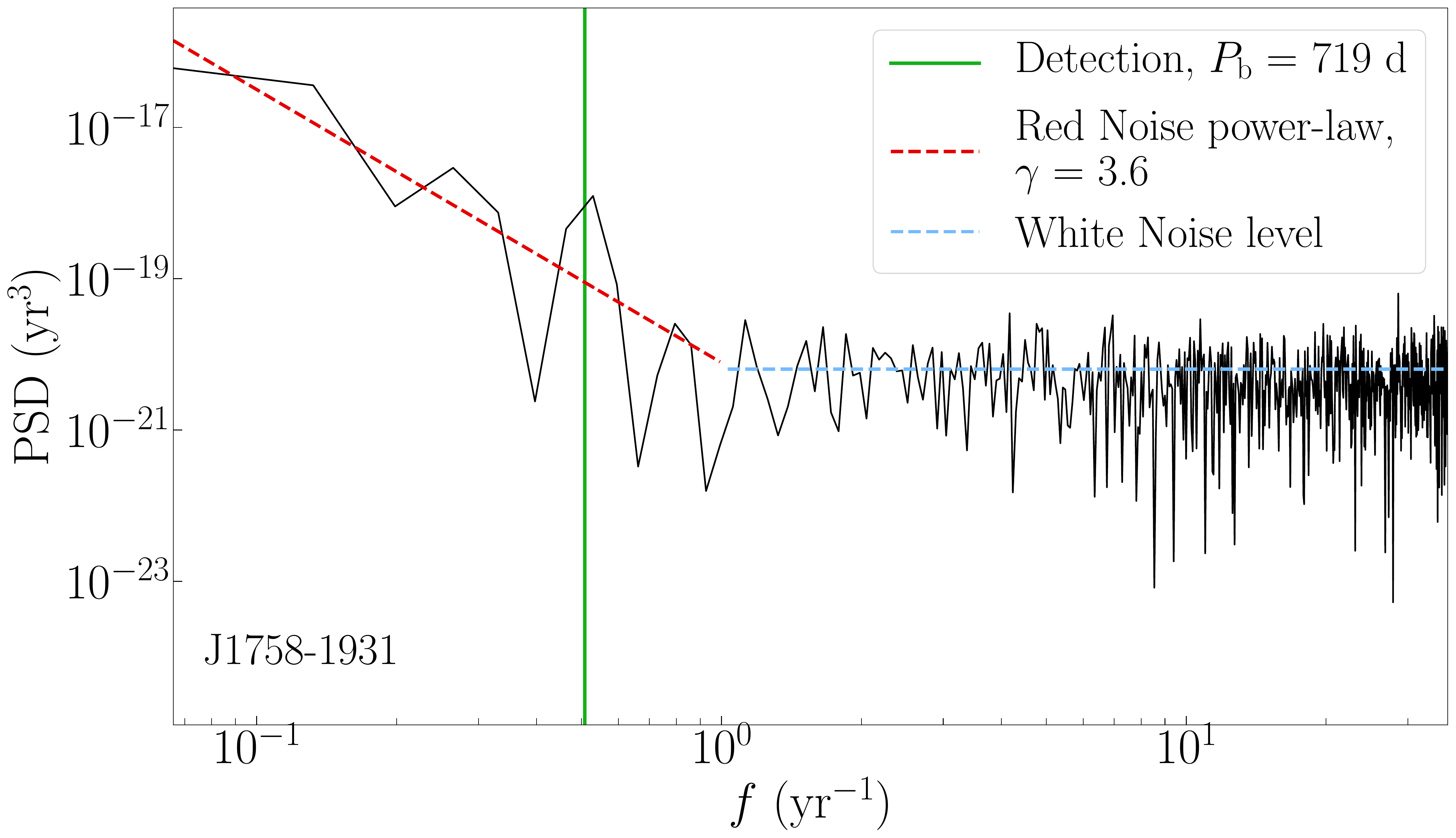}
	\includegraphics[width=.49\linewidth]{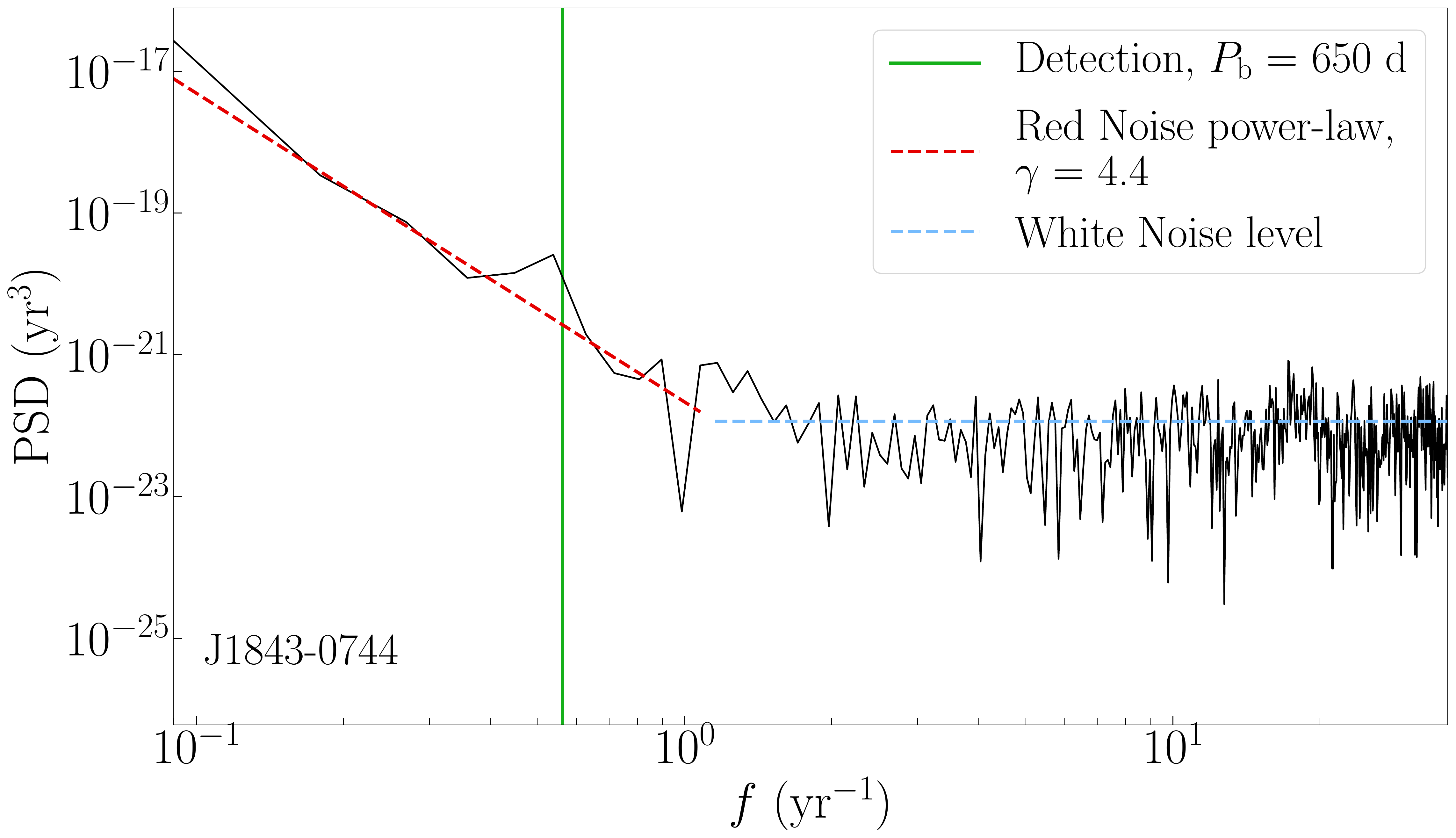}
    \\
	\includegraphics[width=.49\linewidth]{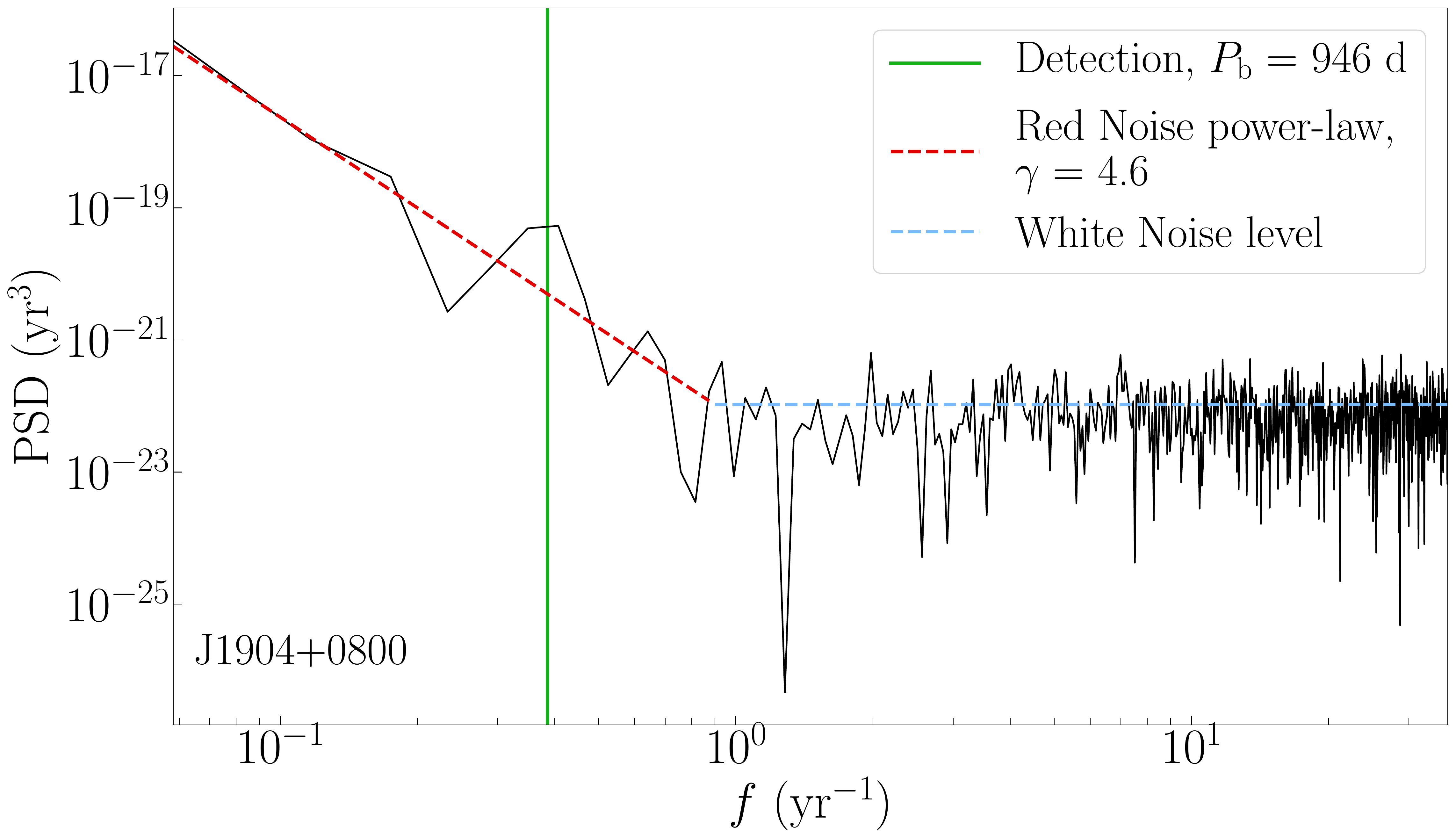}
	\includegraphics[width=.49\linewidth]{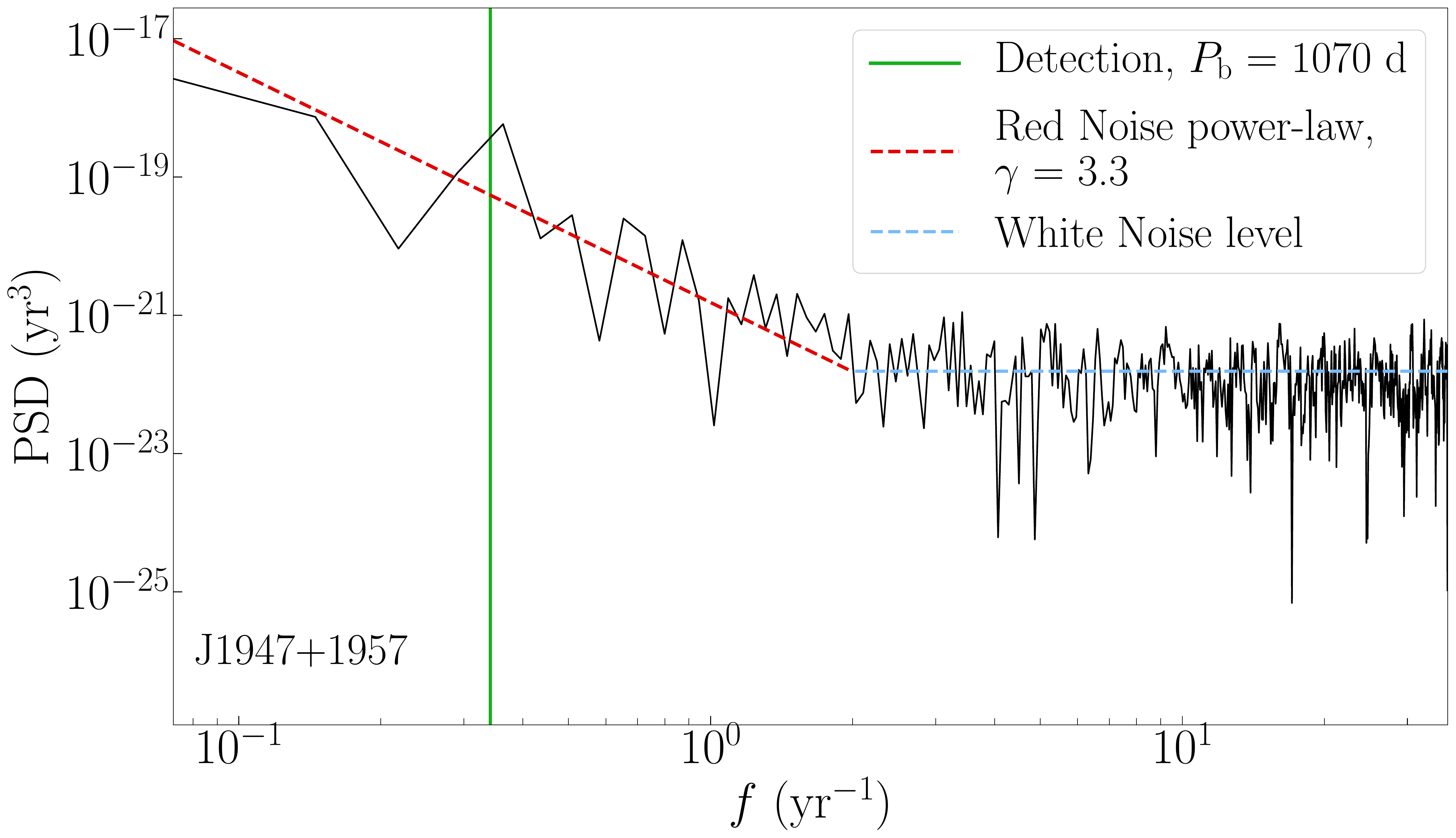}
	\\
	\includegraphics[width=.49\linewidth]{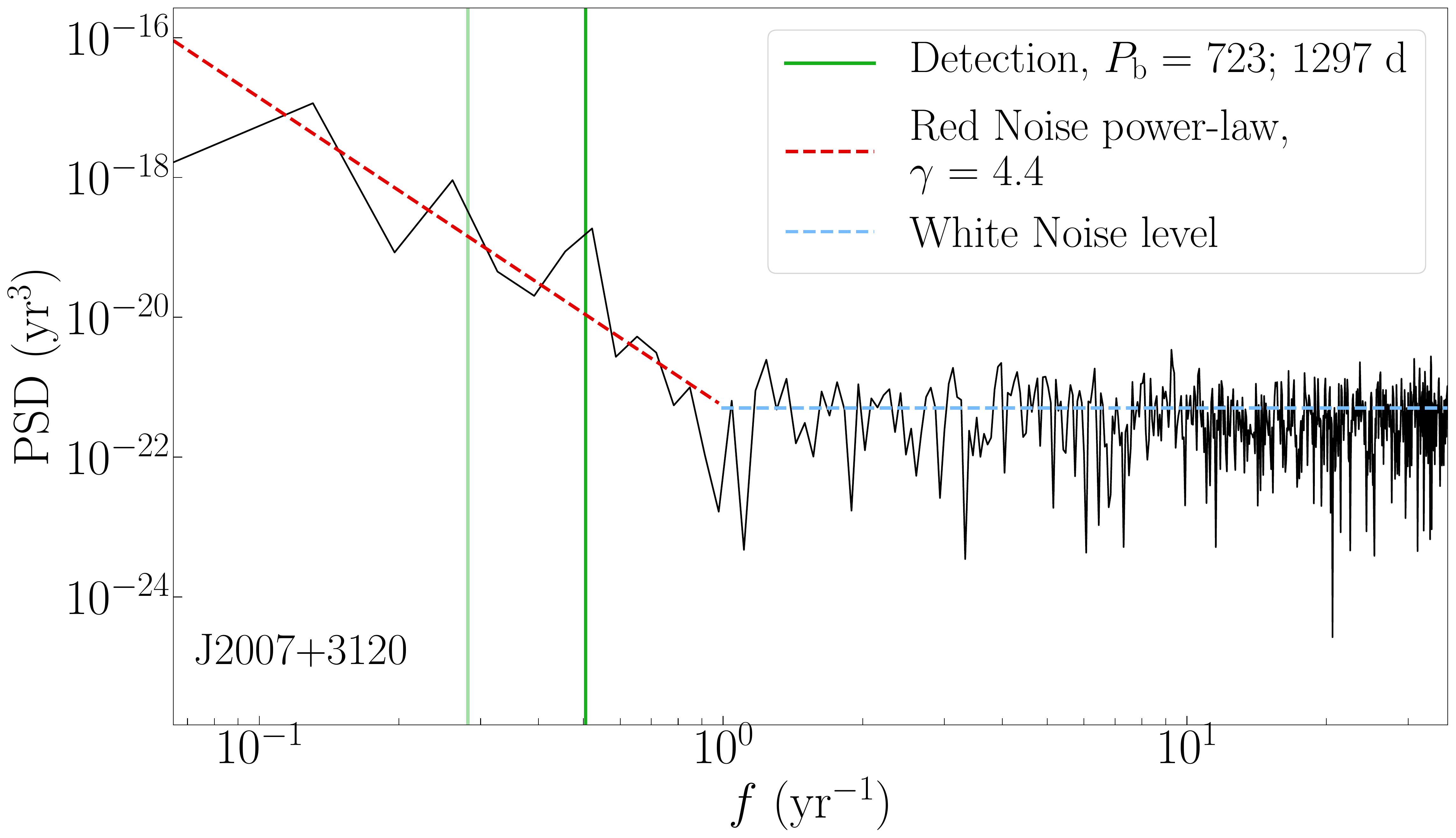}
	\includegraphics[width=.49\linewidth]{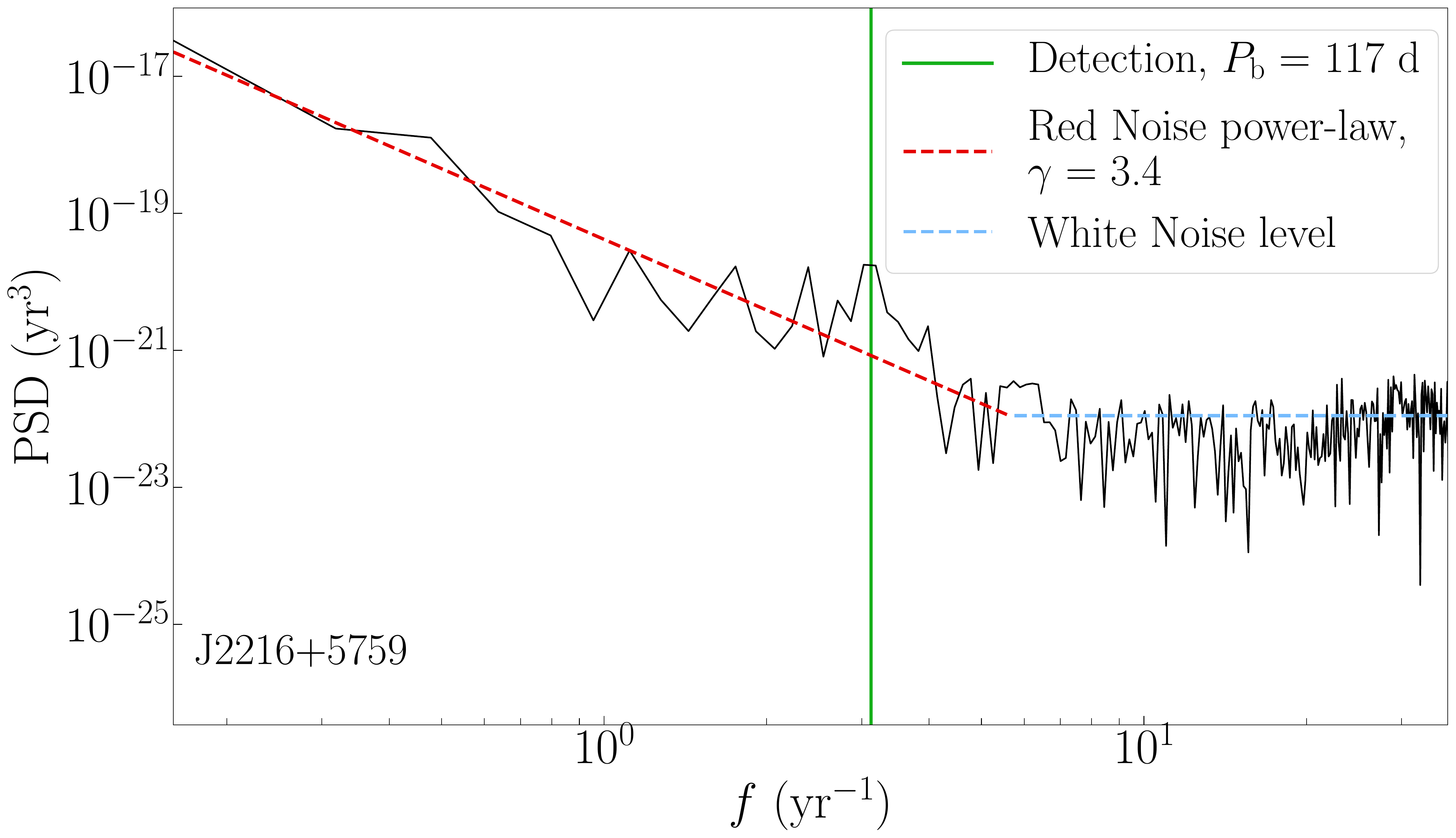}

	\caption{(contd.)}
\end{figure*}


\bsp	
\label{lastpage}
\end{document}